\documentclass[a4paper,twocolumn,11pt,accepted=2025-05-01]{quantumarticle}
\pdfoutput=1
\usepackage[utf8]{inputenc}
\usepackage[english]{babel}
\usepackage[T1]{fontenc}
\usepackage[bbgreekl]{mathbbol}
\usepackage{amsmath,dsfont,mathtools,amssymb}
\usepackage{stmaryrd}
\usepackage{amsthm}
\usepackage{physics}
\usepackage{amsfonts}
\usepackage{bm}
\usepackage{enumitem}
\usepackage{chngcntr}

\usepackage{titletoc}
\newcommand\DoToC{%
  \startcontents
  \printcontents{}{1}{\textbf{Contents}\vskip3pt\hrule\vskip5pt}
  \vskip3pt\hrule\vskip5pt
}

\usepackage{csquotes}
\MakeOuterQuote{"}

\usepackage{tabularx}
\usepackage{algorithm}
\usepackage[noend]{algpseudocode}

\usepackage{thmtools}
\usepackage{thm-restate}

\usepackage{hyperref}
\usepackage[capitalise]{cleveref}
\usepackage{xcolor}
\definecolor{quantumviolet}{HTML}{53257F}
\definecolor{navy}{RGB}{47,60,126}

\hypersetup{
    colorlinks=true,
    linkcolor=navy,
    citecolor=quantumviolet,
    urlcolor=quantumviolet,
    breaklinks=true,
}

\usepackage{tikz}
\usepackage{lipsum}

\declaretheorem[name=Theorem]{thm}
\declaretheorem[name=Lemma]{lem}
\declaretheorem[name = Corollary]{cor}
\declaretheorem[name = Conjecture]{conj}
\declaretheorem[name = Definition]{definition}

\makeatletter
\newcommand{\IfRestatedTF}[2]{\ifthmt@thisistheone #2\else #1\fi}
\makeatother

\begin{document}

\title{Trainability and Expressivity of Hamming-Weight Preserving Quantum Circuits for Machine Learning}

\author{Léo Monbroussou}
\email{leo.monbroussou@lip6.fr}
\affiliation{Laboratoire d’Informatique de Paris 6, CNRS, Sorbonne Université, 4 Place Jussieu, 75005 Paris, France}
\affiliation{CEMIS, Direction Technique, Naval Group, 83190 Ollioules, France}
\author{Eliott Z. Mamon}
\affiliation{Laboratoire d’Informatique de Paris 6, CNRS, Sorbonne Université, 4 Place Jussieu, 75005 Paris, France}
\affiliation{School of Informatics, University of Edinburgh, 10 Crichton Street, Edinburgh, United Kingdom}
\author{Jonas Landman}
\affiliation{School of Informatics, University of Edinburgh, 10 Crichton Street, Edinburgh, United Kingdom}
\affiliation{QC Ware, Palo Alto, USA and Paris, France}
\author{Alex B. Grilo}
\affiliation{Laboratoire d’Informatique de Paris 6, CNRS, Sorbonne Université, 4 Place Jussieu, 75005 Paris, France}
\author{Romain Kukla}
\affiliation{CEMIS, Direction Technique, Naval Group, 83190 Ollioules, France}
\author{Elham Kashefi}
\affiliation{Laboratoire d’Informatique de Paris 6, CNRS, Sorbonne Université, 4 Place Jussieu, 75005 Paris, France}
\affiliation{School of Informatics, University of Edinburgh, 10 Crichton Street, Edinburgh, United Kingdom}
\maketitle

\begin{abstract}  
Quantum machine learning (QML) has become a promising area for real world applications of quantum computers, but near-term methods and their scalability are still important research topics. In this context, we analyze the trainability and controllability of specific Hamming weight preserving variational quantum circuits (VQCs).
These circuits use qubit gates that preserve subspaces of the Hilbert space, spanned by basis states with fixed Hamming weight $k$.

In this work, we first design and prove the feasibility of new heuristic data loaders, performing quantum amplitude encoding of $\binom{n}{k}$-dimensional vectors by training an $n$-qubit quantum circuit. These data loaders are obtained using controllability arguments, by checking the Quantum Fisher Information Matrix (QFIM)'s rank. Second, we provide a theoretical justification for the fact that the rank of the QFIM of any VQC state is almost-everywhere constant, which is of separate interest. Lastly, we analyze the trainability of Hamming weight preserving circuits, and show that the variance of the $l_2$ cost function gradient is bounded according to the dimension $\binom{n}{k}$ of the subspace. This proves conditions of existence/lack of Barren Plateaus for these circuits, and highlights a setting where a recent conjecture on the link between controllability and trainability of variational quantum circuits does not apply.
\end{abstract}

\section{Introduction}

Variational quantum circuits (VQCs) are promising candidates for near term quantum computing \cite{Cerezo2020}, but existing and near-term quantum devices still offer limited resources to implement important tasks for quantum machine learning (QML), such as encoding and training. Considering fault-tolerant quantum computation, more advanced QML algorithms that present potential to achieve a quantum advantage exist. The key step of some of these "quantum linear algebra" algorithms~\cite{Harrow2008, Kerenidis2020, Lloyd2013} is amplitude encoding, where the input vector's components become the quantum amplitudes of the input state in the computational basis. Amplitude encoding on the entire Hilbert space is unlikely to be achieved in the near term, and recent work proposed to use amplitude encoding in small subspaces \cite{Johri2020, Landman2022, Cherrat2022}. In other methods for variational QML, the data encoding is done by putting directly the vector component as the gate parameters, which leads to some limitations \cite{Schuld2020, RFF_LIP6}. 

\begin{figure}[h]
    \centering
    \includegraphics[width=0.48\textwidth]{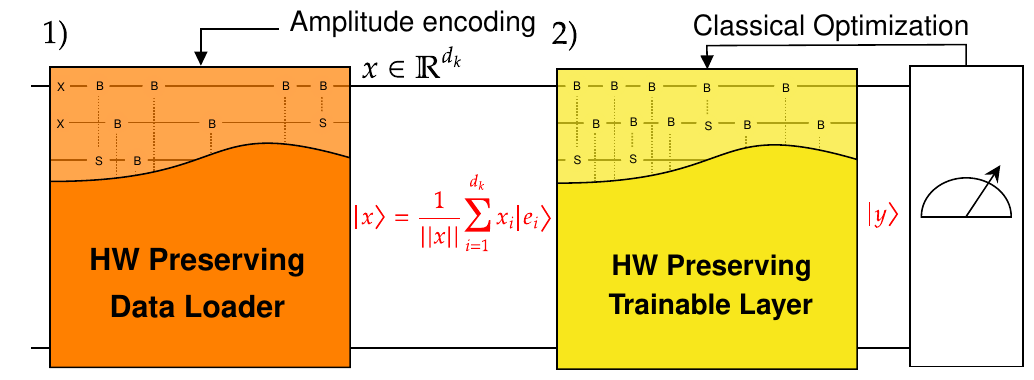}
    \caption{HW preserving quantum circuit for QML: (1) is the encoding part trained to represent the classical input, and (2) is the trainable layer or quantum neural network. The gates represented with B and S are RBS.}
    \label{fig:figure_intro}
\end{figure}

In this work, we propose a method to achieve the amplitude encoding of any $\binom{n}{k}$-dimensional vector using $n$ qubits by training a VQC made of Hamming weight (HW) preserving gates. In recent work \cite{Johri2020}, a $n$-qubit quantum data loader using HW-preserving gates was presented to encode any $n$-dimensional vector using a VQC, but without training. HW-preserving quantum circuits allow one to restrict the state created to a superposition of states of the same HW as the input, i.e., to maintain the number of qubits in state $\ket{1}$ at the same time. Such subspace invariant quantum circuits can tackle an important scaling problem with QML methods called Barren Plateaus (BPs) \cite{McClean2018}. It has been recently shown that one can avoid BPs while using input data on an invariant subspace of low dimension under certain conditions on their controllability and expressivity \cite{Larocca2021}. However, knowing if HW-preserving VQCs are prone to BPs without those assumptions is still an open question that we tackle in this work.

One of our main results can be informally stated as follows: for a $n$-qubit circuit made of specific HW-preserving gates, namely Reconfigurable Beam Splitter (RBS) and Fermionic Beam Splitter (FBS), in the subspace spanned by Hamming weight $k$, the gradient of the cost function vanishes as $O(1/\binom{n}{k})$. This is neither exponentially decreasing with $n$ (Barren Plateau), nor quadratically decreasing as one could have expected for the FBS case, despite its lower controllability.

We first develop in Section~\ref{chap:Encoding} a framework to design a quantum data loader on any invariant subspace of dimension $\binom{n}{k}$, with a $n$-qubit circuit and $k$ the chosen HW (see Fig.~\ref{fig:figure_intro}). {Those quantum data loaders are classicaly simulable, and can be trained classically or with a quantum computer with a speed-up advantage restricted to a polynomial one.} In addition, we propose a study of the trainability for HW-preserving VQCs in Section~\ref{chap:trainability_RBS_circuit} (this concerns both blocks in Fig.~\ref{fig:figure_intro}). We show that one could avoid BPs using HW-preserving gates according to the choice of the subspace used for the encoding without any assumption on the controllability or expressivity of the VQC. 

\subsection*{Related work}

The preservation of the HW is a symmetry that we use in this work in order to propose encoding and trainable layers with theoretical guarantees on their trainability. Quantum machine learning models that present symmetries have been proposed as potential more efficiently trainable than common models \cite{Larocca2022, Meyer2022, Cohen2016,You2022, Anschuetz2022_2}. 
More recently, problem-inspired ansatz have been study using tools from quantum optimal control to highlight a link between the dimension of their corresponding dynamical Lie algebra and their trainability \cite{Larocca2021} using a certain set of assumptions. 
One main conjecture left open in \cite{Larocca2021} concerned the link between trainability and controllability \emph{in a subspace}.
Recent works just prove that this conjecture applies to many commonly used ansatzes \cite{Ragone2023, Fontana2023} such as the Hamiltonian Variational Ansatz \cite{Wecker2015}, Quantum Alternating Operator Ansatz \cite{Farhi2014,Hadfield2017}, and many equivariant quantum neural networks. 
However, in our work, we show that this conjecture is not respected in the case of HW-preserving ansatzes, which received increasing attention \cite{Johri2020, Landman2022, Cherrat2022, Kerenidis2022, monbroussou2024, raj2025}. Our results are consistent and independent of the two works recently released \cite{Ragone2023, Fontana2023} as HW-preserving ansatzes don't respect the assumptions these papers are relying on.
One can notice that \cite{Fontana2023} studies the same ansatzes as we do (see their Appendix C), and propose an upper-bound on their controllability in a specific setting. In addition, \cite{diaz2023showcasing} provided some more theory around the scenario that appears to evade the DLA-based framework. In Section~\ref{chap:trainability_RBS_circuit}, we give tighter theoretical guarantees on the trainability of those ansatzes, using a different analytical proof. See Section \ref{sec:discussion} for discussion.

\section{Space-efficient amplitude encoding.}
\label{chap:Encoding}

We first define the Amplitude Encoding and HW-preserving quantum data loaders. Then we show how to achieve it efficiently using HW-preserving gates and their subspace preserving properties. {In this section, we study the controllability of HW-preserving quantum circuits through the lens of designing an amplitude encoder. However, since our data loader is classically simulable, the potential quantum speed-up for training is limited to a polynomial advantage. Nevertheless, the controllability of HW-preserving quantum circuits—analyzed through their DLA dimension and QFIM rank—is of independent interest.}

\subsection{Hamming weight preserving quantum data loaders}\label{chap:HW_QDL}

Let us first define the Amplitude Encoding scheme and explain what are HW-preserving quantum data loaders. 

\begin{definition}[Amplitude Encoding]\label{def:amplitude_encoding}
    An amplitude encoding data loader is a parametrized $n$-qubit quantum circuit that, given a classical vector $x = (x_1, \dots, x_{d}) \in \mathbb{R}^d$ (of a certain fixed length $d \leq 2^n$), prepares the quantum state:
    \begin{equation}
        \ket{x} = \frac{1}{||x||} \sum_{i=1}^{d} x_i \ket{e_i},
    \end{equation}
    where $\left\{\ket{e_1}, \dots, \ket{e_d}\right\}$ is a fixed family of $d$ orthonormal quantum states, and $||\cdot||$ denotes the $2$-norm of $\mathbb{R}^d$.  
\end{definition}

\begin{definition}[Data Loader in a HW-preserving subspace]\label{def:HW_basis}
    We define a n-qubit data loader in the subspace of HW $k$, to be any quantum circuit that performs amplitude encoding on the subspace basis:
    \begin{equation}\label{eq:Basis_HW_k}
        B_k^n = \left\{ \ket{e} \middle| e \in \{0,1\}^n \text{ and } \text{HW} (e) = k \right\}
    \end{equation}
    with $d_k = |B_k^n| = \binom{n}{k}$.
\end{definition}

For example, when considering $n = 3$ qubits and a HW $k=2$, the basis states are:
\begin{equation*}
    B_2^3 = \left\{ \ket{110}, \ket{101}, \ket{011} \right\}
\end{equation*}

One could use the Reconfigurable Beam Splitter (RBS) gate to perform such an encoding. This HW-preserving gate is easy to implement or native on many quantum devices. Notice that our results hold for another HW-preserving gate named the Fermionic Beam Splitter (FBS) which was already used for QML application in \cite{Kerenidis2022} but has less good properties in term of controllability (see Section~\ref{chap:Existence_QDL} and Appendix~\ref{chap:properties_FBS}). 

\begin{definition}[Reconfigurable Beam Splitter gate]
    The Reconfigurable Beam Splitter (RBS) gate is a 2-qubit gate that corresponds to a $\theta$-planar rotation between the states $\ket{01}$ and $\ket{10}$:
    \begin{equation}\label{eq:RBS_2_qubit_gate}
    RBS(\theta) = e^{i \theta H_{RBS}} = \begin{pmatrix}
        1 & 0 & 0 & 0 \\
        0 & \cos(\theta) & \sin(\theta) & 0 \\
        0 & -\sin(\theta) & \cos(\theta) & 0 \\
        0 & 0 & 0 & 1 \\
        \end{pmatrix} \, ,
    \end{equation}
    with its corresponding Hamiltonian:
    \begin{equation}
        H_{RBS} = \begin{pmatrix} 
        0 & 0 & 0 & 0 \\
        0 & 0 & -i & 0 \\
        0 & i & 0 & 0 \\
        0 & 0 & 0 & 0
        \end{pmatrix}\, . 
    \end{equation}
    
\end{definition}

\begin{definition}[Fermionic Beam Splitter]\label{def:FBS}
    Let $i,j \in [n]$ be qubits and $S = s_1 \dots s_n \in \{0,1\}^n$ a binary word corresponding to a basis state of fixed HW $\ket{S}$ with $n$ the total number of qubits. Then the Fermionic Beam Splitter (FBS) acts on the qubits $i$ and $j$ as the following unitary:
    \begin{equation}\label{eq:RBF_2_qubit_gate}
        \begin{pmatrix}
            1 & 0 & 0 & 0 \\
            0 & \cos(\theta) & (-1)^{f}\sin(\theta) & 0 \\
            0 & (-1)^{f+1}\sin(\theta) & \cos(\theta) & 0 \\
            0 & 0 & 0 & 1 \\
        \end{pmatrix}\, ,
    \end{equation}
    with $f = f_{i,j,S} = \sum_{i<l<j} s_l$
\end{definition}

The Hilbert space associated with a subspace preserving VQC decomposes into a direct sum of invariant subspaces. For HW preserving VQCs, a mere rearrangement of the order of the computational basis leads to the corresponding block-diagonal form for their unitary matrices. Since the global matrix $W$ is orthogonal, each block matrix $W^k$ (corresponding to the action of $W$ in the subspace of HW $k$) is an orthogonal matrix as well.

\begin{figure}[H]
    \centering
    \includegraphics[width=0.52\textwidth]{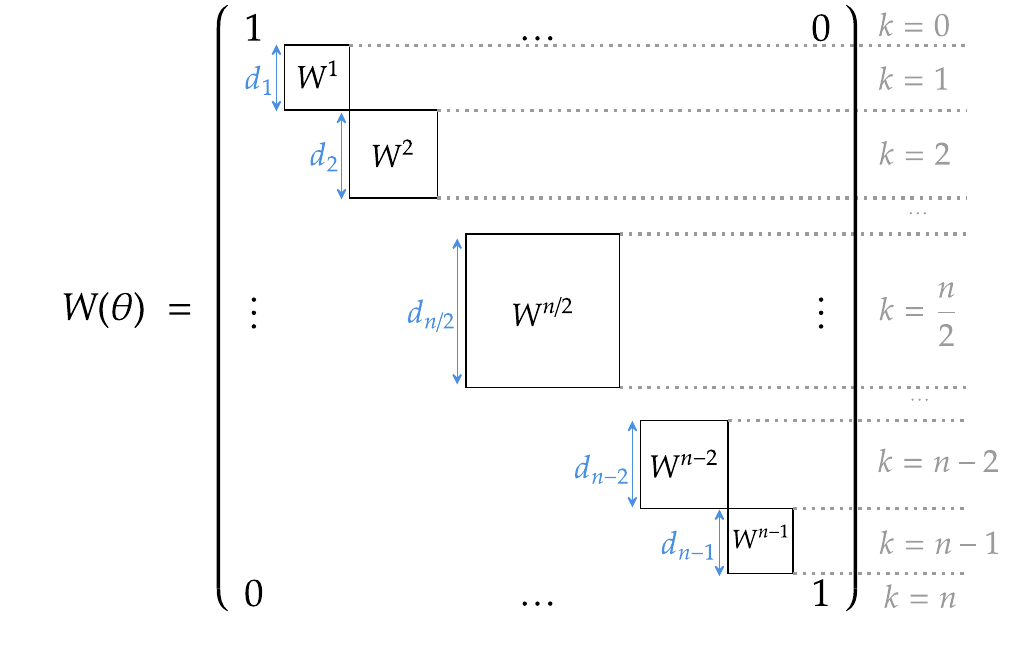}
    \caption{Block representation of the HW-preserving unitaries. $W$ is the $2^n \times 2^n$ unitary corresponding to a n-qubit HW-preserving quantum circuit. Each block $k$ is the unitary matrix corresponding to the preserved subspace of HW $k$, and the state basis $B_k^n$. Their size are $d_k\times d_k$ where $d_k = \binom{n}{k}$.}
    \label{fig:RBS_circuit_block_unitary}
\end{figure}

We now explain our data loading scheme. First, we initialize the quantum state to be $\ket{e_s}$, a basis state of HW $k$. Then we split off this state onto the states in $B_k^n$ using RBS. In \cite{Johri2020}, the authors used a similar method on the unary basis $B_1^n$. Notice that achieving an amplitude encoding with such a basis would allow us to encode many more parameters, namely $\binom{n}{k} \gg n$, in an $n$ qubit state.
To design our quantum data loader, we need to ensure that any $\binom{n}{k}$-dimensional real vector $x$ can be encoded, i.e., that there exists a corresponding set of RBS gate parameters $\bm{\theta} = \{\theta_1, \dots, \theta_D\}$ (depending on $x$) such that:
\begin{equation}\label{eq:encoding_particular_equivalent_eq}
    W^{k}(\bm{\theta})\ket{e_{s}} - \frac{1}{||x||}\sum_{i=1}^{\binom{n}{k}} x_i \ket{e_i} = 0 \, ,
\end{equation}
Finding the corresponding set of variational parameters or proving their existence is generally very hard when $k > 1$, see Appendix~\ref{chap:QDL_equation}) for more discussion. 
In this work, we focus rather on the existence of an \textit{approximate} solution, to the following related optimization problem,
\begin{equation}\label{eq:encoding_optimization_pb}
    \bm{\theta}^\ast = \arg \min_{\bm{\theta}} ||\frac{1}{||x||}\sum_{i=1}^{\binom{n}{k}} x_i \ket{e_i} - W^k(\bm{\theta})\ket{e_{s}}||_2^2\, ,
\end{equation}
which can be addressed using gradient-based optimizers. Theoretical arguments on the amenability of the loss of Eq.~\eqref{eq:encoding_optimization_pb} to gradient-based optimizers are provided later in Section~\ref{chap:trainability_RBS_circuit}. Namely, this cost function does not induce a Barren Plateaus even though it is a type of global cost function \cite{CerezoSone2020}, as the Hilbert space corresponding to the state of HW $k$ is not exponentially large for small choice of $k$. We confirm in Section~\ref{chap:trainability_RBS_circuit} that a large choice of $k$ results in the existence of Barren Plateaus.

Subspace preserving quantum circuits are easier to simulate in small subspaces than random quantum circuits over the entire Hilbert space \cite{Anschuetz2022}. In the case of a HW-preserving VQC, the speedup of using a quantum computer grows with $k$. Classical simulability of the encoding part itself is not an issue, if it is later combined with a trainable layer that is hard to simulate.

    \subsection{Existence of Quantum Data Loader} \label{chap:Existence_QDL}

Previously, we proposed the use of HW-preserving ansatz such as a n-qubit RBS based VQC to create a quantum data loader. This circuit will be trained to perform the amplitude encoding of a $\binom{n}{k}$-dimensional vector with $k$ the chosen HW. In this Section, we give tools to study the existence of a circuit that always presents a solution for our optimization problem given by Eq.~\eqref{eq:encoding_optimization_pb}. More precisely, we show how to know if, given a qubit connectivity and RBS gates, one can design a quantum data loader. If so, we give a method to design the circuit in Section~\ref{chap:Finding_QDL}.

\begin{figure}[h!]
    \centering
    \includegraphics[width=0.4\textwidth]{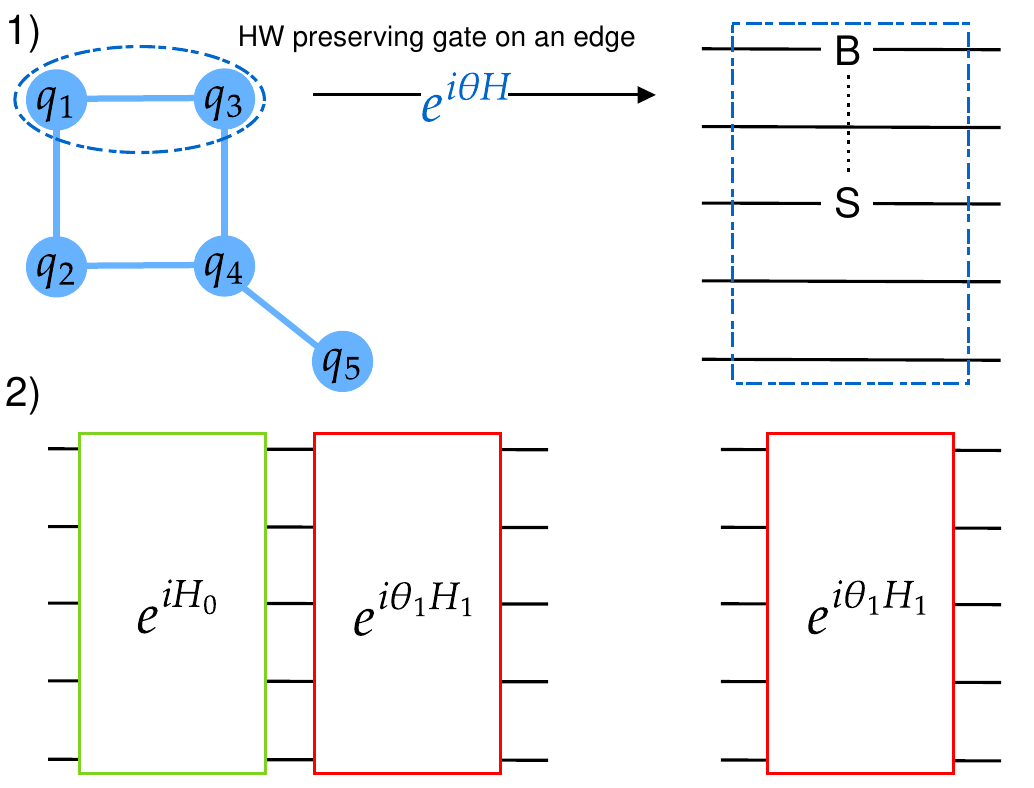}
    \caption{Hamiltonian representation of a HW-preserving VQC: (1) highlights the link between the qubit connectivity and the generators of the circuit, and (2) represents our quantum data loader. The bit flips used to prepare the initial state are represented in green, and RBS are represented red.}
    \label{fig:Hamiltonian_system_expression}
\end{figure}

We propose to use quantum optimal control tools to show that, according to the circuit connectivity, we can prove the existence of a quantum data loader. This method can be generalized to find quantum data loaders for any subspace invariant ansatz. We first define an essential tool to study the controllability of a quantum circuit in the unitary space.

\begin{definition}[Dynamical Lie Algebra]
    Let us consider a circuit made from controllable Hamiltonian matrices in a set $\{ H_k \}_{k \in \llbracket 0, G \rrbracket}$ that we call the set of generators of our quantum system. The Dynamical Lie algebra is defined as:
    \begin{equation}
        \mathfrak{g} = \text{span} <iH_0, \dots, iH_G>_{Lie} \subseteq su(d)\, ,
    \end{equation}
    with $\left< S \right>_{Lie}$ the Lie closure, i.e., the set of all nested Lie commutators between the elements in $S$.
\end{definition}

In the case of RBS based quantum circuits, the generators are given by the qubit connectivity as we restrict ourselves to the use of a unique 2-qubit gate (see Fig.~\ref{fig:Hamiltonian_system_expression}). We show in Appendix~\ref{chap:DLA} how to compute the dimension of the DLA. We can restrict this study of the DLA to a particular subspace of HW $k$. Then, its dimension indicates 

\newpage

\onecolumngrid

\begin{figure}[t]
    \centering
    \includegraphics[width=1.0\textwidth]{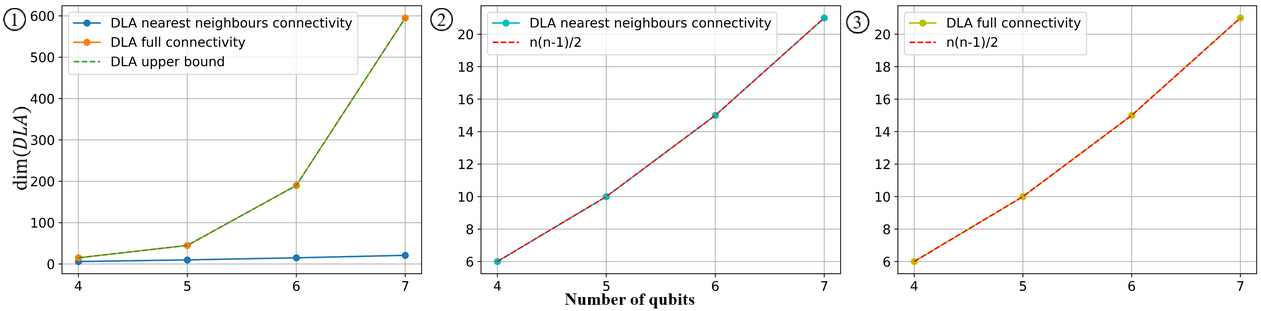}
    \caption{Evolution of the dimension of the DLA in the subspace of HW $k = \lfloor \frac{n}{2} \rfloor$ for: (1) the use of RBS gates; (2) the use of FBS gates with nearest neighbors connectivity; (3) the use of FBS gates with full connectivity. This plot highlights the difference of controllability potential between RBS and FBS based quantum circuit.}
    \label{fig:DLA_evolution_connectivity}
\end{figure}

\twocolumngrid

the maximal number of coefficients we can independently fix in $W^k$. As this matrix is orthogonal, the dimension of the DLA is upper bounded by $\frac{1}{2}d_k(d_k - 1)$. We can thus compute the DLA corresponding to our circuit in the subspace of the chosen HW $k$.

If the dimension of the DLA in the subspace of HW $k$ is lower than $d_k - 1$ (with $d_k = \binom{n}{k}$), we know we cannot control enough coefficients to achieve the encoding described in Eq.~\eqref{eq:encoding_particular_equivalent_eq}.

On the other hand, if the dimension is maximal (equal to $\frac{1}{2}d_k(d_k-1)$), it indicates, provided that the circuit were deep enough to reach all of its Dynamical Lie Group, that each independent coefficient of the orthogonal matrix $W^k$, and hence one of its columns, can be controlled to some degree (over some non-empty range); thus providing support in favor of the solvability of Eqs.~\eqref{eq:encoding_particular_equivalent_eq} and \eqref{eq:encoding_optimization_pb} for any $x$.

However, if the DLA dimension is between those two values, we cannot argue in the above manner about the existence of the data loader , as we may control at least $d_k - 1$ coefficients but not those in the columns corresponding to $\ket{e_s}$.

In practice, one could either choose to reduce the choice of $k$ in hopes of achieving the full controllability for a smaller subspace, or alternatively, proceed with our method to find the data loader given in Section~\ref{chap:Finding_QDL}. Indeed, one can use the rank of the Quantum Fisher Information matrix (QFIM) to ensure this sufficient controllability in the state space (see Section~\ref{chap:Finding_QDL}).

We know that having enough control to design a quantum data loader requires considering a connected graph in order to reach any state in our encoding basis. Therefore, the nearest neighbor's connectivity is the worst as it has a minimal number of edges, and full connectivity maximizes the dimension of the DLA. The scaling of the DLA dimension according to a specific type of graph needs to be tackled in future work, but Fig.~\ref{fig:DLA_evolution_connectivity} gives numerical evidence of good scaling in terms of controllability for an RBS based quantum circuit.

We observe in the Fig.~\ref{fig:DLA_evolution_connectivity} that the evolution of the DLA dimension in the nearest neighbors connectivity setting seems to follow the maximum controllability of the first subspace, which is given by $\frac{1}{2}n(n-1)$. The DLA dimension for the maximum connectivity seems to evolve according to the upper bound given by the maximum controllability of a $d_{n/2} \times d_{n/2}$ orthogonal matrix and equal to $\frac{1}{2}d_{n/2}(d_{n/2}-1)$ when considering RBS gates.  In the case of the FBS gates, we know that each block $W^k$ is perfectly determined by the first one $W^1$ \cite{Kerenidis2022} (see Appendix~\ref{chap:properties_FBS}). As a result, the dimension of the DLA of FBS based quantum circuit is upper-bounded by  $\frac{1}{2} n(n-1)$ as observed in the previous figure. The limitation of the controllability of FBS gates is reminded in Appendix~\ref{chap:properties_FBS}. 

In this section we show that the DLA study gives us tools to know if we can design a quantum data loader from a qubit connectivity and a specific quantum gate. 

\subsection{Finding the quantum data loader} \label{chap:Finding_QDL}

Using previous results, we know that if we have a quantum hardware such that the DLA dimension is high enough, there exists a quantum data loader circuit. A remaining question is how to design the quantum data loader from a given subspace and the connectivity that induces the existence of such a circuit. In this Section, we will present two algorithms to design the quantum data loader based on the study of controllability in the state space.

\begin{figure}[h!]
    \centering
    \includegraphics[width=0.5\textwidth]{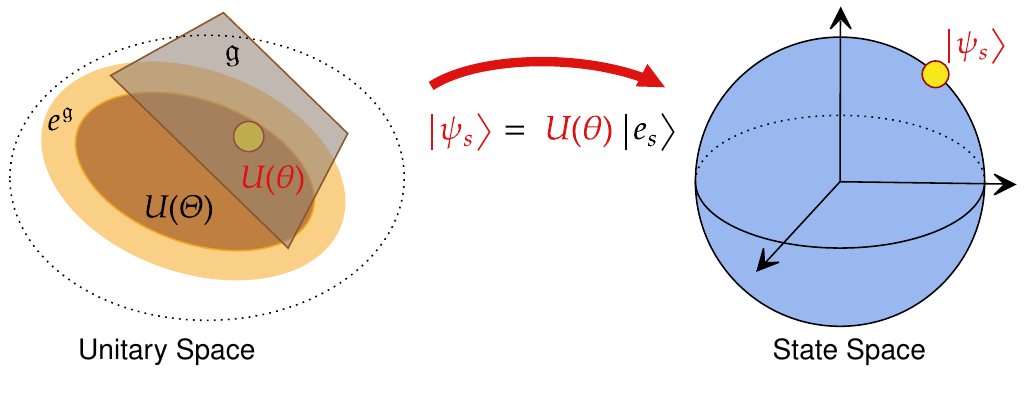}
    \caption{Representation of the unitary and output state spaces. The DLA is the tangent space of the unitary space. The possible directions for the evolution of the output state are given by the Quantum Fisher Information Matrix eigenvectors.}
    \label{fig:relevant_mathematical_spaces_for_QML}
\end{figure}

A subspace preserving circuit's ability to achieve amplitude encoding on one of its preserved subspaces is equivalent to this circuit perfectly controlling the state space created by its output. In particular, an RBS based VQC would achieve perfect amplitude encoding (see Definition~\ref{def:amplitude_encoding}) on the subspace of HW $k$ if its output state could be any (real, normalized) superposition of states in $B_k^n$; that is, if the space of kets that its output explores were the entirety of a certain sphere of (real) dimension $d_k - 1$, noted $S^{d_k - 1}$ and illustrated in Fig.~\ref{fig:relevant_mathematical_spaces_for_QML}. 

Now we define an essential tool to study the controllability of a quantum circuit in the state space.

\begin{restatable}[Quantum Fisher Information Matrix]{definition}{QFIMdefinition}
\label{def:QFIM}

The \textit{Quantum Fisher Information Matrix} (QFIM) associated to any parametrized pure state $\ket{\psi(\bm{\theta})}$ that uses $p$ continuous parameters $\bm{\theta}=(\theta_1,\dots,\theta_p)$, is the following $p\times p$ real matrix assigned to each parameter vector $\bm{\theta}$:
    \begin{equation}\label{eq:QFIM}
    \IfRestatedTF{
        [\mathrm{QFIM}(\bm{\theta})]_{i,j} = 4\,\mathrm{Re}\big[\, \big\langle \partial_{\theta_i} \psi(\bm{\theta}) \big| \partial_{\theta_j} \psi(\bm{\theta}) \big\rangle\,-\,\big\langle \partial_{\theta_i} \psi(\bm{\theta}) \big| \psi(\bm{\theta}) \big\rangle \big\langle \psi(\bm{\theta}) \big|  \partial_{\theta_j}\psi(\bm{\theta}) \big\rangle \,\big]\,.}
        {
        \begin{aligned}
           [\mathrm{QFIM}(\bm{\theta})]_{i,j} &= 4\,\mathrm{Re}\big[\, \big\langle \partial_{\theta_i} \psi(\bm{\theta}) \big| \partial_{\theta_j} \psi(\bm{\theta}) \big\rangle\,-\\
           & \big\langle \partial_{\theta_i} \psi(\bm{\theta}) \big| \psi(\bm{\theta}) \big\rangle \big\langle \psi(\bm{\theta}) \big|  \partial_{\theta_j}\psi(\bm{\theta}) \big\rangle \,\big]\,.
        \end{aligned}
        }
    \end{equation}
\end{restatable}

Given a basis state $\ket{e_s}$ of HW $k$ as the input state to our RBS/FBS circuit $U(\bm{\theta})$, calculating the QFIM of the output state $\ket{\psi_s(\bm{\theta})} := U(\bm{\theta}) \ket{e_s}$ has a computational cost which depends on the subspace dimension $d_k$ considered. Each state $\ket{\psi_s(\bm{\theta})}$ and $\ket{\partial_{\theta_i} \psi_s(\bm{\theta})}$ can be simulated as vector of dimension $d_k$, and the overall computational cost of calculating the matrix $\mathrm{QFIM}_s(\bm{\theta})$ is $\mathcal{O}(p^2 \, d_k^2$).

The maximal rank (over parameter space) of the QFIM is a metric of controllability in the state space \cite{Liu2019}, as it gives us the number of independent directions that can be taken by the state when tuning the gate parameters $\bm{\theta}$. For our encoding method in the subspace of HW $k$, a consequence of the fact that the kets are constrained to belong to $S^{d_k - 1}$ is that the QFIM ranks are upper-bounded by $d_k -1$ (for any parameter values):
\begin{equation}
    \max_{\bm{\theta}} \; \rank [\mathrm{QFIM}_s(\bm{\theta})] \leq  d_k-1 .  
\end{equation}

\begin{figure}[H]
    \centering
    \includegraphics[width=0.45\textwidth]{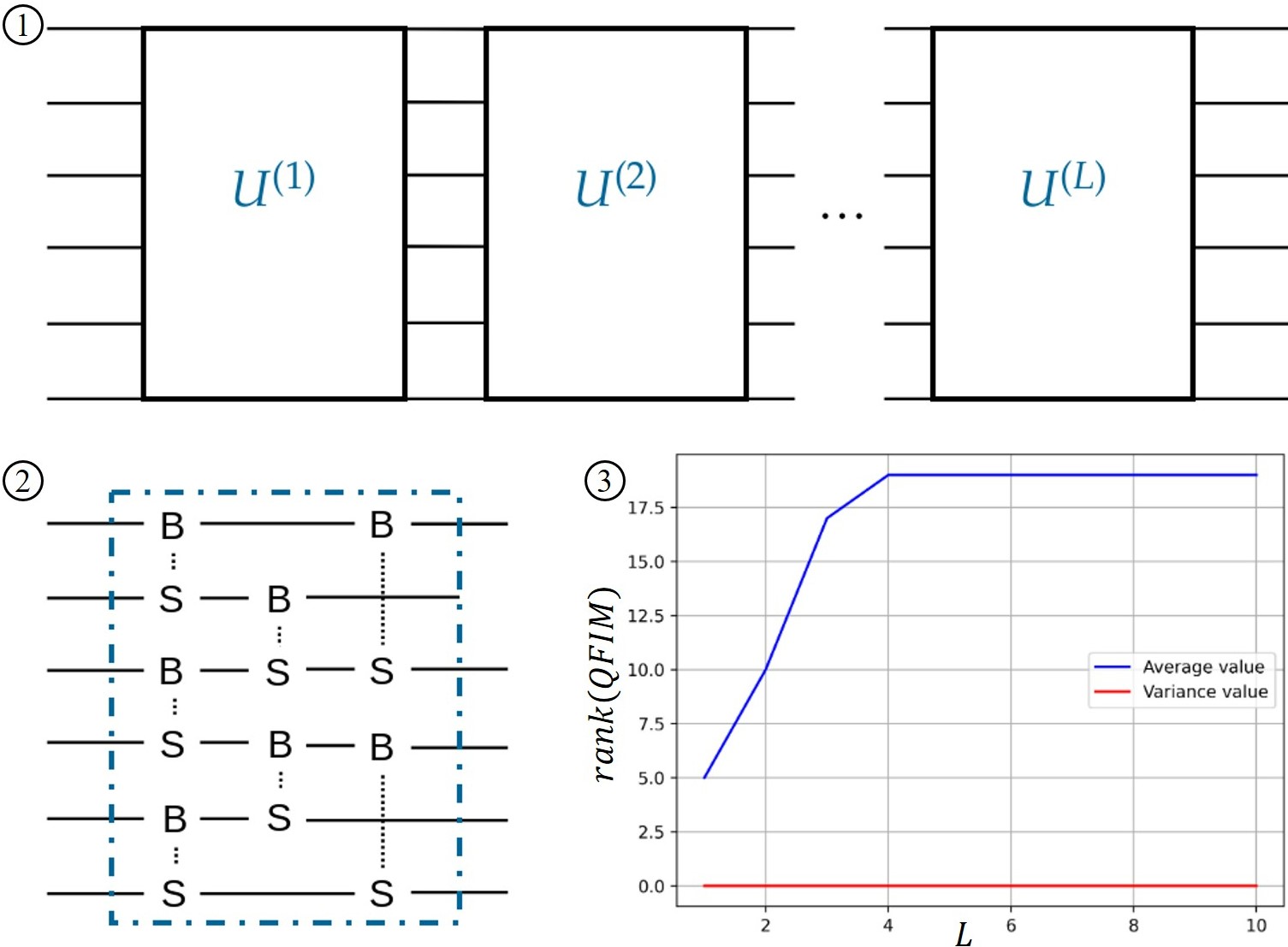}
    \caption{Evolution of the rank of the QFIM for a periodic structure ansatz presented in (1). Each block is represented in (2), and the evolution of the rank of the corresponding QFIM is given by (3). The derivation of the QFIM rank is done in the largest subspace ($k=n/2=3$).}
    \label{fig:QFIM_rank_evolution}
\end{figure}

As in \cite{Haug2021}, we find numerical evidence (see Fig.~\ref{fig:QFIM_rank_evolution}) that upon randomly sampling parameter values $\bm{\theta} \in [0,2 \pi]^p$, the value of $\rank [\mathrm{QFIM}_s(\bm{\theta})]$ is independent of $\bm{\theta}$.
In fact, this property can be justified theoretically for any standard VQC, which we state here as Theorem \ref{thm:QFIMrankthm}.

\begin{restatable}[Almost-constant property of QFIM rank]{thm}{QFIMrankthm}
\label{thm:QFIMrankthm}
A VQC's output state $\ket{\psi(\bm{\theta})}$ always has the following property: \emph{almost everywhere} on the considered parameter space $\Theta$, the rank of their QFIM is constant, equal to $r_{\mathrm{max}}:=\max\limits_{\bm{\theta} \in \bm{\theta}}r(\bm{\theta})$. Consequently, drawing a point $\bm{\theta} \in \Theta$ uniformly at random and calculating its QFIM rank value $r(\bm{\theta})$ yields $r(\bm{\theta})=r_{\mathrm{max}}$ with probability $1$.
\end{restatable}

The reason behind the validity of Theorem \ref{thm:QFIMrankthm} is attributed to the \emph{analyticity} (a strong type of smoothness) of the functions at play in VQCs, as well as to the nature of the QFIM. We provide more detail and prove this result in Appendix~\ref{chap:proof_QFIMrankthm} (c.f. Lemma \ref{lemma:rankQFIM-VQC}).

Using the QFIM on a given subspace of HW $k$, we propose a first algorithm to design a quantum data loader in this subspace from an initial state created using bit-flips, and the possible generators $\mathcal{G}$ given by the qubit connectivity and the RBS gate Hamiltonian. When the maximal rank (over parameter space) of the QFIM of a quantum data loader circuit is equal to $\dim(S^{d_{k}-1}) = d_k - 1$, we take it as evidence that it may achieve any state in $S^{d_{k}-1}$, i.e., achieve the amplitude encoding on the subspace of HW $k$. Following this idea, Algorithm~\ref{alg:QDL_1} creates iteratively a circuit by adding RBS gates one at a time, while making sure that each added gate has actually incremented the QFIM rank, and only stops when the QFIM rank attains the dimension of the sphere $S^{d_{k}-1}$, suggesting that the data loader capability has been obtained.

\begin{algorithm}[H]
\caption{to design a HW-preserving quantum data loader}\label{alg:QDL_1}
\begin{algorithmic}[1]
\Require $\mathcal{G}$ the generators, $\ket{e_s}$ the initial state 
\State circuit = $\emptyset$
\While{$(\max_{\bm{\theta}} \rank[\mathrm{QFIM}(\text{circuit}, \bm{\theta})] < d_k - 1)$}
    \For{$RBS \in \mathcal{G}$}
        \State $\text{circuit'} = \text{circuit} + RBS$
        \If{($\max_{\bm{\theta}'} \rank [\mathrm{QFIM}(\text{circuit'},\bm{\theta}')] > \max_{\bm{\theta}} \rank [\mathrm{QFIM}(\text{circuit},\bm{\theta})]$)}
            \State $\text{circuit} = \text{circuit'}$
        \EndIf
    \EndFor
\EndWhile\label{euclidendwhile}
\State \textbf{return} \text{circuit}
\end{algorithmic}
\end{algorithm}
Using Theorem \ref{thm:QFIMrankthm}, it suffices to calculate the QFIM rank at just one randomly sampled $\bm{\theta}$ to obtain the maximum rank over parameter space. Another more heuristic approach is given by Algorithm~\ref{alg:QDL_2}, using the concept of overparametrization introduced in \cite{Larocca2021}:

\begin{definition}[Overparametrization]\label{def:Overparametrization}
    A VQC overparametrized if the number of parameters $D$ is such that the QFIM, for all the states in the training set, simultaneously saturate their rank $r_{\mathrm{max}}$:
    \begin{equation}
        \max_{D \geq D_{c}, \bm{\theta}} \rank[\mathrm{QFIM}_s(\bm{\theta})] = r_{\mathrm{max}}.
    \end{equation}
\end{definition}

The authors showed that for a general type of periodic-structured VQCs, we have:
\begin{equation}\label{eq:critical_threshold_nbr_parameters}
    D_c \sim \dim(DLA) 
\end{equation}

The quantum circuit that performs the full rank of the QFIM can be easily completed using overparametrization according to Eq.~\eqref{eq:critical_threshold_nbr_parameters}. We propose another algorithm based on the overparametrization phenomenon where the idea is to remove gates to reduce the circuit depth while preserving the controllability of the output state.

\begin{algorithm}[H]
\caption{to design a HW-preserving quantum data loader}\label{alg:QDL_2}
\begin{algorithmic}[1]
\Require circuit, flag = True
\While{flag}
    \State flag = False
    \For{$RBS \in \text{circuit}$}
        \State $\text{circuit'} = \text{circuit} - RBS$
        \If{($\max_{\bm{\theta}'} \rank [\mathrm{QFIM}(\text{circuit'},\bm{\theta}')] = \max_{\bm{\theta}} \rank [\mathrm{QFIM}(\text{circuit},\bm{\theta})]$)}
            \State circuit, flag $=$ circuit', True
            \State \textbf{break for}
        \EndIf
    \EndFor
\EndWhile
\State \textbf{return} \text{circuit}
\end{algorithmic}
\end{algorithm}

The reason we can naively increase the rank of the QFIM to its maximal in Algorithm~\ref{alg:QDL_1} is because of the results from \cite{Larocca2021} on the Theory of Overparametrization, reminded in Definition~\ref{def:Overparametrization}. Algorithm~\ref{alg:QDL_2} must be initialized by considering a quantum circuit made of a high number of gates according to the dimension of its DLA as explained in Eq.~\eqref{eq:critical_threshold_nbr_parameters}. Those gates can be chosen randomly or in such a way to reduce the circuit depth. Using Algorithm \ref{alg:QDL_2} allows to first design a circuit with a number of gate slightly larger than the optimal, and then to reduce the circuit by removing some gates according to the QFIM. This method can be useful to first choose a circuit that corresponds to other figure of merits (favoring the use of qubits of better quality, reducing the depth) and then to avoid to derive too many time the rank of the QFIM which can be costly.

In practice and with both algorithms, one must pay particular attention to the order of the generators that are tested in the algorithm with regard to the circuit depth. The computational cost of both algorithms depends on the difficulty to calculate the rank of the QFIM, which has computational complexity of $\mathcal{O}(p^2 \, d_k^2 + p^3)$ on a classical simulation, with $p$ the number of parameters and $d_k$ the dimension of the chosen subspace (the $p^3$ term corresponds to the complexity of calculating the matrix rank). Algorithm \ref{alg:QDL_1} and \ref{alg:QDL_2} are thus easy to run for small subspaces, which correspond to the case where HW-preserving quantum circuits are trainable (see the following Section). {Since such data loaders are classically simulable, the potential quantum speed-up for training is limited to a polynomial advantage. However, one could train them classically to represent classical data and then associate them with a quantum circuit that is harder to simulate—for example, by increasing the number of qubits, the Hamming weight, or incorporating gates that do not preserve Hamming weight.}

\section{Trainability of HW-preserving quantum circuits}\label{chap:trainability_RBS_circuit}

In the previous Section, we discussed how to prove the existence of a quantum data loader using RBS gates and how to design such an encoding circuit for a specific subspace. In this Section, we show a study of the trainability of HW-preserving quantum circuits that could be used for our encoding but also for other types of trainable layers, such as represented in Fig.~\ref{fig:figure_intro}. It is known that some QML proposals suffer from negative optimization landscape properties \cite{McClean2018} that lead to strong limitation in their trainability. In this Section, we will give strong results on the gradient of the cost function for VQCs made of RBS or FBS gates. First, we present in Section~\ref{chap:backpropagation} the backpropagation formalism applied to RBS and FBS based VQCs. Then we present the resulting theorems on the variance and expectation value of the cost function gradient in Section~\ref{chap:Avoid_BP}.

     \subsection{Backpropagation for gradient calculus}\label{chap:backpropagation}

We describe a HW-preserving quantum circuit only made of RBS gates, or only made of FBS gates. We decompose the quantum circuit as a series of such gates, for which we denote their unitary matrices in the basis $B_k^n$ by $w^{\lambda}(\theta_\lambda)$, for $\lambda=1,\dots,\lambda_{\mathrm{max}}$, with $\theta_\lambda$ denoting the a gate's angle parameter. In the subspace of HW $k$, we denote respectively the initial, intermediate, and final quantum states by $\zeta^0$, $\zeta^\lambda$ (for $\lambda=1,\dots,\lambda_{\mathrm{max}}$), and $z$ --- they are all normalized vectors in $\mathbb{R}^{d_k}$. We also denote the inner error associated to the state $\lambda$ by $\delta^{\lambda} := \partial \mathcal{C} \slash \partial \zeta^{\lambda}$.
The cost function we consider in this work is the \textit{squared Euclidean distance} between the output state $z$ of the circuit and fixed target output $y$: 
\begin{equation}\label{eq:square-euclidean-loss-function}
    \mathcal{C}(\bm{\theta}) = ||z(\bm{\theta}) - y||^2_2\,.
\end{equation}
We focus on this cost function because it is ubiquitous in classical machine learning tasks (where it is usually termed the \textit{$l_2$ loss}). The reader might notice that the cost function in Eq.~\eqref{eq:square-euclidean-loss-function} depends on the phase on the output state $z$ (i.e. changing $z$ to $-z$ generally changes the cost function value), and thus it may not be written as an expectation value $\bra{z}O\ket{z}$ of some hermitian $O$ (as those latter functions are phase invariant). While this is true, it is possible to extend the system with a single ancillary qubit such that $\mathcal{C}(\bm{\theta})$ may be estimated through a quantum observable on the extended system (see the second tomography procedure described in \cite[Appendix A.5]{Landman2022} for more details).\footnote{This in essence switches the status of the $\pm$ sign from a global phase to a local one, making it is physically observable on the extended system.}

\begin{figure}[H]
    \centering
    \includegraphics[width=0.5\textwidth]{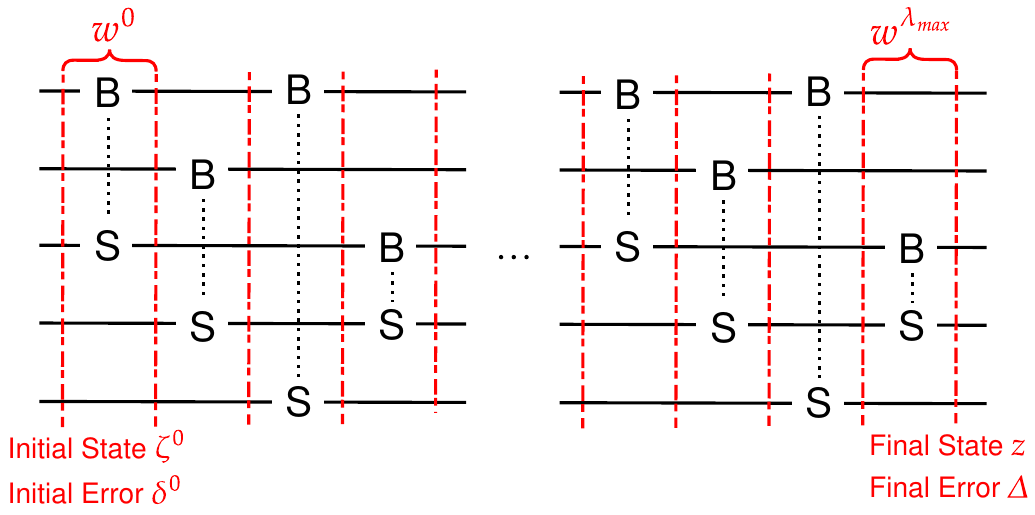}
    \caption{Decomposition of the HW-preserving quantum circuit for the backpropagation method.}
    \label{fig:particule_preserving_backprop}
\end{figure}

The equivalent weight matrix of our VQC is $W^k = w^{\lambda_{\mathrm{max}}} \dots w^1 w^0$. To train our circuit, we want to update each RBS parameter $\theta_\lambda$ with respect to the gradient of the cost function $\mathcal{C}$. Using the chain rule, we may decompose the cost function's derivative with respect to a gate's parameter $\theta_\lambda$ in terms of the components of the subsequent quantum state $\zeta^{\lambda+1}$:  
\begin{equation}
    \frac{\partial \mathcal{C}}{\partial \theta_\lambda} = \sum_p \frac{\partial \mathcal{C}}{\partial \zeta_p^{\lambda +1}} \frac{\partial \zeta_p^{\lambda +1}}{\partial \theta_\lambda} = \sum_p \delta_p^{\lambda + 1} \frac{\partial (w_p^{\lambda} \cdot \zeta^{\lambda})}{\partial \theta_\lambda}
\end{equation}

Each parameter $\theta_\lambda$ corresponds to applying a $\theta_\lambda$-planar rotation between two qubits. Such a rotation between two qubits corresponds, at the level of the subspace of HW $k$, to multiple pairs of basis directions $(l,j)$ that are undergoing a rotations in the subspace. For a circuit made of RBS gates, we have:

\begin{equation}\label{eq:backpropagation_equation}
    \begin{split}
    \frac{\partial \mathcal{C}}{\partial \theta_\lambda} =  \sum_{(l,j)} &\delta_l^{\lambda} (-\sin(\theta_\lambda) \zeta_l^{\lambda} + \cos(\theta_\lambda)\zeta_j^{\lambda}) + \\ &\delta_j^{\lambda} (-\cos(\theta_\lambda) \zeta_l^{\lambda} - \sin(\theta_\lambda)\zeta_j^{\lambda}),
    \end{split}
\end{equation}
where the sum is over all pairs $(l,j)$ of basis state indices which are undergoing a planar rotation of angle $\theta$ by the layer $\lambda$.

Similarly, for a circuit made of FBS gates, we have:
\begin{equation}\label{eq:backpropagation_equation_FBS}
    \begin{split}
    \frac{\partial \mathcal{C}}{\partial \theta_\lambda} = \sum_{(l,j)} &\delta_l^{\lambda} (-\sin(\theta_\lambda) \zeta_l^{\lambda} + (-1)^{f(a,b,\zeta_j^{\lambda})} \cos(\theta_\lambda)\zeta_j^{\lambda}) + \\
    &\delta_j^{\lambda} ((-1)^{f(a,b,\zeta_l^{\lambda})+1}\cos(\theta_\lambda) \zeta_l^{\lambda} - \sin(\theta_\lambda)\zeta_j^{\lambda}),
    \end{split}
\end{equation}
with $f(a,b,\zeta_\lambda^{\lambda}) = \sum_{a<p<b} s_p$, where $s \in \{0,1\}^{n}$ is the binary word corresponding to the state given by the index $\lambda$: $\ket{\zeta_\lambda} = \ket{s_1 \cdots s_n}$ ($a$ and $b$ are the qubits affected by the FBS).

    \subsection{Avoiding Barren Plateaus}\label{chap:Avoid_BP}

We can use the backpropagation analytic definition of the cost function gradient (Eqs.~\eqref{eq:backpropagation_equation},\eqref{eq:backpropagation_equation_FBS}) to study the phenomenon of \emph{Barren Plateaus} (BPs), a detrimental situation in which gradients of the cost function are exponentially suppressed. We recall their definition:

\begin{definition}[Barren Plateau]
    The cost function $\mathcal{C}(\bm{\theta})$ landscape of a $n$-qubit VQC is said to exhibit a Barren Plateau (BP) if for all $\lambda$:
    \begin{equation}
        \mathbb{E}_{\bm{\theta}}[\partial_{\theta_\lambda} \mathcal{C}(\bm{\theta})] = 0, \;\;\; \mathrm{Var}_{\bm{\theta}}[\partial_{\theta_\lambda} \mathcal{C}(\bm{\theta})] = O(\frac{1}{b^n}),
    \end{equation}
    with $b > 1$.
\end{definition}

It is possible to determine the existence of BPs under the assumption that the ensemble of parametrized unitaries form an approximate 2-design \cite{holmes2022connecting}, for in that case the quantity  $\mathrm{Var}_{\bm{\theta}}[\partial_{\theta_\lambda} \mathcal{C}(\bm{\theta})]$ may be evaluated using the standard Weingarten calculus integration formulas (see e.g. \cite{Puchala2011}) and found to be inversely proportional to the dimension of the Hilbert space. In recent work \cite{Larocca2021}, authors have shown that if a  subspace-preserving VQC satisfies the assumption of full controllability of the subspace (meaning that the dimension of the DLA is maximal, i.e. the same as the dimension of all unitaries on that subspace), as well as the 2-design assumption on that subspace,then the variance of the cost gradient scales with the inverse of the dimension of the subspace. As a result, one could avoid BPs using a subspace invariant quantum circuit with a subspace of small dimension.

Here, we show that one can indeed avoid BPs for subspace invariant quantum circuits based on RBS or FBS gates, \emph{without} making a 2-design assumption or any assumption on the controllability, provided one employs the circuit only in one given HW $k$ subspace with a fixed $k$, and some assumption on the qubit connectivity.
The central result which enables this claim is the following Lemma \ref{lemma:VarianceHWPreserving}, which by leveraging the specific form of RBS/FBS circuits, provides an exact analytic expression for the the variance of the cost gradient (given an initial state and a target state), for our cost function of interest. From this result it is then possible to derive absence of BP results depending on the situation of interest. We propose two such applications in that regard, presented as Theorems \ref{thm:NoBPPSA} and \ref{thm:NoBPgeneralcase}, that deal respectively with the case of a periodic connected ansatz with any input/target states, and an arbitrary circuit with randomly sampled input/target states accordng to a family of distributions. 

\begin{restatable}[Variance of RBS and FBS based VQCs]{lem}{VarianceHWPreserving}
\label{lemma:VarianceHWPreserving}
Let us consider an $n$-qubit HW-preserving VQC made of $D\geq1$ RBS or FBS gates only, that is employed in the subspace of HW $k$ (i.e. both the initial state $\zeta^0$ and the target state $y$ are normalized real superpositions of the basis $B_k^n$), along with the cost function $\mathcal{C}(\bm{\theta})$ taken as the squared Euclidean distance between the final state $\zeta^{\lambda_{\mathrm{max}}}$ and the target state $y$. If $\bm{\theta}$ is distributed uniformly in $\bm{\theta}:=[0,2\pi]^D$, then we have for all $\lambda \in \llbracket1, \lambda_{\mathrm{max}}\rrbracket$:
\begin{equation}
    \mathbb{E}_{\bm{\theta}}[\partial_{\theta_{\lambda}} \mathcal{C}(\bm{\theta})] = 0 \,
\end{equation}
\begin{equation}\label{eq:VarianceHWPreserving-Var}
    \begin{split}
        \mathrm{Var}_{\bm{\theta}}[\partial_{\theta_{\lambda}} \mathcal{C}(\bm{\theta})] = 2 & \sum_{l,j} \left( \frac{1}{\left(2 \pi\right)^D} \int_{\bm{\theta}} (\zeta_l^{\lambda})^2 + (\zeta_j^{\lambda})^2 d\bm{\theta} \right) \\
        & \cdot \left( \frac{1}{\left(2 \pi\right)^D} \int_{\bm{\theta}} (\tilde{y}_l^{\lambda})^2 + (\tilde{y}_j^{\lambda})^2 d\bm{\theta} \right)\,,
    \end{split}
\end{equation}
with $\zeta^{\lambda} = \omega^{\lambda - 1} \dots \omega^{1} \cdot \zeta^0$ the intermediate state (before inner layer $\lambda$), $\tilde{y}^{\lambda} = (\omega^{\lambda+1})^\intercal \dots (\omega^{\lambda_{\mathrm{max}}})^\intercal \cdot y$ the back-propagated target state, and where the sum is over all pairs $(l,j)$ of basis state indices which are undergoing a planar rotation of angle $\theta$ by the layer $\lambda$.
\end{restatable}
The proof of Lemma \ref{lemma:VarianceHWPreserving} is given in Appendix \ref{chap:proof_Lemma_Var}.

To state the next theorem, we introduce the following assumption on our circuits.

\begin{definition}[CPSA]\label{def:CPSA}

We say that an RBS/FBS circuit is a \emph{Connected Periodic Structure Ansatz} (CPSA) if it is composed of $L\geq1$ parametrized repetitions of a pattern $U_0$ of RBS or FBS gates, i.e. a circuit where the equivalent unitary $U(\bm{\theta})$ in the subspace of fixed HW $k$ is of the form:
\begin{equation}\label{eq:PeriodicStructureAnsatz}
    U(\bm{\theta}) = \prod_{l=1}^{L} U_0(\bm{\theta}_l), \quad U_0(\bm{\theta}_l) = \prod_{j=1}^J e^{-i \theta_{l,j} H^j_{RBS/FBS}},
\end{equation}
and if furthermore the repeated pattern $U_0(\cdot)$ \emph{connects all qubits}, i.e. a path between any two qubits may be traced on $U_0(\cdot)$'s RBS/FBS circuit diagram.

\end{definition}
Of course, CPSA's are only possible on quantum architectures that have a \emph{connected} qubit connectivity graph. The simplest example of such a $U_0$ that connected all qubits is the diagonal line of $n-1$ RBS/FBS gates connecting qubits $1$ and $2$, $2$ and $3$, and so on.

The following holds:

\begin{restatable}[Absence of Barren Plateaus, informal]{thm}{NoBPPSA}
\label{thm:NoBPPSA}
Under the same assumptions as Lemma \ref{lemma:VarianceHWPreserving}, if additionally the gates are arranged in a CPSA (Definition \ref{def:CPSA}),
then there exists an integer $q\geq1$ such that if the number of repetitions $L$ grows at least as fast as $n^q$, then for all $j$, and for any $0<\alpha<1$, setting $l=\lfloor \alpha \, L \rfloor$ implies
\begin{equation}
        \begin{split}
            &\mathrm{Var}_{\bm{\theta}}[\partial_{\theta_{l,j}} \mathcal{C}(\bm{\theta})] = \frac{k(n-k)}{n(n-1)}\frac{8}{d_k} + \varepsilon\,,
        \end{split}
    \end{equation}
where $\varepsilon$ decays exponentially with $n$.
Thus, after some polynomial amount of repetitions, and for angles located at any constant fraction of the depth, there is an absence of Barren Plateaus for CPSA ansätze.

\end{restatable}

The fact that this statement concerns gates located at constant ratios of the circuit depth may be interpreted as the condition that the gates are not too close to either extremities of the circuit.
The formal and more general version of \cref{thm:NoBPPSA} is given as \cref{thm:concluding-back-to-variance} in \cref{subchap:precised-theorem-2}. We concede that the proof of this theorem depends on the validity of a small conjecture that we make about spectral gaps of certain stochastic matrices (\cref{conj:spectral-gap}), for which we present numerical evidence in \cref{subsec:numerical-evidence-spectral-gap}.
We refer to \cref{sec:ProofNoBPPSA} for further details, but the proof of that remaining conjecture is left for future work.
Since \cref{thm:NoBPPSA} states a requirement of a polynomial number of repetitions $L$ to reach its conclusion of absence of Barren Plateaus, one may wonder if it can generally be subsumed by the usual approximate $2$-design argument, since usually polynomial repetitions produce approximate $2$-designs.
A discussion on the subtleties behind this is given in \cref{subchap:design-discussion}.

The next theorem does not require any lower limit on the circuit depth:

\begin{restatable}[Evolution of the variance for RBS and FBS based quantum circuits]{thm}{NoBPgeneralcase}
\label{thm:NoBPgeneralcase}
Under the same assumptions as Lemma \ref{lemma:VarianceHWPreserving}, if additionally the initial state $\zeta^0$ and the target state $y$ are each independently distributed on the sphere $S^{d_{k}-1}$ in such a way that:
\begin{equation}\label{eq:hyp_distribution_NoBPGeneral}
    \forall r \in [d_k], \quad \begin{cases} \mathbb{E}[\zeta_r^0] = \mathbb{E}[y_r] = 0\,, \\ \mathbb{E}[(\zeta_r^0)^2] = \mathbb{E}[(y_r)^2] = \frac{1}{d_k}\,, \end{cases}
\end{equation}
then we have for all $\lambda \in \llbracket1, \lambda_{\mathrm{max}}\rrbracket$:
    \begin{equation}\label{eq:var_NoBPgeneralcase}
        \begin{aligned}
            \mathbb{E}_{\zeta^0,y} \mathrm{Var}_{\bm{\theta}}[\partial_{\theta_{\lambda}} \mathcal{C}(\bm{\theta})] &= \frac{k(n-k)}{n(n-1)}\frac{8}{d_k}\,.
        \end{aligned}
    \end{equation}

\end{restatable}

The complete proof of this Theorem is presented in Appendix~\ref{chap:proof_BP_general}. Note that the assumption of Eq.~\eqref{eq:hyp_distribution_NoBPGeneral} on the distributions lies in between the assumptions of spherical $t$-designs of $t=1$ and $t=2$ (the first line of Eq.~\eqref{eq:hyp_distribution_NoBPGeneral} imposes a spherical 1-design, while the second line only enforces values of the \emph{homogeneous} $2^{\mathrm{nd}}$ order moments, but leaves \emph{correlations} between components like $\mathbb{E}[y_1 y_2]$ unconstrained).  
 We illustrate this result in Fig.~\ref{fig:Gradient_RBS_FBS_study} and Fig.~\ref{fig:Gradient_RBS_FBS_study_Encoding} in the additional simulations given in Appendix~\ref{chap:Additional_Simulations}.

We can conclude from these results that there is no Barren Plateaus for the subspace invariant RBS and FBS based quantum circuits for a fixed choice of subspace $k$.
We emphasize that, unlike recent related works \cite{Larocca2021,Ragone2023,Fontana2023, diaz2023showcasing} (which apply for quite general circuits), our Theorem~\ref{thm:NoBPPSA} on RBS/FBS circuits does not make use of a 2-design assumption on the global unitary of the circuit.
We also emphasize that while our Theorem \ref{thm:NoBPgeneralcase} considered particular assumptions on the input/target state distributions and cost function that do not encompass all possible learning tasks, the proof in Appendix \ref{chap:proof_BP_general} could very well be adapted to other tasks.

\subsection{Complexity and simulation of subspace preserving RBS circuit}\label{chap:HW_Complexity}

If a HW-preserving circuit is employed entirely in a subspace of fixed HW $k$ (independent of the number of qubits $n$), then it may be classically simulated directly, i.e. by constructing all the polynomially-sized unitary matrices ($d_k \times d_k$) and state vectors ($d_k$) classically and performing matrix multiplications.
However in the case where $k = k(n)$ grows with $n$ such that $d_k = \Omega(\exp(n))$, this method has an exponential cost. Besides, if the circuit is used with initial/target states that do not lie in a fixed HW subspace, then this method may not apply either.
In those cases, other simulation techniques such as \textit{matchgate simulation} and \textit{Lie-algebraic simulation} may in some cases be considered. Indeed, HW-preserving gates are a particular type of matchgates \cite{Valiant2002QuantumCT}, a well studied type of gates, and those can be efficiently simulated classically in the case of nearest neighbours connectivity \cite{Jozsa_2008}. In addition, recent work \cite{goh2023liealgebraic} showed that subspace preserving quantum circuits can also be simulated if the Lie algebra is of polynomial dimension in $n$, and if the observable lies inside the Lie algebra. In cases where the Lie algebra generated by the RBS/FBS architecture is of polynomial dimension while at the same time being associated to a subspace dimension $d_k$ that is exponential, the Lie-algebraic simulation technique may prove useful; although the circuits that we consider do not directly fullfill the assumption of having an observable lying inside the Lie algebra, and regardless the simulation method requires knowing a basis of the Lie algebra which could be exponentially costly to obtain. In fact, in \cite{Kerenidis2022}, the authors have proposed algorithms based on FBS gates that present exponential complexity advantages, while the FBS based quantum circuits have a DLA of low dimension.
To sum up, it is hard to give a statement on the simulability of RBS based quantum circuits for any connectivity.

In the subspace of HW $k$, the effect of a RBS gate is a rotation between $k$ pairs of states in the basis $B_k^n$. Therefore, a RBS is easy to simulate classically in the unary basis but can require an exponential number of operations when $k$ is close to $n/2$ as $d_{n/2}$ grows exponentially with $n$ the number of qubits. In Table~\ref{table:Running_Time_RBS}, we illustrate the differences in operation count for running and training an RBS/FBS VQC, using a quantum implementation versus a classical simulation.

Nevertheless, using a greater subspace may be a wrong choice according to the circuit depth. For example, achieving the full controllability of a larger subspace requires a more significant amount of gates but the number of qubits is limited. It results in a rise in the circuit depth. In addition, using a greater subspace results in the apparition of BPs, as shown in section \ref{chap:trainability_RBS_circuit}.

\begin{table}[h!]
    \begin{center}
    \begin{tabular}{|c|c|c|} 
        \hline
        \textbf{Algorithm} & \textbf{Feedforward} & \textbf{Training} \\
        \hline
        RBS VQCs & $\mathcal{O}(D/(p*\delta^2))$ & $\mathcal{O}((D/p)^2)$ \\
        \hline
        Simulation & $\mathcal{O}(D\binom{n-2}{k-1})$ & $\mathcal{O}((D\binom{n-2}{k-1})^2)$ \\
       \hline
    \end{tabular}
    \caption{Running time summary. $n$ is the number of qubits, $k$ the chosen Hamming weight corresponding to the selected subspace, $D$ is the number of RBS/FBS gates, $\delta$ is the error parameter in the quantum implementation, and we call $p$ the average number of gates we can parallelize at the same time (upper-bounded by $\lfloor \frac{n}{2} \rfloor$).}
    \label{table:Running_Time_RBS}
    \end{center}
\end{table}

Note that working in subspace of large HW $k$ may still be envisioned, if the circuit depth (and thus the number of parameters) is relatively low.

However, the use of subspace preserving RBS quantum circuits is keen on giving a speedup advantage for an application that requires to process a large amount of information but with a limited number of degrees of freedom (and thus of RBS gates). In Section~\ref{chap:RBS_QONN}, we give an example of such an application for a fully connected orthogonal neural network that is not fully controllable. Problem-inspired architectures using symmetries are perfect candidates for this type of application.

\section{Additional Simulations}\label{chap:Additional_Simulations}

In this section, we present additional simulations. We propose to illustrate the use of Algorithm~\ref{alg:QDL_1} for an existing hardware and the corresponding simulations in Section~\ref{chap:Encoding_existing_hardware}. In Section~\ref{chap:Additional_Simu_HW_training}, we propose additional simulations to illustrate the results in Section~\ref{chap:trainability_RBS_circuit}. 

    \subsection{Using our encoding method for an existing hardware}\label{chap:Encoding_existing_hardware}

\begin{figure}[h!]
    \centering
    \includegraphics[width=0.4\textwidth]{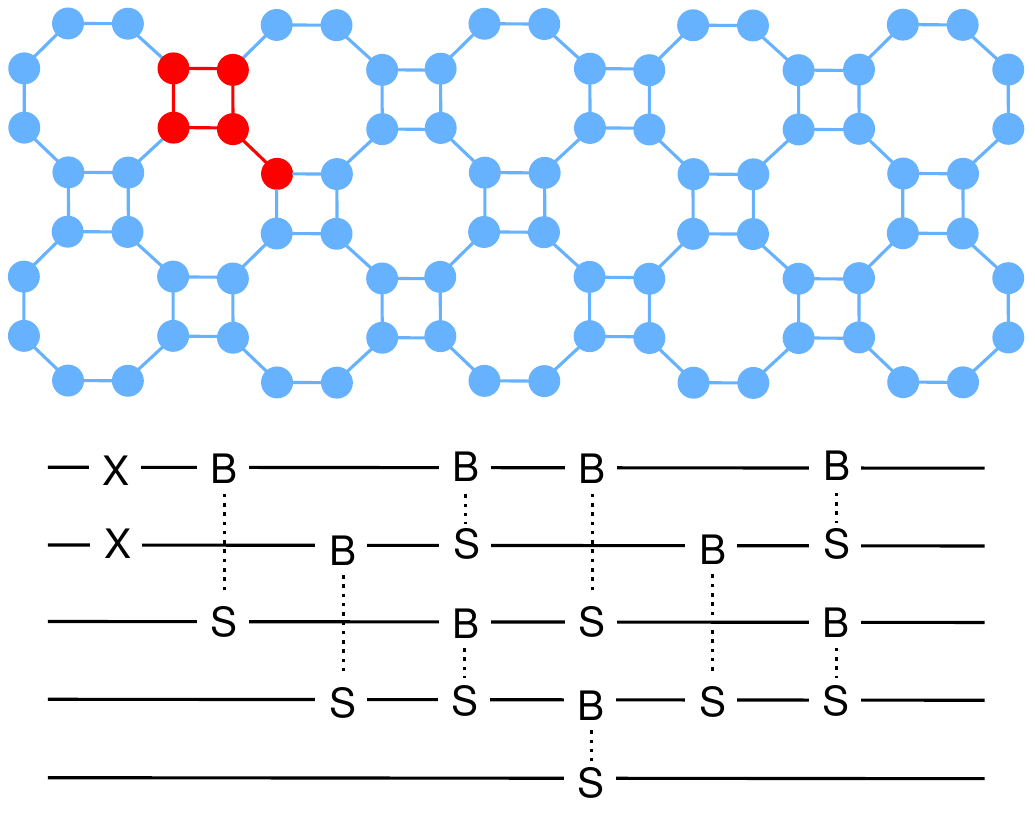}
    \caption{Rigetti ASPEN M2 connectivity graph and the Quantum Data loader quantum circuit obtained using our Algorithm~\ref{alg:QDL_1}.}
    \label{fig:Rigetti_ASPEN_M2_connectivity_graph}
\end{figure}

In this part, we propose to use the Algorithm~\ref{alg:QDL_1} to design a 5-qubit quantum data loader that fits the Rigetti ASPEN M2 connectivity given in Fig.~\ref{fig:Rigetti_ASPEN_M2_connectivity_graph}. This hardware is keen on implementing such a data loader thanks to its high connectivity and to the fact that the RBS gate is native to this platform (and called the XY gate). 

Using the Algorithm~\ref{alg:QDL_1}, we can design a quantum data loader for this connectivity. To minimize the depth, one can use in practice a variation of this algorithm that minimizes the depth by testing in priority the gates that are more likely to parallelize themselves. As a result, the circuit obtained is given in Fig.~\ref{fig:Rigetti_ASPEN_M2_connectivity_graph}.

In order to highlight the performance of our quantum data loader, we simulate our method using Python/Numpy but also using Qiskit \cite{qiskit2024}, a quantum instruction language developed as a Python library by IBM. We test our data loader for a well-known data set called the Fashion MNIST dataset \cite{Fashion-MNIST}. 

In Table~\ref{Encoding_performance}, we compare our simulation in a specific subspace of HW $k$ using Numpy, with a simulation in the full $2^n$-dimensional Hilbert space using Qiskit. Our simulation codes are more efficient as we only consider the desired subspace, but we propose this comparison to show that the results are equivalent.

\begin{table}[h!]
    \begin{center}
    \begin{tabular}{|c|c|c|} 
        \hline
        \textbf{Simulation} & \textbf{Average Error} & \textbf{Variance} \\
        \hline
        Numpy & 0.0097 & 0.00030 \\
        \hline
        Qiskit & 0.0094 & 0.00028 \\
       \hline
    \end{tabular}
    \caption{Performance of the quantum data loader presented in Fig.~\ref{fig:Rigetti_ASPEN_M2_connectivity_graph}.}
    \label{Encoding_performance}
    \end{center}
\end{table}

To achieve this simulation, we use our codes after applying a Principal Component Analysis (PCA) to reduce our dataset to $\binom{n}{k}$-dimensional vector, with $n=5$ the number of qubits, and $k=2$ the chosen HW. We used 1000 samples to derive those values.

The approximation error is measured using the following cost function between the real state that we are supposed to have $x^{\ast}$ and the output of our quantum system given by $x = W^{k}(\theta) \cdot e_s$ (with $e_s$ the vector representation of the initial state in our method):
\begin{equation}\label{eq:cost_function_encoding_simulation}
    \mathcal{C}(x) = || W^{k}(\bm{\theta}) \cdot e_s - x^{\ast} ||_2^2
\end{equation}

For an actual implementation on a Quantum Processor Unit, one can generalize the tomography procedure described in \cite{Landman2022}.

    \subsection{Trainability of Hamming weight preserving quantum circuits}\label{chap:Additional_Simu_HW_training}

In this part, we illustrate the results obtained from section~\ref{chap:trainability_RBS_circuit} on the trainability of RBS and FBS gate based quantum circuits in a specific subspace corresponding of a choice of Hamming weight $k$. \newline

\begin{figure}[h!]
    \centering
    \includegraphics[width=0.48\textwidth]{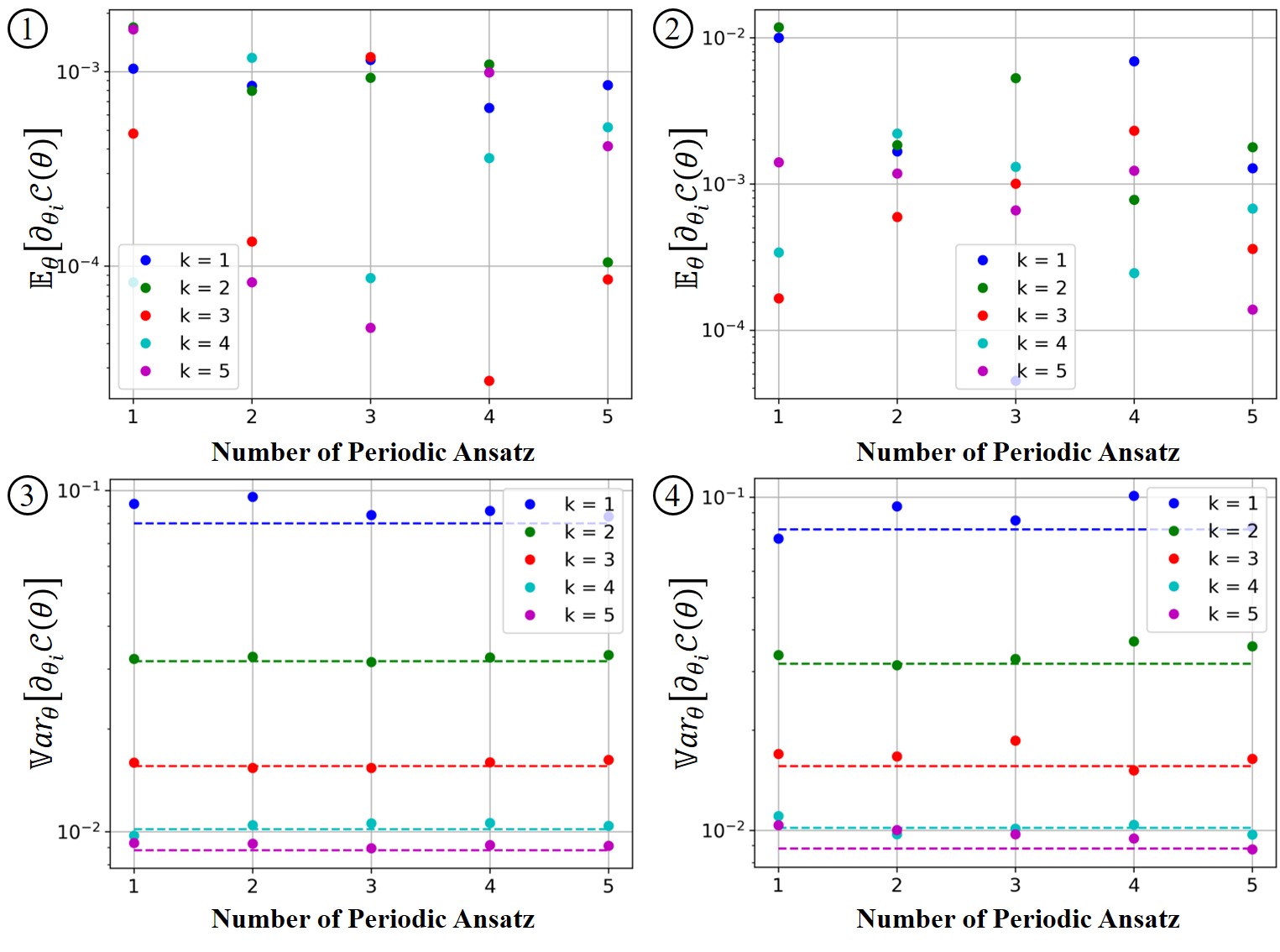}
    \caption{Gradient study for the same Periodic Structure Ansatz as in Fig.~\ref{fig:QFIM_rank_evolution}. Using $100$ random choices of input, target output, and parameters, we plot the average value of the gradient for each number of periodic ansatz L in (1). In (2), we plot the average gradient in the case where the RBS are replaced by FBS. In (3), we derive the variance of the gradient for the RBS case, and in (4), the variance for the FBS case. The dotted lines correspond to the theoretical values from Theorem~\ref{thm:NoBPgeneralcase}.}
    \label{fig:Gradient_RBS_FBS_study}
\end{figure}

In Fig.~\ref{fig:Gradient_RBS_FBS_study}, we plot the average value and the variance of the gradient for quantum circuits made only of RBS and FBS for parameters sampled uniformly in $[0,2\pi]$, and for input/target states sampled according to the uniform measure on $S^{d_{k}-1}$. This setting fits in the case of Theorem~\ref{thm:NoBPgeneralcase}. We observe that the average gradient is close to zero and its variance corresponds to the theoretical values.

In the previous Fig.~\ref{fig:Gradient_RBS_FBS_study}, one can notice that the value of the variance of the gradient is unchanged with the number of periodic ansatz repetitions $L$. As the number of repetitions increases, the number of gates increases but also the controllability in the state space. Indeed, Fig.~\ref{fig:QFIM_rank_evolution} shows the evolution of the controllability in the state space by presenting the evolution of the QFIM rank with $L$ (the maximal controllability is achieved for $L\geq4$). Therefore, Fig.~\ref{fig:Gradient_RBS_FBS_study} highlights a in-dependency between the controllability in the state space and the trainability.

\begin{figure}[h!]
    \centering
    \includegraphics[width=0.48\textwidth]{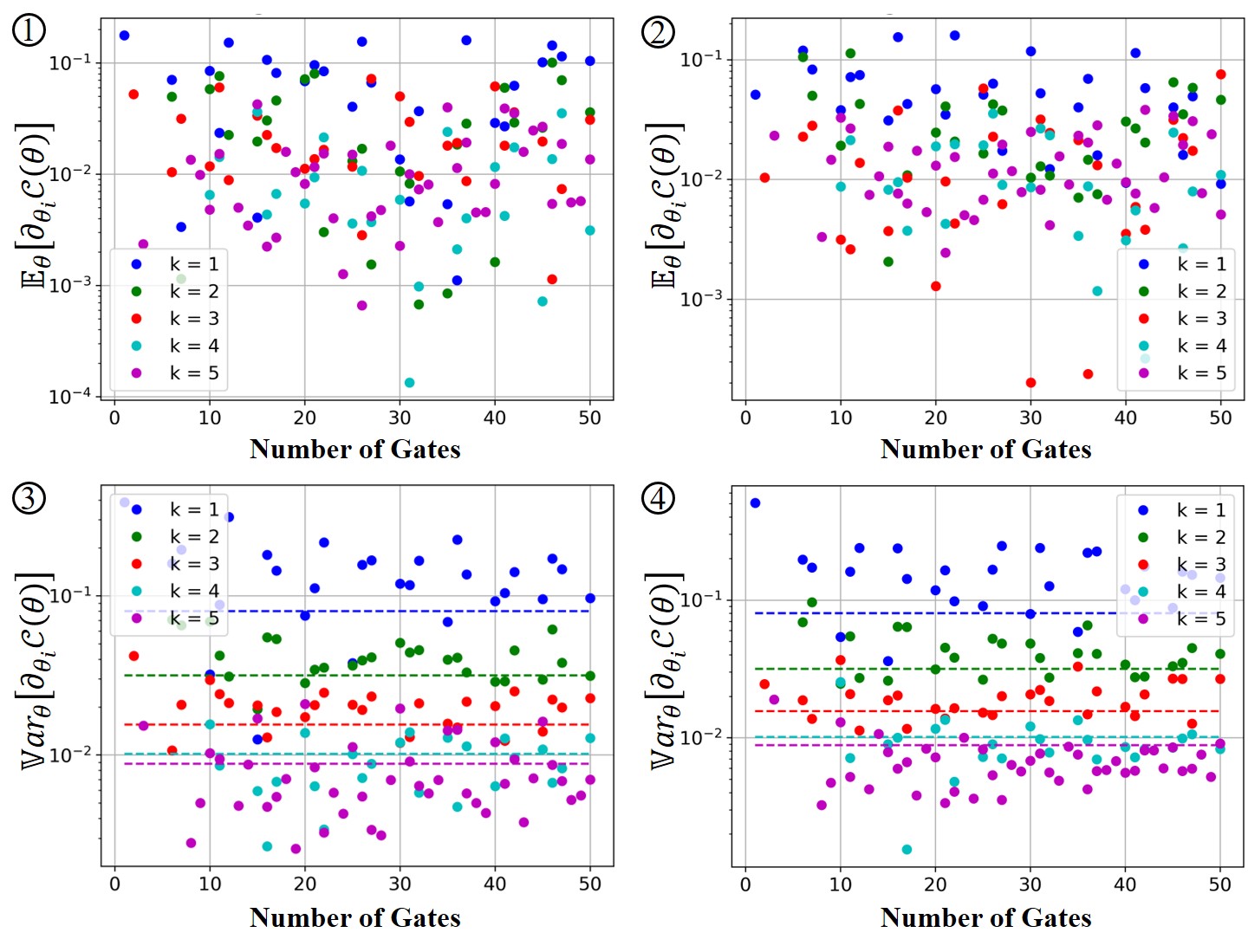}
    \caption{Gradient study for a quantum circuit made of 5 consecutive block circuits as described in Fig.~\ref{fig:QFIM_rank_evolution}. The plots are numerical evidences of Theorem~\ref{thm:NoBPgeneralcase} as the expectation values of the gradient are close to $0$ and the variances follow the theoretical values given by the doted lines. The initial state is fixed as a basis state in $B_k^n$ for any subspace choice $k$. We use $10000$ random target output and set of parameters.}
    \label{fig:Gradient_RBS_FBS_study_Encoding}
\end{figure}

In Fig.~\ref{fig:Gradient_RBS_FBS_study_Encoding}, we conduct the same study as in Fig.~\ref{fig:Gradient_RBS_FBS_study}, except that this time the input state is not randomly sampled but instead fixed to $\zeta^0:=\ket{1\dots1\,0\dots0}$ (the $k$ first qubits are in state $\ket{0}$ and the rest are in state $\ket{0}$).
This is a setting of encoding (described in Section~\ref{chap:Encoding}), where the target state is uniformly distributed on the sphere. It is possible to conduct a similar proof technique as that of Theorem~\ref{thm:NoBPgeneralcase} but for this precise setting, and one would find the gradient variances to be $\propto 1/d_k$ only for parameters far from the left boundary of the circuit, but not necessarily so near it. This is the reason why the points in Fig.~\ref{fig:Gradient_RBS_FBS_study_Encoding} only roughly follow the dashed $\propto 1/d_k$ values.

\subsection{HW-preserving VQC for Neural Networks}\label{chap:HW_QNN}

In our work, we introduce an encoding method using subspace invariant quantum circuits with a focus on the HW-preserving systems, the RBS, and the FBS gates. In this particular setting, it is not an issue if the circuit can be classically simulated as one may use our encoding method with an additional circuit hard to classically simulate. However, the results given in section~\ref{chap:trainability_RBS_circuit} highlights the interest of using such circuit as neural networks. In addition, the use of Hamming weigh preserving quantum circuits for machine learning purposes has already been proposed in \cite{Johri2020, Landman2022, Cherrat2022, Kerenidis2022}, and we could generalize those methods for larger subspace with better speedups. 

In this Section, we present how to use our results for Neural Networks. First, we introduce in \ref{chap:HW_Complexity} the complexity of a RBS based quantum circuit. Then we present in \ref{chap:RBS_QONN} the use of a RBS based VQC for image classification for a toy example.

    \subsection{RBS based Quantum Orthogonal Neural Network}\label{chap:RBS_QONN}

\begin{figure}[H]
    \centering
    \includegraphics[width=0.5\textwidth]{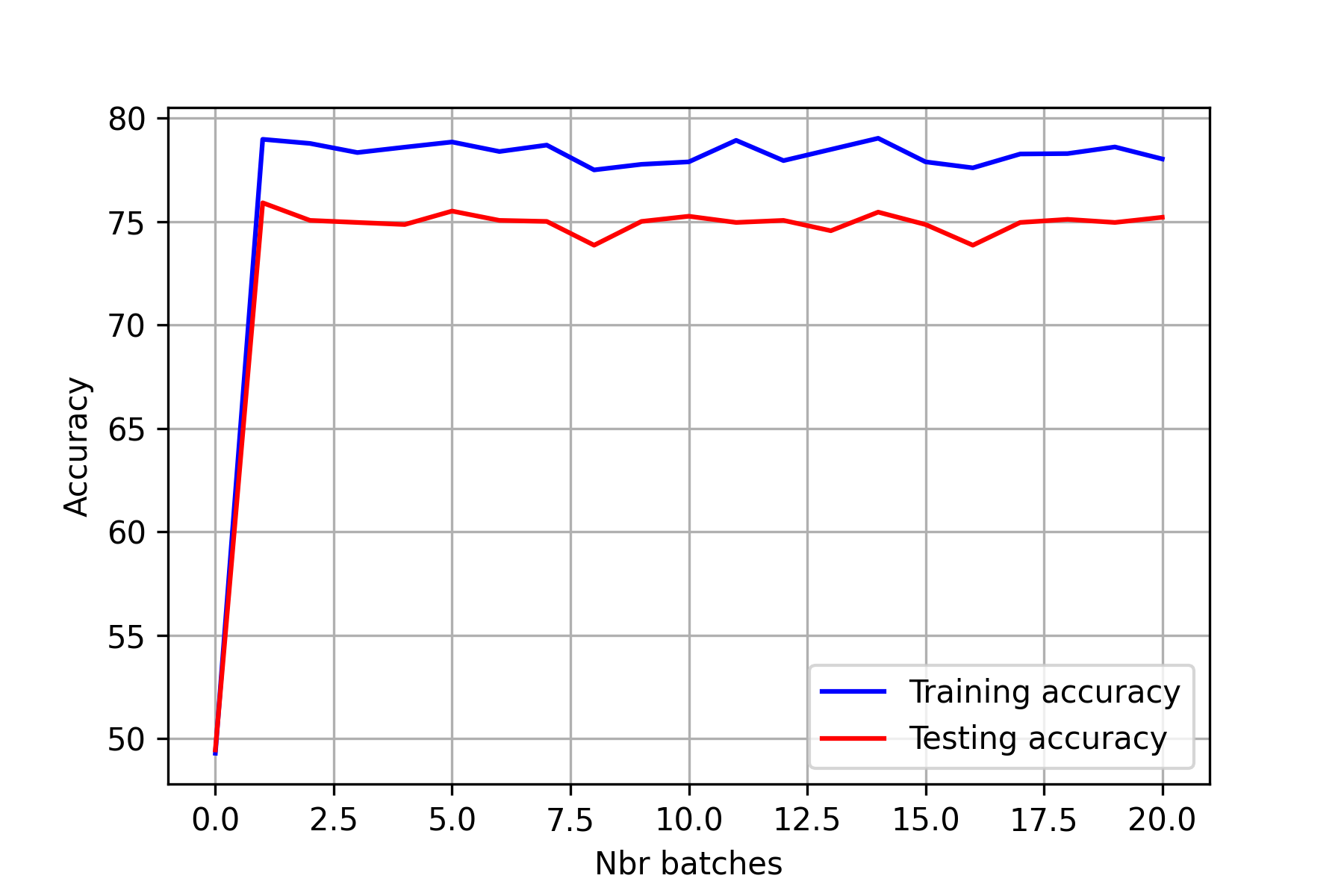}
    \caption{Binary classification on the Fashion MNIST data set for $10000$ training samples and $5000$ testing samples. The optimization method is ADAM with a batch size of $5$, and we used is the Cross Entropy Cost.}
    \label{fig:RBS_QNN_acc}
\end{figure}

In this section, we illustrate the fact that we can use our results for Quantum Neural Network (QNN) applications. HW-preserving gates have been previously used for such applications while considering a HW of $1$ in \cite{Landman2022}, but in every subspace of fixed HW, the equivalent unitary of a RBS based circuit is orthogonal, and thus, we can use those circuit as Orthogonal Neural Networks in similar way as in \cite{Landman2022}. In Fig.~\ref{fig:RBS_QNN_acc}, we plot the training and the testing accuracy for binary classification on the Fashion MNIST dataset, where we use the quantum circuit described in Fig.~\ref{fig:Rigetti_ASPEN_M2_connectivity_graph} as a QNN.

We do not claim that this plot exhibits any advantage to use RBS based quantum circuits as neural networks, but it illustrates that we can easily use such an architecture. Large simulation for more complex HW-preserving quantum neural networks for large value of $k$ must be tackled in future work.   

According to the complexity of such RBS models described in Table~\ref{table:Running_Time_RBS}, there is no exponential advantage in using a quantum orthogonal neural network. In \cite{Landman2022}, the authors present the use of a specific case of such a RBS model in the unary basis and show a quadratic advantage for a fully controllable QNN in the unary basis called the Pyramidal Quantum Neural Network (PQNN). For specific use cases that require less controllability but higher-dimensional input data, one can prefer to use an RBS based orthogonal neural network in a larger Hamming weight basis and can achieve a more significant speedup. 

\section{Discussion}
\label{sec:discussion}

In this work, we study the controllability and the trainability of a subspace-preserving variational quantum circuit through the prism of designing a quantum data loader using RBS gates. We show that the encoding capacity is linked with the controllability of the quantum system and that using only a certain type of gates, the controllability is linked with the connectivity of the hardware. In addition, we show how to study mathematical tools such as the Dynamical Lie Algebra (DLA) or the Quantum Fisher Information Matrix (QFIM) in order to design a data loader. Finally, we show conditions to avoid Barren Plateaus for variational quantum circuits only made of RBS or FBS gates.

The results we show for the trainability of Hamming weight preserving quantum circuits must be compared with our results on the controllability. Indeed, it is now clear that there is a trade-off between the expressivity of a variational quantum circuit and its trainability \cite{holmes2022connecting}. In recent work \cite{Larocca2021}, authors have shown the connection between the dimension of the DLA and the cost function's gradient variance, for subspace preserving variational quantum circuits in the specific setting of the subspace full controllability and with a 2-design assumption on the used ansatz. As a result, they show that such gradient variance evolves in an inverse manner with the dimension of the DLA. They also left open Conjecture \ref{conj:SubspacePreservingVar}.
More recently, \cite{Ragone2023} and \cite{Fontana2023} 
have shown that this evolution is true in a more general setting and hence proved the Conjecture \ref{conj:SubspacePreservingVar} to be true, under some assumptions on the initial state and the observable. In \cite{Fontana2023}, the authors also claim that RBS and FBS based quantum circuits are not part of their framework and show a coherent result to ours by given an upper-bound on the abstract variance of such circuit. Our results are independent from those papers, and show theoretical guarantees for RBS and FBS based VQCs in a general setting.

\begin{restatable}[from \cite{Larocca2021}]{conj}{Subspace_Preserving_Var}
\label{conj:SubspacePreservingVar}
Let the state $\rho$ belong to a subspace $\mathcal{H}_k$ associated with a subspace DLA $\mathfrak{g}_k$ (or sub-DLA, the subrepresentation in $\mathfrak{g}$ where $\rho$ has support on). Then, the scaling of the cost function partial derivative is inversely proportional to the scaling of the dimension of the DLA, i.e.
\begin{equation}
    \mathrm{Var}_{\bm{\theta}}[\partial_{\mu} \mathcal{C}(\bm{\theta})] = \mathcal{O}\left(\frac{1}{\mathrm{poly}(\dim (\mathfrak{g}_k))}\right)\,.
\end{equation}
\end{restatable}

In our work, we show that for the specific cases of RBS/FBS circuits and preserved subspace based on Hamming weight, the cost gradient variance evolves in an inverse manner with the dimension $d_k$ of the subspace, and not as stated in Conjecture \ref{conj:SubspacePreservingVar}.
Furthermore, the mentioned works \cite{Larocca2021,Ragone2023,Fontana2023} all make use of a 2-design assumption to derive the variance, but it is a huge assumption on the expressivity of the quantum system, that often leads to considering that the system is fully controllable. In the specific case of RBS and FBS gates, the expressions of those gates are easy to manipulate analytically, and we are able to circumvent the use of such assumptions.

Although our work gives theoretical guarantees on the trainability of a specific type of variational quantum circuit (VQC), we show that those methods could be useful for near term QML. In addition, we would like to insist on the fact that one cannot state in general that the trainability of a VQC is perfectly defined by its controllability through the dimension of the corresponding DLA. First because the dimension of the DLA is only an upper-bound on the controllability, but also because we show an example of a VQC where the previous conjecture from \cite{Larocca2021} is refuted in its full generality. 

\section{Author Contributions}

L.M. initiated the project, proposed the space-efficient encoding, expressivity study, the study of trainability, and contributed on all aspects. E.M. offered the Theorem~\ref{thm:QFIMrankthm} on the almost-constant property of QFIM rank and formalized the proof of Theorem~\ref{thm:NoBPPSA} using stochastic matrices and a spectral gap conjecture. J.L. contributed on all aspects of the paper. A.G., R.K., and E.K. supervised this work.

\section{Acknowledgment}

This work is supported by the H2020-FETOPEN Grant PHOQUSING (GA no.: 899544), the Engineering and Physical Sciences Research Council (grants EP/T001062/1), and the Naval Group Centre of Excellence for Information Human factors and Signature Management (CEMIS). ABG is supported by ANR JCJC TCS-NISQ ANR-22-CE47-0004, and by the PEPR integrated project EPiQ ANR-22-PETQ-0007 part of Plan France 2030.  This work is also part of HQI initiative (www.hqi.fr) and is supported by France 2030 under the French National Research Agency award number ANR-22-PNCQ-0002.

\bibliographystyle{quantum}
\bibliography{refs}

\clearpage

\onecolumn
\appendix
\section*{Appendix}
\DoToC

\counterwithin{equation}{section}
\counterwithin{thm}{section}
\counterwithin{lem}{section}
\counterwithin{cor}{section}
\counterwithin{conj}{section}
\counterwithin{definition}{section}
\counterwithin{figure}{section}
\renewcommand{\thethm}{\Alph{section}\arabic{thm}}
\renewcommand{\thelem}{\Alph{section}\arabic{lem}}
\renewcommand{\thecor}{\Alph{section}\arabic{cor}}
\renewcommand{\theconj}{\Alph{section}\arabic{conj}}
\renewcommand{\thedefinition}{\Alph{section}\arabic{definition}}
\renewcommand{\thefigure}{\Alph{section}\arabic{figure}}
    
\section{Quantum Data Loading equation}\label{chap:QDL_equation}

In this section, we study the difficulty of finding a good set of variational parameters to perform the data loading.

In Eq.~\eqref{eq:encoding_particular_equivalent_eq} the equation that we need to solve to perform the amplitude encoding of $x \in \mathbb{R}^{d_k}$ on the basis $B_k^n$ of states of Hamming weight $k$. We can express this encoding equation as the following set of equations (here, we suppose that the classical vector $x$ has been normalized):
\begin{equation}\label{eq:encoding_particular_equivalent_system}
    \begin{cases}
        w_{s, 1}(\bm{\theta}) - x_1 = 0 \\
        \dots \\
        w_{s, \binom{n}{k}}(\bm{\theta}) - x_{\binom{n}{k}} = 0 
    \end{cases}\
\end{equation}
with $w_{i,j}$ the matrix coefficient from $W^k(\bm{\theta})$ corresponding to the amplitude to go from state $\ket{e_j}$ to $\ket{e_i}$.

In general, finding the solution (or proving the existence of one solution) of a parametric set of non-linear equations is very hard. However, according to the expression of the RBS gate \eqref{eq:RBS_2_qubit_gate}, each coefficient $w_{i,j}(\bm{\theta})$ is a sum of product of cosines and sines of the variational parameters. Thus, using the change of variables:
\begin{equation}\label{eq:change_of_variables_Grobner_basis}
    \cos(\theta_i) = c_i \quad \text{and} \quad \sin(\theta_i) = s_i
\end{equation}
we can express the parametric set of nonlinear equations as a parametric set of polynomial equations:
\begin{equation}\label{eq:encoding_param_set_polynomials}
    \begin{cases}
        f_1(c_1,\dots, c_{\binom{n}{k}}, s_1,\dots, s_{\binom{n}{k}}) - x_1 = 0 \\
        \dots \\
        f_{\binom{n}{k}}(c_1,\dots, c_{\binom{n}{k}}, s_1,\dots, s_{\binom{n}{k}}) - x_{\binom{n}{k}} = 0 \\
        c_1^2 + s_1^2 - 1 = 0 \\
        \dots \\
        c_{\binom{n}{k}}^2 + s_{\binom{n}{k}}^2 - 1 = 0
    \end{cases}
\end{equation}
Several tools exist to solve such a polynomial system, for example, Gröbner basis \cite{Faugère1993, Faugère1999}, but they may come with a very high computational cost. This is reason why we prefer focus on an equivalent optimization problem as defined by Eq~\eqref{eq:encoding_optimization_pb}.

\section{Dynamical Lie Algebra derivation}\label{chap:DLA}

In this section, we remind the main properties of the Dynamical Lie Algebra (DLA). First, we give a more formal definition of the DLA. Then, we highlight the link between the dimension of the DLA and the dimension of the unitary equivalent manifold to show the link between the DLA and the degrees of freedom introduce in Section~\ref{chap:Existence_QDL}. 

\begin{definition}[Dynamical Lie Algebra]
    Let us consider the question of controllability for the Schrödinger operator equation in the general form:
    \begin{equation}\label{eq:Schrodinger}
        \dot{X} = -i H(u) X, \quad X(0) = \mathbb{1}_{d \times d}
    \end{equation}
    The Dynamical Lie algebra is defined as 
    \begin{equation}
        \mathcal{L} = span_{u \in \mathcal{U}} \{ -i H(u) \} \subseteq \mathfrak{su}(d)
    \end{equation}
\end{definition}

The set of reachable states from system~\eqref{eq:Schrodinger} is the connected Lie group associated with the DLA and called Dynamical Lie Group: $\mathcal{R} = e^{\mathcal{L}}$.

In general, the system is said controllable when $\mathcal{L} = \mathfrak{su}(d)$ (we can achieve any equivalent unitary from the system considered), which is in general equivalent to having $\dim(\mathcal{L}) = d^2 - 1$ with $d$ the dimension of the Hilbert space. It tells us that we can choose each coefficient of the equivalent unitary if we respect the condition that the corresponding matrix is unitary.

The Algorithm~\ref{alg:rank_Lie_algebra} \cite{Schirmer2000} allows us to compute the rank of the Lie Algebra generated by a control system. Basically, this algorithms derive the dimension of the Lie Algebra by updating a matrix $W$ with potential generators of the Lie Algebra as columns and derive the rank of this matrix after every test to ensure that it increases the dimension of the spanned space.

In what follows, the input matrices $\hat{H}_0, \dots, \hat{H}_M$ are understood in the subspace $k$ (i.e. as Hermitians of size $d_k \times d_k$. Given a $d_k \times d_k$ complex matrix $X$, the operation "$\mathrm{vec}(X)$" outputs the corresponding $2(d_{k})^2$-sized real column vector obtained by flattening and splitting into real and imaginary parts; while conversely given a $2(d_{k})^2$-sized real column vector $x$, the operation "$\mathrm{mat}(x)$" outputs back the associated complex matrix $X$. 
\begin{algorithm}[H]
\caption{Rank of the Lie Algebra generated by a control system $\{ \hat{H}_0, \dots, \hat{H}_M \}$.}\label{alg:rank_Lie_algebra}
\begin{algorithmic}[1]
\State $W = \mathrm{vec}(i H_0)$
\State $r = 1$
\For{$m = 1,2,\dots, M$} 
    \If{$\mathrm{rank}([W;\mathrm{vec}(i H_m)])>r$}
    \State $W = [W;\mathrm{vec}(i H_m)]$
    \State $r$ = $r$ + 1
    \EndIf
\EndFor
\State $r_{\mathrm{old}} = 0$
\State $r_{\mathrm{new}} = \mathrm{rank}(W)$
\While{$r_{\mathrm{new}} \neq r_{\mathrm{old}}$\ \ and\ \ $r_{\mathrm{new}} < d_k(d_k -1)/2$}
    \For{$l=1, \dots, r_{\mathrm{new}}$}
        \For{$j=l+1, \dots, r_{\mathrm{new}}$}
            \State $H = [\mathrm{mat}(W_{\cdot,l}), \mathrm{mat}(W_{\cdot,j})]$
            \If{$\mathrm{rank}([W;\mathrm{vec}(H)]) > r$}
                \State $W = [W;\mathrm{vec}(H)]$
                \State $r = r + 1$
            \EndIf
        \EndFor
    \EndFor
    \State $r_{\mathrm{old}} = r_{\mathrm{new}}$
    \State $r_{\mathrm{new}} = \mathrm{rank}(W)$
\EndWhile
\State \textbf{return} $r_{\mathrm{new}}$
\end{algorithmic}
\end{algorithm}
Note that in lines 4,5 and 13,14, the square brackets $[W;x]$ mean "the matrix obtained from appending to the matrix $W$ an extra right-most column given as the column vector $x$"; while in line 12 the square brackets refer to the matrix commutator $[A,B]:=AB - BA$.

\section{Properties of the FBS gate}\label{chap:properties_FBS}

In this section, we remind some of the properties of FBS gates introduced in \cite{Kerenidis2022} to show this gate limitation in term of controllability. 

According to Definition~\ref{def:FBS}, the application of a FBS between the qubits $i$ and $j$, $FBS_{i,j}$, acts as $RBS_{i,j}$ if the parity of the qubits 'between' $i$ and $j$ is even, and is the conjugate gate $RBS_{i,j}(-\theta)$ otherwise. The $FBS_{i,j}$ is a non local gate that can be implemented using an RBS gate together with $\mathcal{O}(|i-j|)$ additional two qubit parity gates with a circuit of depth $\mathcal{O}(log(|i-j|))$. 

Althought similar to the RBS gate, the FBS is less controllable. As a HW-preserving gate, one can express the FBS gate as a block diagonal matrix (see Fig.~\ref{fig:RBS_circuit_block_unitary}), where each block $W^k$ is an unitary that represents the action of the FBS in the subspace of state of HW $k$. However, each block $W^k$ for $k>1$ is perfectly determined by $W^1$ as it is the k-compound matrix of $W^1$. Therefore, the controllability of the FBS unitary is upper-bounded by the controllability of the first block that represents the effect of the gate in the subspace of unitary states, i.e., states of HW 1. In the unitary basis, the FBS is equivalent of the RBS gate. We can thus conclude that the maximal dimension of the DLA for the FBS is $\frac{1}{2}n(n-1)$.

\begin{definition}[Compound matrix]
    Given a matrix $A \in \mathbb{R}^{n \times n}$, the compound matrix $\mathcal{A}^k$ for $k \in [n]$ is the $\binom{n}{k}$ dimensional matrix with entries $\mathcal{A}^k_{I J} = det(A_{I J})$ where $I$ and $J$ are subsets of rows and columns of $A$ with size $k$.  
\end{definition}

The limitation in its controllability is illustrated in Fig.~\ref{fig:DLA_evolution_connectivity}, and prevents the use of FBS for amplitude encoding on a subspace of HW $k$ with $k > 1$. However, FBS can be used to perform Clifford loader, a loader determined on the entire Hilbert space and restricted to the direct sum of subspace produce by the FBS. In \cite{Kerenidis2022}, algorithms for quantum determinant sampling, singular value estimation for compound matrices, and for Topological data analysis are presented using FBS gates and Clifford loader.

\section{Proof of Theorem~\ref{thm:QFIMrankthm}}
\label{chap:proof_QFIMrankthm}

To prove Theorem~\ref{thm:QFIMrankthm}, we first need a lemma from linear algebra, asserting that the rank associated to a list of vectors is the same as the rank of a Gram matrix of these vectors:

\begin{lem}
\label{lemma:rankGram-eq-rank}
Let $\mathcal{V}$ be a real vector space of finite dimension $d$, and let $\langle \cdot,\cdot\cdot  \rangle$ be any inner-product on $\mathcal{V}$. For a finite list of vectors $\mathcal{F}=(\vec{v}_{1},\dots,\vec{v}_{p})$ in $\mathcal{V}$, let $G_{\langle \cdot,\cdot\cdot  \rangle}(\mathcal{F})$ denote the corresponding $p \times p$ Gram-Matrix, i.e. $\big(G_{\langle \cdot,\cdot\cdot  \rangle}(\mathcal{F})\big)_{ij}:=\langle \vec{v}_{i} , \vec{v}_{j} \rangle$. Then, for all basis $\mathcal{B}$ of $\mathcal{V}$ (not necessarily orthonormal), it holds that: 
	\begin{equation}
		\mathrm{rank}[\, G_{\langle \cdot,\cdot\cdot  \rangle}(\mathcal{F}) \,] = \mathrm{rank}[\, F_{\mathcal{B}}\,],
	\end{equation}
	where $F_{\mathcal{B}}$ denotes the $d \times p$ matrix of the components of the vectors of $\mathcal{F}$ in basis $\mathcal{B}$.
\end{lem}

\begin{proof}
One can readily verify that for any basis $\mathcal{B}$ of $\mathcal{V}$: 
	\begin{equation}\label{eq:lemma-rankGram-eq-rank--eq1}
		G_{\langle \cdot,\cdot\cdot  \rangle}(\mathcal{F}) = (F_{\mathcal{B}})^{T} \cdot G_{\langle \cdot,\cdot\cdot  \rangle}(\mathcal{B}) \cdot (F_{\mathcal{B}}).
	\end{equation}
	Thus, 
	\begin{equation}
		\mathrm{rank}[\, G_{\langle \cdot,\cdot\cdot  \rangle}(\mathcal{F}) \,] = \mathrm{rank}[\, (F_{\mathcal{B}_{\mathrm{ON}}})^{T} (F_{\mathcal{B}_{\mathrm{ON}}}) \,] = \mathrm{rank}[\, F_{\mathcal{B}_{\mathrm{ON}}}\,] = \mathrm{rank}[\, F_{\mathcal{B}}\,],
	\end{equation}
	where the first equality is Eq.~\eqref{eq:lemma-rankGram-eq-rank--eq1} applied to a basis $\mathcal{B}_{\mathrm{ON}}$ of $\mathcal{V}$ that is orthonormal with respect to  $\langle \cdot,\cdot\cdot  \rangle$, the second equality uses the general fact $\mathrm{rank}[A^{T} A]=\mathrm{rank}[A]$ holding for any real-valued rectangular matrix $A$, and the third equality, in which $\mathcal{B}$ is another arbitrary basis of $\mathcal{V}$, is straight-forward (e.g. since $F_{\mathcal{B}_{\mathrm{ON}}} = P \cdot F_{\mathcal{B}}$ for some invertible matrix $P$).
\end{proof}

Let us recall the expression of the QFIM:

\QFIMdefinition*

We know explain in the following lemma why, when a ket is parametrized by means of an \textit{analytic} map, the rank of the QFIM must be almost-everywhere constant.
We recall that, for a map $f:\mathbb{R}^a \to \mathbb{R}^b$, $f$ is said to be \textit{analytic} if: 1) $f$ is \textit{smooth} ($C^\infty$) and 2) for each point $x \in \mathbb{R}^a$, $f$ coincides locally on a neighborhoud of $x$ with its Taylor series calculated at $x$. Note that when we refer to analyticity for a complex valued map $f:\mathbb{R}^a \to \mathbb{C}^b$, we mean the analycity of the corresponding map $\tilde{f}:\mathbb{R}^a \to \mathbb{R}^{2b}$ that splits the outputs into real and imaginary part. Lastly, we denote the Lebesgue measure on $\mathbb{R}^{p}$ by $\mu_{p}$, and if a measure or an "almost-everywhere" statement is made in the following, it is always with respect to the Lebesgue measure on the Euclidean space in question. 

\begin{lem}[Almost constant QFIM rank for analytic maps]
\label{lemma:rankQFIM-abstract}
	Consider an abstract map $\ket{\psi}:\Theta\to \mathcal{H}$ defining a parametrized ket with $p$ continuous parameters, where $\Theta$ is assumed to be an open-connected subset of $\mathbb{R}^p$, $\mathcal{H}$ is a finite-dimensional Hilbert space of dimension $d_{\mathcal{H}}$, and the outputs kets $\ket{\psi}(\bm{\theta})$ of the map are assumed to be normalized.
	Denote $r(\bm{\theta}):=\mathrm{rank}[\,\mathrm{QFIM}(\ket{\psi})(\bm{\theta})\,]$ and $r_{\mathrm{max}}:=\max\limits_{\bm{\theta} \in \Theta}\,r(\bm{\theta})$.
	
	If the map $\ket{\psi}$ is analytic, then $\mu_{p}(\{\bm{\theta} \in \Theta \,|\, r(\bm{\theta}) < r_{\mathrm{max}} \})=0$.
\end{lem}

\begin{proof}

	This follows from the fact that analytic maps have their jacobian be of constant rank almost-everywhere \cite[Prop. B.4]{Bamber-HowMany-1985}, along with the fact that the QFIM is the Gram matrix of the Jacobian's columns (for a certain inner-product). Let us explain this in more detail.
	
	For now, we consider a fixed $\bm{\theta} \in \Theta$. The vectors $\frac{\partial \ket{\psi}}{\partial \theta_{1}}(\bm{\theta}) ,\dots, \frac{\partial \ket{\psi}}{\partial \theta_{p}}(\bm{\theta})$ live in the \textit{tangent space} of $\mathcal{H}$ at the base point $\ket{\psi}(\bm{\theta}) \in \mathcal{H}$, which we denote as $\mathcal{V_{\bm{\theta}}}$. Working with only real dimensions/components, $\mathcal{V}_{\bm{\theta}}$ is  a real vector space (of dimension $2 d_{\mathcal{H}}$). Now, we recall the standard fact that the pure-state expression of $\mathrm{QFIM}(\ket{\psi})(\bm{\theta}) \,/\, 4$ (where $\mathrm{QFIM}(\ket{\psi})(\bm{\theta})$ is given by Eq.~\eqref{eq:QFIM}) is a matrix that is precisely the $\{\partial_{\theta_{i}}\}$-coordinate expression of the pullback along the map $\Theta \overset{\ket{\psi}}{\rightarrow} \mathcal{H} \overset{\Pi}{\rightarrow} \mathcal{PH}$ of a certain Riemmanian metric tensor on the projective Hilbert space $\mathcal{PH}$\footnote{(where the projective Hilbert space $\mathcal{PH}$ is the space of rays --- i.e. in which the points in $\mathcal{H}$ have been regrouped if they only differ by a global phase factor ---, and $\Pi$ denotes the map that send a point $\ket{\psi}\in\mathcal{H}$ to its corresponding ray)}, called the real-part of the \textit{Fubini-Study metric} (for more information, see e.g. \cite{Heydari-GeometricFormulation-2016,Facchi-ClassicalQuantum-2010,Liu2019}). From this, we only need to retain the fact that: there exists a certain inner-product  $\langle \cdot,\cdot\cdot  \rangle_{\bm{\theta}}$ on $\mathcal{V}_{\bm{\theta}}$ such that:
	\begin{equation}\label{eq:lemma-rankQFIM--eq1}
		\frac{1}{4}\mathrm{QFIM}(\ket{\psi})(\bm{\theta}) = G_{\langle \cdot,\cdot\cdot  \rangle_{\bm{\theta}}}(\mathcal{F}_{\bm{\theta}}),
	\end{equation} for $\mathcal{F}_{\bm{\theta}}:=\big(\frac{\partial \ket{\psi}}{\partial \theta_{1}}(\bm{\theta}) ,\dots, \frac{\partial \ket{\psi}}{\partial \theta_{p}}(\bm{\theta})\big)$. By lemma \ref{lemma:rankGram-eq-rank}, the fact alone that there exists an inner-product such that Eq.~\eqref{eq:lemma-rankQFIM--eq1} holds implies (without needing to make that inner-product explicit) that: 
	\begin{equation}\label{eq:lemma-rankQFIM--eq2}
		\text{rank}[\, \mathrm{QFIM}(\ket{\psi})(\bm{\theta}) \,] = \text{rank}[\, F_{\bm{\theta},\mathcal{B}}\,],
	\end{equation} with $F_{\bm{\theta},\mathcal{B}}$ being the matrix of the vectors $\mathcal{F}$ in any basis $\mathcal{B}$.
	
	Letting go of a fixed $\bm{\theta}$, and going back to our map $\ket{\psi}:\Theta \to \mathcal{H}$, with the identification of $\mathcal{H}$ as $\mathbb{R}^{2d_{\mathcal{H}}}$ so that the usual representation of the jacobian matrix of $J(\ket{\psi})(\bm{\theta})$ of this map \big($[J(\ket{\psi})(\bm{\theta})]_{ij}:=\frac{\partial \ket{\psi}_{i}}{\partial \theta_{j}}$\big)  is equal to $J(\ket{\psi})(\bm{\theta})=F_{\bm{\theta},\mathcal{B}_{c}}$ (with $\mathcal{B}_{c}$ the canonical basis of $\mathbb{R}^{2d_{\mathcal{H}}}$), we have by Eq.~\eqref{eq:lemma-rankQFIM--eq2} that 
	\begin{equation}\label{eq:lemma-rankQFIM--eq3}
		\forall \in \Theta\ \,r(\bm{\theta}) = \text{rank}[\, J(\ket{\psi})(\bm{\theta})\,].
	\end{equation} But since the map $\ket{\psi}:\Theta \to \mathcal{H}$ is analytic on an open-connected domain, it holds by the result \cite[Prop. B.4]{Bamber-HowMany-1985} that for \textit{almost all} $\bm{\theta}$,  $\text{rank}[\, J(\ket{\psi})(\bm{\theta})\,]$ is constant, equal to its maximal-reached value over $\Theta$. Combining this with Eq.~\eqref{eq:lemma-rankQFIM--eq3}, the claim is established.

\end{proof}

To clarify the statement that will follow, we now precise here what we mean in general by a VQC:
\begin{definition}[VQC]\label{def:VQC}
	We refer to \textit{variational quantum circuit} (VQC) on a Hilbert space $\mathcal{H}$ for any unitary on $\mathcal{H}$ parametrized by $p\geq 1$ continuous parameters in the general form:
	\begin{equation}\label{eq:VQC-general-form-unitary}
		U\big(\bm{\theta}=(\theta_1,\dots,\theta_p)\big)=V_p(e^{-i \theta_p H_{p}}) \cdots V_1 (e^{-i \theta_1 H_{1}})V_0,
	\end{equation}
	where $V_0,\dots V_p$ are fixed $d_\mathcal{H} \times d_\mathcal{H}$ unitary matrices and $H_1\dots H_p$ are fixed $d_\mathcal{H} \times d_\mathcal{H}$ hermitian matrices, with $d_\mathcal{H}:=\dim(\mathcal{H})$. Additionaly, for a chosen input state $\ket{0}\in \mathcal{H}$, the output parametrized ket of this VQC is
	\begin{equation}\label{eq:VQC-general-form-outputstate}
		\ket{\psi}(\bm{\theta}) := U(\bm{\theta})\ket{0}.
	\end{equation}
\end{definition}

Lastly, the following lemma covers the main text Theorem \ref{thm:QFIMrankthm}, and asserts that the almost-constant QFIM rank property holds for output states of a VQC, for any restriction on the parameters, and that the maximum rank observed does not depend on that restriction.

\begin{lem}[Almost constant QFIM rank for VQCs with any parameter ranges]
\label{lemma:rankQFIM-VQC}
	If $\ket{\psi}:\mathbb{R}^p \to \mathcal{H}$ is the output parametrized ket of any VQC with $p$ continuous parameters (Eqs.~\eqref{eq:VQC-general-form-unitary},\eqref{eq:VQC-general-form-outputstate}), then for an arbitrary (measurable, and with positive measure) subset $\Theta \subseteq \mathbb{R}^p$ chosen as a restricted set of parameters, the value of $r_{\mathrm{max}}:=\max\limits_{\bm{\theta} \in \Theta}r(\bm{\theta})$ is indepedent of $\Theta$, and
	$\mu_{p}(\{\bm{\theta} \in \Theta \,|\, r(\bm{\theta}) < r_{\mathrm{max}} \})=0$.
	
	Consequently, if a point $\bm{\theta}\in \Theta$ is drawn randomly according to any (absolutely continuous w.r.t Lebesgue measure) probability distribution over $\Theta$, then the resulting value of $r(\bm{\theta})$ will be equal to $r_{\mathrm{max}}$ with probability one.
\end{lem}
\begin{proof}
	Let $\Theta \subseteq \mathbb{R}^p$ be an arbitrary --- measurable, and of positive measure --- set of parameters. Since the map $\ket{\psi}:\mathbb{R}^p \to \mathcal{H}$ is analytic (as Eqs.~\eqref{eq:VQC-general-form-unitary},\eqref{eq:VQC-general-form-outputstate} are made only of matrix exponentials and products, which are analytic), Lemma \ref{lemma:rankQFIM-abstract} applies to it, giving that the set $\Omega:=\{\bm{\theta} \in \mathbb{R}^p \,|\, r(\bm{\theta}) = \max\limits_{\bm{\theta} \in \mathbb{R}^p}r(\bm{\theta}) \}$ is a full-measure subset of $\mathbb{R}^p$ (i.e. its complement in $\mathbb{R}^p$ has measure zero). Therefore, $\Omega':=\Omega \cap \Theta$ is a full-measure subset of $\Theta$, i.e. $\Theta \setminus \Omega'= \{\bm{\theta} \in \Theta \,|\, r(\bm{\theta}) < \max\limits_{\bm{\theta} \in \mathbb{R}^p}r(\bm{\theta}) \}$, has measure zero. Furthermore since $\Omega'$ has full-measure in $\Theta$, it is in particular non-empty, which implies that $\max\limits_{\bm{\theta} \in \Theta}r(\bm{\theta}) = \max\limits_{\bm{\theta} \in \mathbb{R}^p}r(\bm{\theta})$ --- hence the left-hand-side is independent of $\Theta$, and the claim is established.
\end{proof}

\section{Proof of Lemma ~\ref{lemma:VarianceHWPreserving}}\label{chap:proof_Lemma_Var}

We recall the Lemma~\ref{lemma:VarianceHWPreserving}:

\VarianceHWPreserving*
\begin{proof}

To prove this Lemma, we will first consider the unary case in Section~\ref{chap:proof_lemma_unary}, i.e., the case where the states are in $B^1_n$. In the unary case, RBS and FBS are equivalent. Then, we will show how to extend our result to any HW $k$ for RBS in Section~\ref{chap:proof_lemma_general_RBS}. Finally, we will explain in Section~\ref{chap:proof_lemma_general_FBS} how to adapt this result to the FBS case.

    \subsection{Unary case}\label{chap:proof_lemma_unary}

We consider the case of the squared Euclidean distance cost function. We call $\Delta^L$ the final error: 
\begin{equation}
    \Delta^L = 2(z^L - y) = 2[(w^{\lambda_{\mathrm{max}}} \cdots w^{\lambda + 1}) \cdot w^{\lambda} \cdot \zeta^{\lambda} - y]    
\end{equation}

We consider the case where for each inner layer, there is only one RBS gate considered. The number of parameters $D$ is equal to the number of inner layers $\lambda_{\mathrm{max}}$. We call $\zeta_j$ the amplitude of the $j^{\text{th}}$ state in the state basis considered of $\zeta$. For each inner layer, the action of the gate in the state basis $B_1^n$ results in a rotation of the amplitudes for two states that we call $(l,j)$:
\begin{equation}
    \zeta^{\lambda + 1} = \overrightarrow{cst} + (\cos(\theta_{i}) \cdot \zeta^{\lambda}_{l} + \sin(\theta_{i}) \cdot \zeta^{\lambda}_{j}) \ket{e_l} + (-\sin(\theta_{i}) \cdot \zeta^{\lambda}_{l} + \cos(\theta_{i}) \cdot \zeta^{\lambda}_{j}) \ket{e_j}  
\end{equation}
with $\overrightarrow{cst}^\intercal \cdot \ket{e_j} = \overrightarrow{cst}^\intercal \cdot \ket{e_l} = 0$

For example, the action of a RBS with parameter $\theta_i$ in $B_1^n$ on the two first qubits of a 4-qubit quantum circuit results in the $\theta_i$-planar rotation between the states $\ket{1000}$ and $\ket{0100}$.   

We can define the error according to the final error:
\begin{equation}
    \delta^{\lambda + 1} = (w^{\lambda +1})^{-1} \cdots (w^{\lambda_{\mathrm{max}}})^{-1} \cdot \Delta^l = 2 [w^{\lambda} \cdot \zeta^{\lambda} - (w^{\lambda +1})^{-1} \cdots (w^{\lambda_{\mathrm{max}}})^{-1} \cdot y]
\end{equation}

We have by orthogonality: $\forall \lambda, \quad (w^{\lambda})^{-1} = (w^{\lambda})^{t}$.
    
We call: $\tilde{y} = (w^{\lambda +1})^{t} \cdots (w^{\lambda_{\mathrm{max}}})^{t} \cdot y$ and $\Theta = [0:2\pi]^D$

We use the notation $\tilde{y}_j$ for the amplitude of the $j^{\text{th}}$ state in the state basis considered of $\tilde{y}$.

According to the backpropagation formalism presented in Section~\ref{chap:backpropagation}, we have in the unary case:
\begin{equation}
    \frac{\partial \mathcal{C}}{\partial \theta_i} = \delta_l^{\lambda} (-\sin(\theta_i) \zeta_l^{\lambda} + \cos(\theta_i)\zeta_j^{\lambda}) + \delta_j^{\lambda} (-\cos(\theta_i) \zeta_l^{\lambda} - \sin(\theta_i)\zeta_j^{\lambda})
\end{equation}

Therefore, we can express the variance of the cost function gradient as:

\begin{equation}
    \begin{split}
    \mathrm{Var}_{\bm{\theta}}[\partial_{\theta_i} \mathcal{C}(\bm{\theta})]  &= \mathrm{Var}_{\bm{\theta}} [\delta_l^{\lambda} (-\sin(\theta_i) \zeta_l^{\lambda} + \cos(\theta_i)\zeta_j^{\lambda}) + \delta_j^{\lambda} (-\cos(\theta_i) \zeta_l^{\lambda} - \sin(\theta_i)\zeta_j^{\lambda})] \\
    & = \mathrm{Var}_{\bm{\theta}}[\delta_l^{\lambda} (-\sin(\theta_i) \zeta_l^{\lambda} + \cos(\theta_i)\zeta_j^{\lambda})] + \mathrm{Var}_{\bm{\theta}} [\delta_j^{\lambda} (-\cos(\theta_i) \zeta_l^{\lambda} - \sin(\theta_i)\zeta_j^{\lambda})] \\
    & \;+ 2 \cdot \mathbb{C}ov_{\bm{\theta}} [\delta_l^{\lambda} (-\sin(\theta_i) \zeta_l^{\lambda} + \cos(\theta_i)\zeta_j^{\lambda}) ; \delta_j^{\lambda} (-\cos(\theta_i) \zeta_l^{\lambda} - \sin(\theta_i)\zeta_j^{\lambda})]
    \end{split} 
\end{equation}    

First:
\begin{equation}
    \begin{split}
    \mathrm{Var}_{\bm{\theta}}&[\delta_l^{\lambda} (-\sin(\theta_i) \zeta_l^{\lambda} + \cos(\theta_i)\zeta_j^{\lambda})] = \mathrm{Var}_{\bm{\theta}}[2(\cos(\theta_{i}) \cdot \zeta^{\lambda}_{l} + \sin(\theta_{i}) \cdot \zeta^{\lambda}_{j} - \tilde{y}_l)) \cdot (-\sin(\theta_i) \zeta_l^{\lambda} + \cos(\theta_i)\zeta_j^{\lambda})]\\
    & = 4 \int_{\bm{\theta} \in \Theta} (\frac{1}{2\pi})^D (\delta_l^{\lambda} (-\sin(\theta_i) \zeta_l^{\lambda} + \cos(\theta_i)\zeta_j^{\lambda}) - \mathbb{E}_{\bm{\theta}}[\delta_l^{\lambda} (-\sin(\theta_i) \zeta_l^{\lambda} + \cos(\theta_i)\zeta_j^{\lambda})])^2 d\bm{\theta}  
    \end{split} 
\end{equation}

With:
\begin{equation}
    \begin{split}
    \mathbb{E}_{\bm{\theta}}&[\delta_l^{\lambda} (-\sin(\theta_i) \zeta_l^{\lambda} + \cos(\theta_i)\zeta_j^{\lambda})] =  \int_{\bm{\theta} \in \Theta} (\frac{1}{2\pi})^D (2(\cos(\theta_{i}) \cdot \zeta^{\lambda}_{l} + \sin(\theta_{i}) \cdot \zeta^{\lambda}_{j} - \tilde{y}_l)) \cdot (-\sin(\theta_i) \zeta_l^{\lambda} + \cos(\theta_i)\zeta_j^{\lambda})) d\bm{\theta} \\
    & = 2 \int_{\bm{\theta} \in \Theta} (\frac{1}{2\pi})^D (\cos(\theta_i) \sin(\theta_i) ((\zeta_j^{\lambda})^2 - (\zeta_l^{\lambda})^2) d\bm{\theta} + 2 \int_{\bm{\theta} \in \Theta} (\frac{1}{2\pi})^D (\cos^2(\theta_i) - \sin^2(\theta_i)) \zeta_j^{\lambda} \cdot \zeta_l^{\lambda} d\bm{\theta} \\
    & \quad + 2 \int_{\bm{\theta} \in \Theta} (\frac{1}{2\pi})^D (\sin(\theta_i) \cdot \zeta_l^{\lambda} - \cos(\theta_i) \cdot \zeta_j^{\lambda}) \tilde{y}_l d\bm{\theta}
    \end{split}     
\end{equation}

According to our circuit decomposition into inner layers, the previous inner layer $\zeta^{\lambda}$ does not depend on the parameter $\theta_i$ but only on the previous parameters in the circuit. On the other hand, $\tilde{y}$ does not depend on the parameter $\theta_i$ but only on the following parameters in the circuit. Thus, we can take out the integral according to $\theta_i$:

\begin{equation}\label{eq:Expected_value_nul}
    \mathbb{E}_{\bm{\theta}}[\delta_l^{\lambda} (-\sin(\theta_i) \zeta_l^{\lambda} + \cos(\theta_i)\zeta_j^{\lambda})] = 2 \int_{\bm{\theta} \in \Theta \backslash \theta_i } (\frac{1}{2\pi})^D (\pi - \pi) \zeta_j^{\lambda} \cdot \zeta_l^{\lambda} d\bm{\theta} = 0    
\end{equation}

Therefore:
\begin{equation}
    \begin{split}
    \mathrm{Var}_{\bm{\theta}}&[\delta_l^{\lambda} (-\sin(\theta_i) \zeta_l^{\lambda} + \cos(\theta_i)\zeta_j^{\lambda})] = 4 \int_{\bm{\theta} \in \Theta} (\frac{1}{2\pi})^D (\delta_l^{\lambda} (-\sin(\theta_i) \zeta_l^{\lambda} + \cos(\theta_i)\zeta_j^{\lambda}))^2 d\bm{\theta} \\
    & = 4 \int_{\bm{\theta} \in \Theta} (\frac{1}{2\pi})^D [\cos^2(\theta_i) \sin^2(\theta_i) \cdot ((\zeta_j^{\lambda})^2 - (\zeta_l^{\lambda})^2)^2
    + 2 \cos^3(\theta_i) \sin(\theta_i) \zeta_j^{\lambda} \zeta_l^{\lambda}((\zeta_j^{\lambda})^2 - (\zeta_l^{\lambda})^2) \\
    & - 2 \cos(\theta_i) \sin^3(\theta_i) \zeta_j^{\lambda} \zeta_l^{\lambda}((\zeta_j^{\lambda})^2 - (\zeta_l^{\lambda})^2) + 2 \cos(\theta_i) \sin^3(\theta_i) \zeta_l^{\lambda}((\zeta_j^{\lambda})^2 - (\zeta_l^{\lambda})^2)\tilde{y}_l \\
    & -2 \cos^2(\theta_i) \sin(\theta_i) \zeta_j^{\lambda}((\zeta_j^{\lambda})^2 - (\zeta_l^{\lambda})^2)\tilde{y}_l + [\cos^4(\theta_i) - 2 \cos^2(\theta_i) \sin^2(\theta_i) + \sin^4(\theta_i)](\zeta_l^{\lambda})^2(\zeta_j^{\lambda})^2 \\
    & + 2 \cos^2(\theta_i) \sin(\theta_i)(\zeta_l^{\lambda})^2 \zeta_j^{\lambda} \tilde{y}_l - 2 \sin^3(\theta_i)(\zeta_l^{\lambda})^2 \zeta_j^{\lambda} \tilde{y}_l - 2 \cos^3(\theta_i) \zeta_l^{\lambda} (\zeta_j^{\lambda})^2  \tilde{y}_l + 2 \cos(\theta_i) \sin^2(\theta_i) \zeta_l^{\lambda} (\zeta_j^{\lambda})^2  \tilde{y}_l \\
    & + \sin^2(\theta_i) (\zeta_l^{\lambda})^2 (\tilde{y}_l)^2 - 2 \cos(\theta_i) \sin(\theta_i) \zeta_l^{\lambda} \zeta_j^{\lambda} (\tilde{y}_l)^2 + \cos^2(\theta_i) (\zeta_j^{\lambda})^2 (\tilde{y}_l)^2 d\bm{\theta} \\
    & = 4 \int_{\bm{\theta} \in \Theta \backslash \theta_i} (\frac{1}{2\pi})^D [\frac{\pi}{4} (\zeta_l^{\lambda})^4 + \frac{\pi}{2} (\zeta_l^{\lambda})^2 (\zeta_j^{\lambda})^2 + \frac{\pi}{4} (\zeta_j^{\lambda})^4 + \pi (\zeta_l^{\lambda})^2 (\tilde{y}_l)^2 + \pi (\zeta_j^{\lambda})^2 (\tilde{y}_l)^2] d\bm{\theta}
    \end{split} 
\end{equation}

Finally, we have:
\begin{equation}
    \mathrm{Var}_{\bm{\theta}}[\delta_l^{\lambda} (-\sin(\theta_i) \zeta_l^{\lambda} + \cos(\theta_i)\zeta_j^{\lambda})] = \int_{\bm{\theta} \in \Theta} (\frac{1}{2 \pi})^D [\frac{1}{2}(\zeta_l^{\lambda})^4 + (\zeta_l^{\lambda})^2 (\zeta_j^{\lambda})^2 + \frac{1}{2}(\zeta_j^{\lambda})^4 + 2 ((\zeta_l^{\lambda})^2 + (\zeta_j^{\lambda})^2) (\tilde{y}_l)^2] d\bm{\theta}
\end{equation}
    
With the same methods, it comes:
\begin{equation}
    \mathrm{Var}_{\bm{\theta}}[\delta_j^{\lambda} (-\cos(\theta_i) \zeta_l^{\lambda} - \sin(\theta_i)\zeta_j^{\lambda})] = \int_{\bm{\theta} \in \Theta} (\frac{1}{2 \pi})^D [\frac{1}{2}(\zeta_l^{\lambda})^4 + (\zeta_l^{\lambda})^2 (\zeta_j^{\lambda})^2 + \frac{1}{2}(\zeta_j^{\lambda})^4 + 2 ((\zeta_l^{\lambda})^2 + (\zeta_j^{\lambda})^2) (\tilde{y}_l)^2] d\bm{\theta}
\end{equation}

We can decompose the covariance term:
\begin{equation}
    \begin{split}
    \mathbb{C}ov_{\bm{\theta}} & [\delta_l^{\lambda} (-\sin(\theta_i) \zeta_l^{\lambda} + \cos(\theta_i)\zeta_j^{\lambda}) ; \delta_j^{\lambda} (-\cos(\theta_i) \zeta_l^{\lambda} - \sin(\theta_i)\zeta_j^{\lambda})]\\
    & = \mathbb{E}_{\bm{\theta}}[\delta_l^{\lambda} (-\sin(\theta_i) \zeta_l^{\lambda} + \cos(\theta_i)\zeta_j^{\lambda}) \cdot \delta_j^{\lambda} (-\cos(\theta_i) \zeta_l^{\lambda} - \sin(\theta_i)\zeta_j^{\lambda})]\\
    & - \mathbb{E}_{\bm{\theta}}[\delta_l^{\lambda} (-\sin(\theta_i) \zeta_l^{\lambda} + \cos(\theta_i)\zeta_j^{\lambda})] \cdot \mathbb{E}_{\bm{\theta}}[\delta_j^{\lambda} (-\cos(\theta_i) \zeta_l^{\lambda} - \sin(\theta_i)\zeta_j^{\lambda})]
    \end{split} 
\end{equation}

As shown with Eq.~\eqref{eq:Expected_value_nul}: 
\begin{equation}
    \mathbb{E}_{\bm{\theta}}[\delta_l^{\lambda} (-\sin(\theta_i) \zeta_l^{\lambda} + \cos(\theta_i)\zeta_j^{\lambda})] = \mathbb{E}_{\bm{\theta}}[\delta_j^{\lambda} (-\cos(\theta_i) \zeta_l^{\lambda} - \sin(\theta_i)\zeta_j^{\lambda})] = 0
\end{equation}

\textbf{Therefore, we have that the expectation value of the cost function gradient is null}. In addition, we have:

\begin{equation}
        \begin{split}
            \mathbb{C}ov_{\bm{\theta}} & [\delta_l^{\lambda} (-\sin(\theta_i) \zeta_l^{\lambda} + \cos(\theta_i)\zeta_j^{\lambda}) ; \delta_j^{\lambda} (-\cos(\theta_i) \zeta_l^{\lambda} - \sin(\theta_i)\zeta_j^{\lambda})]\\
            & = \mathbb{E}_{\bm{\theta}}[\delta_l^{\lambda} (-\sin(\theta_i) \zeta_l^{\lambda} + \cos(\theta_i)\zeta_j^{\lambda}) \cdot \delta_j^{\lambda} (-\cos(\theta_i) \zeta_l^{\lambda} - \sin(\theta_i)\zeta_j^{\lambda})] \\
            &= \int_{\bm{\theta} \in \Theta} (\frac{1}{2\pi})^D [\delta_l^{\lambda} (-\sin(\theta_i) \zeta_l^{\lambda} + \cos(\theta_i)\zeta_j^{\lambda}) \cdot \delta_j^{\lambda} (-\cos(\theta_i) \zeta_l^{\lambda} - \sin(\theta_i)\zeta_j^{\lambda})] d\bm{\theta} \\
            & = \int_{\bm{\theta} \in \Theta} (\frac{1}{2\pi})^D [\cos^2(\theta_i) \sin^2(\theta_i)(-(\zeta^{\lambda}_l)^4 + 2(\zeta^{\lambda}_l)^2 (\zeta^{\lambda}_j)^2 - (\zeta^{\lambda}_j)^4)\\
            & + \cos(\theta_i)\sin(\theta_i)(\sin^2(\theta_i)-\cos^2(\theta_i))((\zeta^{\lambda}_j)^2 - (\zeta^{\lambda}_l)^2)\zeta^{\lambda}_l\zeta^{\lambda}_j + \cos^2(\theta_i)\sin(\theta_i)((\zeta^{\lambda}_j)^2 - (\zeta^{\lambda}_l)^2)\zeta^{\lambda}_l \tilde{y}^{\lambda}_j \\
            & - \cos(\theta_i)\sin^2(\theta_i)((\zeta^{\lambda}_j)^2 - (\zeta^{\lambda}_l)^2)\zeta^{\lambda}_j \tilde{y}^{\lambda}_j + \cos(\theta_i)\sin(\theta_i)(\cos^2(\theta_i)-\sin^2(\theta_i))((\zeta^{\lambda}_j)^2 - (\zeta^{\lambda}_l)^2)\zeta^{\lambda}_l\zeta^{\lambda}_j \\
            & - (\cos^4(\theta_i) - 2\cos^2(\theta_i)\sin^2(\theta_i) + \sin^4(\theta_i))(\zeta^{\lambda}_l\zeta^{\lambda}_j)^2 + \cos(\theta_i)(\cos^2(\theta_i)-\sin^2(\theta_i))(\zeta^{\lambda}_l)^2\zeta^{\lambda}_j\tilde{y}^{\lambda}_j \\
            & = - \sin(\theta_i)(\cos^2(\theta_i)-\sin^2(\theta_i))\zeta^{\lambda}_l(\zeta^{\lambda}_j)^2\tilde{y}^{\lambda}_j + \cos(\theta_i)\sin^2(\theta_i)((\zeta^{\lambda}_j)^2 - (\zeta^{\lambda}_l)^2)\zeta^{\lambda}_l \tilde{y}^{\lambda}_l \\
            & + \sin(\theta_i)(\sin^2(\theta_i) - \cos^2(\theta_i))(\zeta^{\lambda}_l)^2\zeta^{\lambda}_j \tilde{y}^{\lambda}_l + \cos(\theta_i) \sin(\theta_i) (\zeta^{\lambda}_l)^2 \tilde{y}^{\lambda}_l \tilde{y}^{\lambda}_j - \sin^2(\theta_i) \zeta^{\lambda}_l \zeta^{\lambda}_j \tilde{y}^{\lambda}_l \tilde{y}^{\lambda}_j \\
            & - \cos^2(\theta_i) \sin(\theta_i) ((\zeta^{\lambda}_l)^2 - (\zeta^{\lambda}_j)^2) \zeta^{\lambda}_j \tilde{y}^{\lambda}_l - \cos(\theta_i)(\sin^2(\theta_i) - \cos^2(\theta_i))\zeta^{\lambda}_l(\zeta^{\lambda}_j)^2 \tilde{y}^{\lambda}_l \\
            & + \cos^2(\theta_i) \zeta^{\lambda}_l \zeta^{\lambda}_j \tilde{y}^{\lambda}_l \tilde{y}^{\lambda}_j + \cos(\theta_i) \sin(\theta_i) \zeta^{\lambda}_j \tilde{y}^{\lambda}_l \tilde{y}^{\lambda}_j ] d\bm{\theta} \\
            & = - \frac{1}{2} \int_{\bm{\theta} \in \Theta} (\frac{1}{2\pi})^D (\zeta^{\lambda}_l)^4 d\bm{\theta} - \frac{1}{2} \int_{\bm{\theta} \in \Theta} (\frac{1}{2\pi})^D (\zeta^{\lambda}_j)^4 d\bm{\theta} - \int_{\bm{\theta} \in \Theta} (\frac{1}{2\pi})^D (\zeta^{\lambda}_l)^2 (\zeta^{\lambda}_j)^2 d\bm{\theta}
        \end{split} 
    \end{equation}

 We can derive the variance of the cost gradient:
    \begin{equation}
        \begin{split}
        \mathrm{Var}_{\bm{\theta}}[\partial_{\theta_i} \mathcal{C}(\bm{\theta})]  =& \mathrm{Var}_{\bm{\theta}}[\delta_l^{\lambda} (-\sin(\theta_i) \zeta_l^{\lambda} + \cos(\theta_i)\zeta_j^{\lambda})] + \mathrm{Var}_{\bm{\theta}} [\delta_j^{\lambda} (-\cos(\theta_i) \zeta_l^{\lambda} - \sin(\theta_i)\zeta_j^{\lambda})] \\
        & \;+ 2 \cdot \mathbb{C}ov_{\bm{\theta}} [\delta_l^{\lambda} (-\sin(\theta_i) \zeta_l^{\lambda} + \cos(\theta_i)\zeta_j^{\lambda}) ; \delta_j^{\lambda} (-\cos(\theta_i) \zeta_l^{\lambda} - \sin(\theta_i)\zeta_j^{\lambda})] \\
        & = \int_{\bm{\theta} \in \Theta} (\frac{1}{2 \pi})^D [\frac{1}{2}(\zeta_l^{\lambda})^4 + (\zeta_l^{\lambda})^2 (\zeta_j^{\lambda})^2 + \frac{1}{2}(\zeta_j^{\lambda})^4 + 2 ((\zeta_l^{\lambda})^2 + (\zeta_j^{\lambda})^2) (\tilde{y}_l)^2] d\bm{\theta} \\
        & + \int_{\bm{\theta} \in \Theta} (\frac{1}{2 \pi})^D [\frac{1}{2}(\zeta_l^{\lambda})^4 + (\zeta_l^{\lambda})^2 (\zeta_j^{\lambda})^2 + \frac{1}{2}(\zeta_j^{\lambda})^4 + 2 ((\zeta_l^{\lambda})^2 + (\zeta_j^{\lambda})^2) (\tilde{y}_l)^2] d\bm{\theta} \\
        & - \int_{\bm{\theta} \in \Theta} (\frac{1}{2\pi})^D (\zeta^{\lambda}_l)^4 d\bm{\theta} - \int_{\bm{\theta} \in \Theta} (\frac{1}{2\pi})^D (\zeta^{\lambda}_j)^4 d\bm{\theta} - 2 \int_{\bm{\theta} \in \Theta} (\frac{1}{2\pi})^D (\zeta^{\lambda}_l)^2 (\zeta^{\lambda}_j)^2 d\bm{\theta}
        \end{split}
    \end{equation}

We find that:
    \begin{equation}
        \mathrm{Var}_{\bm{\theta}}[\partial_{\theta_i} \mathcal{C}(\bm{\theta})]  = 2 \left(\int_{\bm{\theta} \in \Theta} (\frac{1}{2\pi})^D (\zeta^{\lambda}_l)^2 + (\zeta^{\lambda}_j)^2 d\bm{\theta}  \right) \cdot \left(\int_{\bm{\theta} \in \Theta} (\frac{1}{2\pi})^D (\tilde{y}^{\lambda}_l)^2 + (\tilde{y}^{\lambda}_j)^2 d\bm{\theta} \right) 
    \end{equation}

    \subsection{Extension to any HW for RBS based VQC}\label{chap:proof_lemma_general_RBS}

We consider once again the case of the squared Euclidean distance cost function. We call $\Delta^L$ the final error: 
    \begin{equation}
        \Delta^L = 2(z^L - y) = 2[(w^{\lambda_{\mathrm{max}}} \cdots w^{\lambda + 1}) \cdot w^{\lambda} \cdot \zeta^{\lambda} - y]    
    \end{equation}

    We still consider the case where for each inner layer, there is only one RBS gate considered. For each inner layer, the action of the gate results in a rotation of the amplitudes between a set of pair of states that we call $\mathcal{R}_{\lambda}$:
    \begin{equation}
        \zeta^{\lambda + 1} = \overrightarrow{cst} + \sum_{(l,j) \in \mathcal{R}_{\lambda}} (\cos(\theta_{i}) \cdot \zeta^{\lambda}_{l} + \sin(\theta_{i}) \cdot \zeta^{\lambda}_{j}) \ket{e_l} + (-\sin(\theta_{i}) \cdot \zeta^{\lambda}_{l} + \cos(\theta_{i}) \cdot \zeta^{\lambda}_{j}) \ket{e_j}  
    \end{equation}
    with $\forall (l,j)\in \mathcal{R}_{\lambda}, \quad \overrightarrow{cst}^\intercal \cdot \ket{e_j} = \overrightarrow{cst}^\intercal \cdot \ket{e_l} = 0$

    For example, the action of a RBS with parameter $\theta_i$ in $B_2^n$ on the two first qubits of a 4-qubit quantum circuit results in the $\theta_i$-planar rotation between the states $\ket{1010}$ and $\ket{0110}$, but also in the same $\theta_i$-planar rotation between the states $\ket{1001}$ and $\ket{0101}$.   

    We can define the error according to the final error:
    \begin{equation}
        \begin{split}
            \delta^{\lambda + 1} &= (w^{\lambda +1})^{-1} \cdots (w^{\lambda_{\mathrm{max}}})^{-1} \cdot \Delta^l = 2 [w^{\lambda} \cdot \zeta^{\lambda} - (w^{\lambda +1})^{-1} \cdots (w^{\lambda_{\mathrm{max}}})^{-1} \cdot y] \\
            &= 2[\overrightarrow{cst} + \sum_{(l,j) \in \mathcal{R}_{\lambda}} (\cos(\theta_{i}) \cdot \zeta^{\lambda}_{l} + \sin(\theta_{i}) \cdot \zeta^{\lambda}_{j}) \ket{e_l} + (-\sin(\theta_{i}) \cdot \zeta^{\lambda}_{l} + \cos(\theta_{i}) \cdot \zeta^{\lambda}_{j}) \ket{e_j} - \tilde{y}^{\lambda}]
        \end{split}
    \end{equation}
    A unique RBS affects $\binom{n-2}{k-1}$ different pairs of states with a rotation, and all the affected states are different. As before, we have:
    \begin{equation}
        \forall (l,j) \in \mathcal{R}_{\lambda}, \quad \delta^{\lambda + 1}_l = 2[\cos(\theta_{i}) \cdot \zeta^{\lambda}_{l} + \sin(\theta_{i}) \cdot \zeta^{\lambda}_{j} - \tilde{y}^{\lambda}_l]
    \end{equation}
    and,
    \begin{equation}
        \forall (l,j) \in \mathcal{R}_{\lambda}, \quad \delta^{\lambda + 1}_j = 2[-\sin(\theta_{i}) \cdot \zeta^{\lambda}_{l} + \cos(\theta_{i}) \cdot \zeta^{\lambda}_{j} - \tilde{y}^{\lambda}_j]
    \end{equation}
    
    We have by orthogonality: $\forall \lambda, \quad (w^{\lambda})^{-1} = (w^{\lambda})^{t}$.
    
    We call: $\tilde{y} = (w^{\lambda +1})^{t} \cdots (w^{\lambda_{\mathrm{max}}})^{t} \cdot y$ and $\Theta = [0:2\pi]^D$
    In the following, we will omit to note the set $\mathcal{R}_{\lambda}$.
    
    \begin{equation}
        \begin{split}
            \mathrm{Var}_{\bm{\theta}}&[\partial_{\theta_i} \mathcal{C}] = \mathrm{Var}_{\bm{\theta}}[\sum_{(l,j)} \delta_l^{\lambda} (-\sin(\theta_i) \zeta_l^{\lambda} + \cos(\theta_i)\zeta_j^{\lambda}) + \delta_j^{\lambda} (-\cos(\theta_i) \zeta_l^{\lambda} - \sin(\theta_i)\zeta_j^{\lambda})] \\
            & = \sum_{(l,j)} \mathrm{Var}_{\bm{\theta}}[\delta_l^{\lambda} (-\sin(\theta_i) \zeta_l^{\lambda} + \cos(\theta_i)\zeta_j^{\lambda}) + \delta_j^{\lambda} (-\cos(\theta_i) \zeta_l^{\lambda} - \sin(\theta_i)\zeta_j^{\lambda})] \\
            & + 2 \sum_{(a,b) \neq (c,d)} \mathbb{C}ov_{\bm{\theta}} [\delta_a^{\lambda} (-\sin(\theta_i) \zeta_a^{\lambda} + \cos(\theta_i)\zeta_b^{\lambda}) + \delta_b^{\lambda} (-\cos(\theta_i) \zeta_a^{\lambda} - \sin(\theta_i)\zeta_b^{\lambda}), \\
            & \delta_c^{\lambda} (-\sin(\theta_i) \zeta_c^{\lambda} + \cos(\theta_i)\zeta_d^{\lambda}) + \delta_d^{\lambda} (-\cos(\theta_i) \zeta_c^{\lambda} - \sin(\theta_i)\zeta_d^{\lambda})]
        \end{split}        
    \end{equation}

    Using previous result from the previous Section~\ref{chap:proof_lemma_unary}, we have:
    \begin{equation}
        \mathrm{Var}_{\bm{\theta}}[\partial_{\theta_i} \mathcal{C}(\bm{\theta})]  = 2 \left(\int_{\bm{\theta} \in \Theta} (\frac{1}{2\pi})^D (\zeta^{\lambda}_l)^2 + (\zeta^{\lambda}_j)^2 d\bm{\theta}  \right) \cdot \left(\int_{\bm{\theta} \in \Theta} (\frac{1}{2\pi})^D (\tilde{y}^{\lambda}_l)^2 + (\tilde{y}^{\lambda}_j)^2 d\bm{\theta} \right)
    \end{equation}

    Then,
    \begin{equation}
        \begin{split}
            \mathbb{C}ov_{\bm{\theta}}&[\delta_a^{\lambda} (-\sin(\theta_i) \zeta_a^{\lambda} + \cos(\theta_i)\zeta_b^{\lambda}) + \delta_b^{\lambda} (-\cos(\theta_i) \zeta_a^{\lambda} - \sin(\theta_i)\zeta_b^{\lambda}), \\
            & \delta_c^{\lambda} (-\sin(\theta_i) \zeta_c^{\lambda} + \cos(\theta_i)\zeta_d^{\lambda}) + \delta_d^{\lambda} (-\cos(\theta_i) \zeta_c^{\lambda} - \sin(\theta_i)\zeta_d^{\lambda})] \\
            & = \mathbb{E}_{\bm{\theta}}[\left( \delta_a^{\lambda} (-\sin(\theta_i) \zeta_a^{\lambda} + \cos(\theta_i)\zeta_b^{\lambda}) + \delta_b^{\lambda} (-\cos(\theta_i) \zeta_a^{\lambda} - \sin(\theta_i)\zeta_b^{\lambda}) \right) \cdot \\
            & \left( \delta_c^{\lambda} (-\sin(\theta_i) \zeta_c^{\lambda} + \cos(\theta_i)\zeta_d^{\lambda}) + \delta_d^{\lambda} (-\cos(\theta_i) \zeta_c^{\lambda} - \sin(\theta_i)\zeta_d^{\lambda}) \right)] \\
            & - \mathbb{E}_{\bm{\theta}}[\delta_a^{\lambda} (-\sin(\theta_i) \zeta_a^{\lambda} + \cos(\theta_i)\zeta_b^{\lambda}) + \delta_b^{\lambda} (-\cos(\theta_i) \zeta_a^{\lambda} - \sin(\theta_i)\zeta_b^{\lambda})] \cdot \\
            & \mathbb{E}_{\bm{\theta}}[\delta_c^{\lambda} (-\sin(\theta_i) \zeta_c^{\lambda} + \cos(\theta_i)\zeta_d^{\lambda}) + \delta_d^{\lambda} (-\cos(\theta_i) \zeta_c^{\lambda} - \sin(\theta_i)\zeta_d^{\lambda})]
        \end{split}        
    \end{equation}

    Using previous result from the previous Section~\ref{chap:proof_lemma_unary} we have:
    \begin{equation}
        \begin{split}
            \mathbb{E}_{\bm{\theta}}&[\delta_a^{\lambda} (-\sin(\theta_i) \zeta_a^{\lambda} + \cos(\theta_i)\zeta_b^{\lambda}) + \delta_b^{\lambda} (-\cos(\theta_i) \zeta_a^{\lambda} - \sin(\theta_i)\zeta_b^{\lambda})] \\
            & = \mathbb{E}_{\bm{\theta}}[\delta_c^{\lambda} (-\sin(\theta_i) \zeta_c^{\lambda} + \cos(\theta_i)\zeta_d^{\lambda}) + \delta_d^{\lambda} (-\cos(\theta_i) \zeta_c^{\lambda} - \sin(\theta_i)\zeta_d^{\lambda})] \\
            & = 0
        \end{split}        
    \end{equation}

    We can now derive the value of the covariance term:
    \begin{equation}
        \begin{split}
            \mathbb{C}ov_{\bm{\theta}}&[\delta_a^{\lambda} (-\sin(\theta_i) \zeta_a^{\lambda} + \cos(\theta_i)\zeta_b^{\lambda}) + \delta_b^{\lambda} (-\cos(\theta_i) \zeta_a^{\lambda} - \sin(\theta_i)\zeta_b^{\lambda}), \\
            & \delta_c^{\lambda} (-\sin(\theta_i) \zeta_c^{\lambda} + \cos(\theta_i)\zeta_d^{\lambda}) + \delta_d^{\lambda} (-\cos(\theta_i) \zeta_c^{\lambda} - \sin(\theta_i)\zeta_d^{\lambda})] \\
            & = \mathbb{E}_{\bm{\theta}}[\left( \delta_a^{\lambda} (-\sin(\theta_i) \zeta_a^{\lambda} + \cos(\theta_i)\zeta_b^{\lambda}) + \delta_b^{\lambda} (-\cos(\theta_i) \zeta_a^{\lambda} - \sin(\theta_i)\zeta_b^{\lambda}) \right) \cdot \\
            & \left( \delta_c^{\lambda} (-\sin(\theta_i) \zeta_c^{\lambda} + \cos(\theta_i)\zeta_d^{\lambda}) + \delta_d^{\lambda} (-\cos(\theta_i) \zeta_c^{\lambda} - \sin(\theta_i)\zeta_d^{\lambda}) \right)] \\
            & = \mathbb{E}_{\bm{\theta}}[\delta_a^{\lambda} (-\sin(\theta_i) \zeta_a^{\lambda} + \cos(\theta_i)\zeta_b^{\lambda}) \cdot \delta_c^{\lambda} (-\sin(\theta_i) \zeta_c^{\lambda} + \cos(\theta_i)\zeta_d^{\lambda})] \\
            & + \mathbb{E}_{\bm{\theta}}[\delta_a^{\lambda} (-\sin(\theta_i) \zeta_a^{\lambda} + \cos(\theta_i)\zeta_b^{\lambda}) \cdot \delta_d^{\lambda} (-\cos(\theta_i) \zeta_c^{\lambda} - \sin(\theta_i)\zeta_d^{\lambda})] \\
            & + \mathbb{E}_{\bm{\theta}}[\delta_b^{\lambda} (-\cos(\theta_i) \zeta_a^{\lambda} - \sin(\theta_i)\zeta_b^{\lambda}) \cdot \delta_c^{\lambda} (-\sin(\theta_i) \zeta_c^{\lambda} + \cos(\theta_i)\zeta_d^{\lambda})] \\
            & + \mathbb{E}_{\bm{\theta}}[\delta_b^{\lambda} (-\cos(\theta_i) \zeta_a^{\lambda} - \sin(\theta_i)\zeta_b^{\lambda}) \cdot \delta_d^{\lambda} (-\cos(\theta_i) \zeta_c^{\lambda} - \sin(\theta_i)\zeta_d^{\lambda})] 
        \end{split}        
    \end{equation}

    Using the integral expression of the expectation value:
    \begin{equation}
        \begin{split}
            \mathbb{E}_{\bm{\theta}}&[\delta_a^{\lambda} (-\sin(\theta_i) \zeta_a^{\lambda} + \cos(\theta_i)\zeta_b^{\lambda}) \cdot \delta_d^{\lambda} (-\cos(\theta_i) \zeta_c^{\lambda} - \sin(\theta_i)\zeta_d^{\lambda})] \\
            & = \int_{\bm{\theta} \in \Theta} (\frac{1}{2\pi})^D[\frac{1}{2}\left( (\zeta^{\lambda}_b)^2(\zeta^{\lambda}_d)^2 - (\zeta^{\lambda}_b)^2(\zeta^{\lambda}_c)^2 -(\zeta^{\lambda}_a)^2(\zeta^{\lambda}_d)^2 + (\zeta^{\lambda}_a)^2(\zeta^{\lambda}_c)^2 \right) \\
            & + 2 \zeta^{\lambda}_a \zeta^{\lambda}_b \zeta^{\lambda}_c \zeta^{\lambda}_d + 2 \zeta^{\lambda}_a \zeta^{\lambda}_c \tilde{y}^{\lambda}_a \tilde{y}^{\lambda}_c + 2 \zeta^{\lambda}_b \zeta^{\lambda}_d \tilde{y}^{\lambda}_a \tilde{y}^{\lambda}_c] d\bm{\theta}
        \end{split}        
    \end{equation}
    
    \begin{equation}
        \begin{split}
            \mathbb{E}_{\bm{\theta}}&[\delta_a^{\lambda} (-\sin(\theta_i) \zeta_a^{\lambda} + \cos(\theta_i)\zeta_b^{\lambda}) \cdot \delta_c^{\lambda} (-\sin(\theta_i) \zeta_c^{\lambda} + \cos(\theta_i)\zeta_d^{\lambda})] \\
            & = \int_{\bm{\theta} \in \Theta} (\frac{1}{2\pi})^D[\frac{1}{2}\left( (\zeta^{\lambda}_b)^2(\zeta^{\lambda}_c)^2 - (\zeta^{\lambda}_a)^2(\zeta^{\lambda}_c)^2 -(\zeta^{\lambda}_b)^2(\zeta^{\lambda}_d)^2 + (\zeta^{\lambda}_a)^2(\zeta^{\lambda}_d)^2 \right) \\
            & - 2 \zeta^{\lambda}_a \zeta^{\lambda}_b \zeta^{\lambda}_c \zeta^{\lambda}_d + 2 \zeta^{\lambda}_a \zeta^{\lambda}_d \tilde{y}^{\lambda}_a \tilde{y}^{\lambda}_d + 2 \zeta^{\lambda}_b \zeta^{\lambda}_c \tilde{y}^{\lambda}_a \tilde{y}^{\lambda}_d] d\bm{\theta}
        \end{split}        
    \end{equation}

    \begin{equation}
        \begin{split}
            \mathbb{E}_{\bm{\theta}}&[\delta_b^{\lambda} (-\cos(\theta_i) \zeta_a^{\lambda} - \sin(\theta_i)\zeta_b^{\lambda}) \cdot \delta_c^{\lambda} (-\sin(\theta_i) \zeta_c^{\lambda} + \cos(\theta_i)\zeta_d^{\lambda})] \\
            & = \int_{\bm{\theta} \in \Theta} (\frac{1}{2\pi})^D[\frac{1}{2}\left( (\zeta^{\lambda}_a)^2(\zeta^{\lambda}_d)^2 - (\zeta^{\lambda}_b)^2(\zeta^{\lambda}_d)^2 -(\zeta^{\lambda}_a)^2(\zeta^{\lambda}_c)^2 + (\zeta^{\lambda}_b)^2(\zeta^{\lambda}_c)^2 \right) \\
            & - 2 \zeta^{\lambda}_a \zeta^{\lambda}_b \zeta^{\lambda}_c \zeta^{\lambda}_d + 2 \zeta^{\lambda}_a \zeta^{\lambda}_d \tilde{y}^{\lambda}_b \tilde{y}^{\lambda}_c + 2 \zeta^{\lambda}_b \zeta^{\lambda}_c \tilde{y}^{\lambda}_b \tilde{y}^{\lambda}_c] d\bm{\theta}
        \end{split}        
    \end{equation}

    \begin{equation}
        \begin{split}
            \mathbb{E}_{\bm{\theta}}&[\delta_b^{\lambda} (-\cos(\theta_i) \zeta_a^{\lambda} - \sin(\theta_i)\zeta_b^{\lambda}) \cdot \delta_d^{\lambda} (-\cos(\theta_i) \zeta_c^{\lambda} - \sin(\theta_i)\zeta_d^{\lambda})]  \\
            & = \int_{\bm{\theta} \in \Theta} (\frac{1}{2\pi})^D[\frac{1}{2}\left( (\zeta^{\lambda}_a)^2(\zeta^{\lambda}_c)^2 - (\zeta^{\lambda}_a)^2(\zeta^{\lambda}_d)^2 -(\zeta^{\lambda}_b)^2(\zeta^{\lambda}_c)^2 + (\zeta^{\lambda}_b)^2(\zeta^{\lambda}_d)^2 \right) \\
            & - 2 \zeta^{\lambda}_a \zeta^{\lambda}_b \zeta^{\lambda}_c \zeta^{\lambda}_d + 2 \zeta^{\lambda}_a \zeta^{\lambda}_c \tilde{y}^{\lambda}_b \tilde{y}^{\lambda}_d + 2 \zeta^{\lambda}_b \zeta^{\lambda}_d \tilde{y}^{\lambda}_b \tilde{y}^{\lambda}_d] d\bm{\theta}
        \end{split}        
    \end{equation}

    By summing the terms:

    \begin{equation}
        \begin{split}
            \mathbb{C}ov_{\bm{\theta}}&[\delta_a^{\lambda} (-\sin(\theta_i) \zeta_a^{\lambda} + \cos(\theta_i)\zeta_b^{\lambda}) + \delta_b^{\lambda} (-\cos(\theta_i) \zeta_a^{\lambda} - \sin(\theta_i)\zeta_b^{\lambda}), \\
            & \delta_c^{\lambda} (-\sin(\theta_i) \zeta_c^{\lambda} + \cos(\theta_i)\zeta_d^{\lambda}) + \delta_d^{\lambda} (-\cos(\theta_i) \zeta_c^{\lambda} - \sin(\theta_i)\zeta_d^{\lambda})] \\
            & = 2 \int_{\bm{\theta} \in \Theta} (\frac{1}{2\pi})^D [(\zeta_a^{\lambda} \zeta_c^{\lambda} - \zeta_b^{\lambda} \zeta_d^{\lambda})(\tilde{y}^{\lambda}_a \tilde{y}^{\lambda}_c + \tilde{y}^{\lambda}_b \tilde{y}^{\lambda}_d) + (\zeta_a^{\lambda} \zeta_d^{\lambda} - \zeta_b^{\lambda} \zeta_c^{\lambda})(\tilde{y}^{\lambda}_a \tilde{y}^{\lambda}_d + \tilde{y}^{\lambda}_b \tilde{y}^{\lambda}_c)] d\bm{\theta}
        \end{split}
    \end{equation}

    Finally,

    \begin{equation}
        \begin{split}
            \mathrm{Var}_{\bm{\theta}}&[\partial_{\theta_i} \mathcal{C}] =  2 \sum_{(l,j)} \left(\int_{\bm{\theta} \in \Theta} (\frac{1}{2\pi})^D (\zeta^{\lambda}_l)^2 + (\zeta^{\lambda}_j)^2 d\bm{\theta}  \right) \cdot \left(\int_{\bm{\theta} \in \Theta} (\frac{1}{2\pi})^D (\tilde{y}^{\lambda}_l)^2 + (\tilde{y}^{\lambda}_j)^2 d\bm{\theta} \right) \\
            & + 4 \sum_{(a,b) \neq (c,d)} \int_{\bm{\theta} \in \Theta} (\frac{1}{2\pi})^D [(\zeta_a^{\lambda} \zeta_c^{\lambda} + \zeta_b^{\lambda} \zeta_d^{\lambda})(\tilde{y}^{\lambda}_a \tilde{y}^{\lambda}_c + \tilde{y}^{\lambda}_b \tilde{y}^{\lambda}_d) + (\zeta_a^{\lambda} \zeta_d^{\lambda} - \zeta_b^{\lambda} \zeta_c^{\lambda})(\tilde{y}^{\lambda}_a \tilde{y}^{\lambda}_d - \tilde{y}^{\lambda}_b \tilde{y}^{\lambda}_c)] d\bm{\theta}
        \end{split}
    \end{equation}

    We now claim that the covariance term is null. To show this, we are about to use an induction proof to prove that $\forall \lambda, \forall a,b \in [d_k], \int_{\bm{\theta} \in \Theta} (\frac{1}{2\pi})^D \zeta_a^{\lambda} \zeta_b^{\lambda} d\bm{\theta} = 0 $ while considering our assumption on the input and target output state distribution. First for the basis:
    \begin{equation}
        \forall a,b \in [d_k], \quad \int_{\bm{\theta} \in \Theta} (\frac{1}{2\pi})^D \zeta_a^{0} \zeta_b^{0} d\bm{\theta} = \zeta_a^{0} \zeta_b^{0} \int_{\bm{\theta} \in \Theta} (\frac{1}{2\pi})^D d\bm{\theta} = \zeta_a^{0} \zeta_b^{0}
    \end{equation}

    Considering the expectation value on the input vectors $\zeta^0$, it comes:
    \begin{equation}
        \forall a,b \in [d_k], \quad \mathbb{E}_{\zeta^0}[\zeta_a^{0} \zeta_b^{0}] = 0
    \end{equation}

    In the following, we consider the expectation value over the input state and output state distribution. For the induction part, we just need to use the recursive formula on the evolution of the inner states. Let us consider the property verified on then layer $\lambda$:
    \begin{equation}
        \forall a,b \in [d_k], \quad \int_{\bm{\theta} \in \Theta} (\frac{1}{2\pi})^D \zeta_a^{\lambda+1} \zeta_b^{\lambda+1} d\bm{\theta} = 
        \begin{cases}
            \frac{1}{2}\int_{\bm{\theta} \in \Theta} (\frac{1}{2\pi})^D \zeta_{a,a'}^{\lambda} \zeta_{b,b'}^{\lambda} d\bm{\theta} + \frac{1}{2}\int_{\bm{\theta} \in \Theta} (\frac{1}{2\pi})^D \zeta_{a',a}^{\lambda} \zeta_{b',b}^{\lambda} d\bm{\theta} = 0 \\
            \int_{\bm{\theta} \in \Theta} (\frac{1}{2\pi})^D (\pm \cos(\theta_i) \zeta_{a,a'}^{\lambda} \pm \sin(\theta_i) \zeta_{a',a}^{\lambda})  \zeta_b^{\lambda} d\bm{\theta} = 0  \\
            \int_{\bm{\theta} \in \Theta} (\frac{1}{2\pi})^D \zeta_a^{\lambda} (\pm \cos(\theta_i) \zeta_{b,b'}^{\lambda} \pm \sin(\theta_i) \zeta_{b',b}^{\lambda}) d\bm{\theta} = 0  \\
            \int_{\bm{\theta} \in \Theta} (\frac{1}{2\pi})^D \zeta_a^{\lambda} \zeta_b^{\lambda} d\bm{\theta} = 0
        \end{cases}
    \end{equation}

    The first case corresponds to the situation where $a$ and $b$ are affected by a rotation in the previous layer $\lambda$. The second and the third case refer to situation where only one state between $a$ and $b$ is affected by a rotation in the previous layer. By integrating over the corresponding parameter, those value are null. Finally, we consider the case where none of them are affected by a rotation in the previous layer. In all the cases if $a \neq b$, the result is zero due to the induction hypothesis. Therefore, we have that:
    \begin{equation}
        \forall \lambda, \forall a,b \in [d_k], \quad a \neq b \implies \quad \int_{\bm{\theta} \in \Theta} (\frac{1}{2\pi})^D \zeta_a^{\lambda} \zeta_b^{\lambda} d\bm{\theta} = 0
    \end{equation}

    Which implies that the covariance term is equal to $0$:
    \begin{equation}
        \sum_{(a,b) \neq (c,d)} \int_{\bm{\theta} \in \Theta} (\frac{1}{2\pi})^D [(\zeta_a^{\lambda} \zeta_c^{\lambda} + \zeta_b^{\lambda} \zeta_d^{\lambda})(\tilde{y}^{\lambda}_a \tilde{y}^{\lambda}_c + \tilde{y}^{\lambda}_b \tilde{y}^{\lambda}_d) + (\zeta_a^{\lambda} \zeta_d^{\lambda} - \zeta_b^{\lambda} \zeta_c^{\lambda})(\tilde{y}^{\lambda}_a \tilde{y}^{\lambda}_d - \tilde{y}^{\lambda}_b \tilde{y}^{\lambda}_c)] d\bm{\theta} = 0
    \end{equation}

    Finally, we have that:
    
    \begin{equation}
            \mathrm{Var}_{\bm{\theta}}[\partial_{\theta_i} \mathcal{C}] =  2 \sum_{(l,j)} \left(\int_{\bm{\theta} \in \Theta} (\frac{1}{2\pi})^D (\zeta^{\lambda}_l)^2 + (\zeta^{\lambda}_j)^2 d\bm{\theta}  \right) \cdot \left(\int_{\bm{\theta} \in \Theta} (\frac{1}{2\pi})^D (\tilde{y}^{\lambda}_l)^2 + (\tilde{y}^{\lambda}_j)^2 d\bm{\theta} \right)
    \end{equation}

We have proved Lemma~\ref{lemma:VarianceHWPreserving} in the specific case of RBS gates. In the following section we show how to adapt the previous proof to FBS gates.

    \subsection{Generalization for FBS gates}\label{chap:proof_lemma_general_FBS}

We consider once again the case of the squared Euclidean cost function. The decomposition of the FBS based quantum circuit for the backpropagation method is very similar to the case of RBS based quantum circuit:
    
    We call $\Delta^L$ the final error: 
    \begin{equation}
        \Delta^L = 2(z^L - y) = 2[(w^{\lambda_{\mathrm{max}}} \cdots w^{\lambda + 1}) \cdot w^{\lambda} \cdot \zeta^{\lambda} - y]    
    \end{equation}

    We consider the case where for each inner layer, there is only one FBS gate considered. For each inner layer, the action of the gate results in a rotation of the amplitudes between a set of pair of states that we call $\mathcal{R}_{\lambda}$:
    \begin{equation}
        \zeta^{\lambda + 1} = \overrightarrow{cst} + \sum_{(l,j) \in \mathcal{R}_{\lambda}} (\cos(\theta_{i}) \cdot \zeta^{\lambda}_{l} + (-1)^{f(l,j, \zeta^{\lambda}_{j})} \sin(\theta_{i}) \cdot \zeta^{\lambda}_{j}) \ket{e_l} + ( (-1)^{f(l,j, \zeta^{\lambda}_{l}) + 1} \sin(\theta_{i}) \cdot \zeta^{\lambda}_{l} + \cos(\theta_{i}) \cdot \zeta^{\lambda}_{j}) \ket{e_j}  
    \end{equation}
    with $\forall (l,j)\in \mathcal{R}_{\lambda}, \quad \overrightarrow{cst}^\intercal \cdot \ket{e_j} = \overrightarrow{cst}^\intercal \cdot \ket{e_l} = 0$

    We can define the error according to the final error:
    \begin{equation}
        \begin{split}
            \delta^{\lambda + 1} &= (w^{\lambda +1})^{-1} \cdots (w^{\lambda_{\mathrm{max}}})^{-1} \cdot \Delta^l = 2 [w^{\lambda} \cdot \zeta^{\lambda} - (w^{\lambda +1})^{-1} \cdots (w^{\lambda_{\mathrm{max}}})^{-1} \cdot y] \\
            &= 2[\overrightarrow{cst} + \sum_{(l,j) \in \mathcal{R}_{\lambda}} (\cos(\theta_{i}) \cdot \zeta^{\lambda}_{l} + (-1)^{f(l,j, \zeta^{\lambda}_{j})} \sin(\theta_{i}) \cdot \zeta^{\lambda}_{j}) \ket{e_l} \\
            & + ((-1)^{f(l,j, \zeta^{\lambda}_{l}) + 1} \sin(\theta_{i}) \cdot \zeta^{\lambda}_{l} + \cos(\theta_{i}) \cdot \zeta^{\lambda}_{j}) \ket{e_j} - \tilde{y}^{\lambda}]
        \end{split}
    \end{equation}
    A unique RBS affects $\binom{n-2}{k-1}$ different pairs of states with a rotation, and all the affected states are different. A unique FBS applied on the same qubits affects the same pairs of states. Therefore:
    \begin{equation}
        \forall (l,j) \in \mathcal{R}_{\lambda}, \quad \delta^{\lambda + 1}_l = 2[\cos(\theta_{i}) \cdot \zeta^{\lambda}_{l} + (-1)^{f(l,j, \zeta^{\lambda}_{j})} \sin(\theta_{i}) \cdot \zeta^{\lambda}_{j} - \tilde{y}^{\lambda}_l]
    \end{equation}
    and,
    \begin{equation}
        \forall (l,j) \in \mathcal{R}_{\lambda}, \quad \delta^{\lambda + 1}_j = 2[(-1)^{f(l,j, \zeta^{\lambda}_{l})+1}\sin(\theta_{i}) \cdot \zeta^{\lambda}_{l} + \cos(\theta_{i}) \cdot \zeta^{\lambda}_{j} - \tilde{y}^{\lambda}_j]
    \end{equation}
    
    Note that by orthogonality: $\forall \lambda, \quad (w^{\lambda})^{-1} = (w^{\lambda})^{t}$.
    
    We call: $\tilde{y} = (w^{\lambda +1})^{t} \cdots (w^{\lambda_{\mathrm{max}}})^{t} \cdot y$ and $\Theta = [0:2\pi]^D$
    In the following, we will omit to note the set $\mathcal{R}_{\lambda}$.
    
    \begin{equation}\label{eq:FBS_developped_expression_var_general_case}
        \begin{split}
            \mathrm{Var}_{\bm{\theta}}[\partial_{\theta_i} \mathcal{C}] &= \mathrm{Var}_{\bm{\theta}}[\sum_{(l,j)} \delta_l^{\lambda} (-\sin(\theta_i) \zeta_l^{\lambda} + (-1)^{f(l,j,\zeta_j^{\lambda})} \cos(\theta_i)\zeta_j^{\lambda}) + \delta_j^{\lambda} ((-1)^{f(l,j,\zeta_l^{\lambda})}\cos(\theta_i) \zeta_l^{\lambda} - \sin(\theta_i)\zeta_j^{\lambda})] \\
            & = \sum_{(l,j)} \mathrm{Var}_{\bm{\theta}}[\delta_l^{\lambda} (-\sin(\theta_i) \zeta_l^{\lambda} + (-1)^{f(l,j,\zeta_j^{\lambda})} \cos(\theta_i)\zeta_j^{\lambda}) + \delta_j^{\lambda} ((-1)^{f(l,j,\zeta_l^{\lambda})}\cos(\theta_i) \zeta_l^{\lambda} - \sin(\theta_i)\zeta_j^{\lambda})] \\
            & + 2 \sum_{(a,b) \neq (c,d)} \mathbb{C}ov_{\bm{\theta}} [\delta_a^{\lambda} (-\sin(\theta_i) \zeta_a^{\lambda} + (-1)^{f(a,b,\zeta_b^{\lambda})} \cos(\theta_i)\zeta_b^{\lambda}) + \delta_b^{\lambda} ((-1)^{f(a,b,\zeta_a^{\lambda})}\cos(\theta_i) \zeta_a^{\lambda} - \sin(\theta_i)\zeta_b^{\lambda}), \\
            & \delta_c^{\lambda} (-\sin(\theta_i) \zeta_c^{\lambda} + (-1)^{f(c,d,\zeta_d^{\lambda})} \cos(\theta_i)\zeta_d^{\lambda}) + \delta_d^{\lambda} ((-1)^{f(c,d,\zeta_c^{\lambda})}\cos(\theta_i) \zeta_c^{\lambda} - \sin(\theta_i)\zeta_d^{\lambda})]
        \end{split}        
    \end{equation}

    The only change in comparison with the RBS case is with the sign of the sine in the previous equations. However, in the proof of Theorem~\ref{thm:NoBPgeneralcase} given in the previous appendix, we derive each term of Eq.~\eqref{eq:FBS_developped_expression_var_general_case} by integrating over the parameter $\theta_i$. The sign of the sine does not change the integration and as a result we find again that: 
    
    \begin{equation}
        \begin{split}
            \mathrm{Var}_{\bm{\theta}}&[\partial_{\theta_i} \mathcal{C}] =  2 \sum_{(l,j)} \left(\int_{\bm{\theta} \in \Theta} (\frac{1}{2\pi})^D (\zeta^{\lambda}_l)^2 + (\zeta^{\lambda}_j)^2 d\bm{\theta}  \right) \cdot \left(\int_{\bm{\theta} \in \Theta} (\frac{1}{2\pi})^D (\tilde{y}^{\lambda}_l)^2 + (\tilde{y}^{\lambda}_j)^2 d\bm{\theta} \right) \\
            & + 4 \sum_{(a,b) \neq (c,d)} \int_{\bm{\theta} \in \Theta} (\frac{1}{2\pi})^D [(\zeta_a^{\lambda} \zeta_c^{\lambda} + \zeta_b^{\lambda} \zeta_d^{\lambda})(\tilde{y}^{\lambda}_a \tilde{y}^{\lambda}_c + \tilde{y}^{\lambda}_b \tilde{y}^{\lambda}_d) + (\zeta_a^{\lambda} \zeta_d^{\lambda} - \zeta_b^{\lambda} \zeta_c^{\lambda})(\tilde{y}^{\lambda}_a \tilde{y}^{\lambda}_d - \tilde{y}^{\lambda}_b \tilde{y}^{\lambda}_c)] d\bm{\theta}
        \end{split}
    \end{equation}

We can show in a similar way that the covariance term is null. Finally, we have that:
 \begin{equation}
            \mathrm{Var}_{\bm{\theta}}[\partial_{\theta_i} \mathcal{C}] =  2 \sum_{(l,j)} \left(\int_{\bm{\theta} \in \Theta} (\frac{1}{2\pi})^D (\zeta^{\lambda}_l)^2 + (\zeta^{\lambda}_j)^2 d\bm{\theta}  \right) \cdot \left(\int_{\bm{\theta} \in \Theta} (\frac{1}{2\pi})^D (\tilde{y}^{\lambda}_l)^2 + (\tilde{y}^{\lambda}_j)^2 d\bm{\theta} \right)
    \end{equation}
    
\textbf{We have proved Lemma~\ref{lemma:VarianceHWPreserving}.}
\end{proof}

\section{Formalization and proof of Theorem~\ref{thm:NoBPPSA}}\label{sec:ProofNoBPPSA}
The goal of this section is to make precise and generalize the informal \cref{thm:NoBPPSA} from the main text, and to prove it.
Let us recall that theorem for convenience:

\NoBPPSA*

In \cref{subchap:InductiveRelation}, we provide a recurrence relation describing how the squared entries of the intermediate quantum states propagate back and forth throughout the circuit, which will be key in obtaining a precise understanding of how the variance evolves. In \cref{subchap:StochasticMatrices}, we introduce the concept of stochastic matrices in order to, in \cref{subchap:variance-formula-recast}, recast the above relation into that language, which will let the final variance quantity be understood through the convergence of certain stochastic matrices powers. In \cref{subchap:irreducible-and-primitive-stochmats}, we introduce special classes of stochastic matrices, and we show in the following \cref{subchap:connected-stochmats-props} that the general ansätze we consider in the main text do belong in one such class called primitive stochastic matrices.
Afterwards, we illustrate in \cref{subchap:convergence-fixed} how the convergence of powers of a fixed stochastic matrix is generally studied, after which we promote, in \cref{subchap:convergence-sequence-and-conjecture}, the case of single stochastic matrix, to the case of a sequence thereof. For this result to carry through properly, we make there a minor conjecture concerning the structure of the stochastic matrices induced from the general RBS/FBS circuit ansätze considered.
All the previous results are put together in \cref{subchap:precised-theorem-2}, to obtain \cref{thm:concluding-back-to-variance}, the precised version of the theorem shown above. In \cref{subsec:numerical-evidence-spectral-gap}, we provide numerical evidence for the previous conjecture, and lastly in \cref{subchap:design-discussion} we discuss the differences of our approach with an approach that argues using 
a closeness to unitary $2$-design assumption.

\subsection{Recurrence relation for squared entries of intermediate states}\label{subchap:InductiveRelation}
    
\begin{lem}[Recurrence relation for squared entries of intermediate states]
\label{lemma:InductiveRelation}
Let us consider a $n$-qubit HW-preserving VQC made of RBS or FBS gates. We consider here a training in the subspace of HW $k$, i.e., corresponding to the basis $B_k^n$.
$\zeta_r^{\lambda}$ denotes the $r^{\mathrm{th}}$ entry (in the basis $B_k^n$) of the intermediate state $\zeta^{\lambda}$.
We have the following recurrence relation between the squared entries of the intermediate state $\zeta^{\lambda+1}$ and those of the previous state $\zeta^\lambda$:
\begin{equation}\label{eq:Inductive_relation}
    \forall r \in [d_k], \quad \int_{\bm{\theta} \in \Theta} (\frac{1}{2\pi})^D (\zeta_r^{\lambda+1})^2 d\bm{\theta} = \begin{cases}
        \frac{1}{2}\int_{\bm{\theta} \in \Theta} (\frac{1}{2\pi})^D (\zeta_{r,r'}^{\lambda})^2 d\bm{\theta} + \frac{1}{2}\int_{\bm{\theta} \in \Theta} (\frac{1}{2\pi})^D (\zeta_{r',r}^{\lambda})^2 d\bm{\theta} \\ \mathrm{or} \quad
        \int_{\bm{\theta} \in \Theta} (\frac{1}{2\pi})^D (\zeta_r^{\lambda})^2 d\bm{\theta}\,,
    \end{cases}
\end{equation}
depending on whether or not the state $\ket{e_r}\in B_k^n$ from the computational basis is undergoing a planar rotation with another state $\ket{e_{r'}}$ due to the layer $\lambda$.
\end{lem}

\begin{proof}
    Eq.~\eqref{eq:Inductive_relation} states that, because the inner states $\zeta^{\lambda+1}$ and $\zeta^{\lambda}$ are connected through the application of one RBS/FBS, and because each RBS/FBS performs $\binom{n-2}{k-1}$ $\theta$-planar rotation between states in the computational basis, a index $r$ of $\zeta^{\lambda+1}$ can take two value:
\begin{itemize}
    \item If the state $\ket{e_r}$ from the computational basis is involved in a planar rotation due to the RBS/FBS, then:
    \begin{equation}\label{eq:Inductive_relation-proof-eq-a}
        \zeta_r^{\lambda+1} = \cos(\theta) \zeta_{r,r'}^{\lambda} \pm \sin(\theta) \zeta_{r',r}^{\lambda}\,,
    \end{equation}
    and thus, we have that:
    \begin{equation}
        \int_{\bm{\theta} \in \Theta} (\frac{1}{2\pi})^D (\zeta_r^{\lambda+1})^2 d\bm{\theta} = \frac{1}{2}\int_{\bm{\theta} \in \Theta} (\frac{1}{2\pi})^D (\zeta_{r,r'}^{\lambda})^2 d\bm{\theta} + \frac{1}{2}\int_{\bm{\theta} \in \Theta} (\frac{1}{2\pi})^D (\zeta_{r',r}^{\lambda})^2 d\bm{\theta}\,.
    \end{equation}
    \item In the opposite case, then we simply have that $\zeta_r^{\lambda+1} = \zeta_{r}^{\lambda}$, and therefore:
    \begin{equation}\label{eq:Inductive_relation-proof-eq-b}
        \int_{\bm{\theta} \in \Theta} (\frac{1}{2\pi})^D (\zeta_r^{\lambda+1})^2 d\bm{\theta} = \int_{\bm{\theta} \in \Theta} (\frac{1}{2\pi})^D (\zeta_r^{\lambda})^2 d\bm{\theta}\,.
    \end{equation}
\end{itemize}
\end{proof}

This relation can also be stated for the propagation of $\int_{\bm{\theta} \in \Theta} (\frac{1}{2\pi})^D (\tilde{y}_r^{\lambda})^2 d\bm{\theta}$, as the backpropagation of the target state only apply planar rotation to this state.

\subsection{Introducing stochastic matrices }\label{subchap:StochasticMatrices}

\paragraph{Stochastic matrices}

Let $T \in \mathbb{R}^{N\times N}$. $T$ is said to be a \emph{column-stochastic} matrix if its columns are probability vectors, i.e. $T_{ij} \geq 0 \ \forall i,j \in [N]$ and $\sum_i T_{ij} =1\ \forall j \in [N]$. Likewise, $T$ is \emph{row-stochastic} if its rows are probability vectors, and is \emph{doubly stochastic} if it is both column-stochastic and row-stochastic.  In the rest of this work, if a matrix is just said to be stochastic (without more precision), we mean to say that is is column-stochastic.

The purpose of defining  column-stochastic matrices is that they are exactly the matrices $M$ for which, when acting on a \textit{probability vector} $\vec{p}$ (i.e. a vector of nonnegative entries that all sum to $1$), the output $M \vec{p}$ remains a probability vector.

Let us note that all three of these subsets of matrices introduced are topologically closed in $\mathbb{R}^{N\times N}$, and therefore if for instance a sequence of doubly stochastic matrices $(T_n)$ converges (with respect to a matrix norm) to a matrix $T_{\infty}$, then $T_{\infty}$ is still doubly stochastic. Furthermore, note that all three of these subsets of matrices are closed under taking a product of two elements.

\paragraph{Eigenvalues, and spectral gap}

Note that $1$ is always an eigenvalue of a column-stochastic matrix $T$, since $T^\intercal$ has the eigenvector $(1,\dots,1)^\intercal$ associated to eigenvalue $1$.

Furthermore, by direct consequence of the Gershgorin circle theorem (\cite[Theorem 6.1.1]{Horn-MatrixAnalysis-2013}), all eigenvalues of a column-stochastic matrix lie in the complex closed unit disk.

Given a column-stochastic matrix $T$, consider studying the behaviour of the sequence of its powers $T^n$ as $n$ increases. A special case of this which comes with the clearest  intuition is when $T$ is diagonalizable. Indeed, since the diagonalization of $T$ provides an expression of the form $T=P \text{diag}(\lambda_1,\dots,\lambda_N) P^{-1}$, so $T^n=P \text{diag}(\lambda_1^n,\dots,\lambda_N^n) P^{-1}$, and thus the behavior of $T^n$ is in that case understood through an exponential vanishing of all its eigenvalues except those of unit modulus. Thus in the case when $1$ is the only unit-modulus eigenvalue, the sequence $(T^n)$ would be converging exponentially fast to a matrix $T_{\infty}$, with the exponential rate being governed by the second largest eigenvalue modulus, denoted $|\lambda_2|$; and the further $|\lambda_2|$ is from $1$, the faster is the rate of the exponential convergence.

In this spirit, we will usually denote by $\Delta:=1-|\lambda_2|$ this gap for the stochastic-matrix $T$, which we will refer to as the \emph{spectral gap} of $T$.

One should keep in mind, though, that in general a stochastic matrix $T$ could have other eigenvalues of unit modulus besides $1$, in which case the sequence $(T^n)$ would not even be necessarily convergent. Besides, a stochastic matrix is not necessarily diagonalizable, either (one such example will be mentioned later in \cref{subsec:numerical-evidence-spectral-gap}).

\subsection{The variance formula cast in terms of stochastic matrices and probability vectors}\label{subchap:variance-formula-recast}

Let us introduce for all $\lambda \in \llbracket0,\lambda_{\text{max}}\rrbracket$, the vectors $\overrightarrow{Z^\lambda},\overrightarrow{W^\lambda} \in \mathbb{R}^{d_k}$ defined by squaring entry-wise the intermediate states $\zeta^{\lambda}$ and the back-propagated target states $z^{\lambda}$, respectively, i.e.
\begin{align}
    (\overrightarrow{Z^\lambda})_u &:= (\zeta^{\lambda}_u)^2\,,\label{eq:proba-vector-forward-lambda}\\
    (\overrightarrow{W^\lambda})_u &:= (z^{\lambda}_u)^2\,,\label{eq:proba-vector-reverse-lambda}
\end{align}
for all $u\in[d_k]$.

Notice that all these vectors $\overrightarrow{Z^\lambda}$ and $\overrightarrow{W^\lambda}$ are probability vectors (i.e. their entries are nonnegative and sum to $1$).

By careful inspection of Lemma \ref{lemma:InductiveRelation}'s recurrence relation \cref{eq:Inductive_relation}, one sees that it consists of a linear recurrence relation relating the probability vectors $\overrightarrow{Z^{\lambda+1}}$ to $\overrightarrow{Z^\lambda}$, through multiplication by a stochastic matrix that directly corresponds to the $\lambda^{\text{th}}$ RBS/FBS gate. Namely:
\begin{equation}\label{eq:stochastic-matrix-recurrence-relation}
     \overrightarrow{Z}^{\lambda+1} = T_{\lambda} \cdot \overrightarrow{Z}^{\lambda} \,,
\end{equation}
where $T_{\lambda}$ is defined as the $d_k \times d_k$ column-stochastic matrix that is constructed by taking the VQC's $\lambda^{\text{th}}$ RBS/FBS unitary is (in the subspace $k$) and replacing its $\pm \cos(\theta_i)$ and $\pm \sin(\theta_i)$ entries with $1/2$.
In other words, if $U_{\lambda}(\theta_\lambda)$ denotes the VQC's $\lambda^{\text{th}}$ RBS/FBS unitary in the subspace $k$, then: 
\begin{equation}\label{eq:stochastic-matrix-Tlambda-def}
     \Big(T_\lambda\Big)_{a,b} := \Big(U_{\lambda}(\theta_\lambda:=\pi/4)\Big)_{a,b}^2\,.
\end{equation}
In fact, Lemma \ref{lemma:InductiveRelation}'s recurrence relation \cref{eq:Inductive_relation} could be shown to hold in full analogy for the back-propagating state vector as well, instead of the forward-propagating one; one would find the "reversed" relation
\begin{equation}\label{eq:stochastic-matrix-recurrence-relation-backward-interm}
     \overrightarrow{Z}^{\lambda-1} = T_{\lambda}^\intercal \cdot \overrightarrow{Z}^{\lambda} \,,
\end{equation}
but since all the stochastic matrices associated to single RBS/FBS gates are symmetric (this follows from their definition in \cref{eq:stochastic-matrix-Tlambda-def}), the reversed recurrence relation writes as
\begin{equation}\label{eq:stochastic-matrix-recurrence-relation-backward}
     \overrightarrow{Z}^{\lambda-1} = T_{\lambda} \cdot \overrightarrow{Z}^{\lambda} \,.
\end{equation}

Now, suppose the VQC has a CPSA architecture (Definition \ref{def:CPSA}). Its unitary (in the subspace $k$) has a periodic structure of the form
\begin{equation}\label{eq:periodic-structure-appendix}
    U(\bm{\theta}) = \prod_{l=1}^{L} U_0(\bm{\theta}_l), \quad U_0(\bm{\theta}_l) = \prod_{j=1}^J e^{-i \theta_{l,j} H^j_{RBS/FBS}}\,. 
\end{equation}
Let us introduce new indices $(\tilde{l},\tilde{j})$ to specify a given single RBS/FBS gate, similarly to a pair $(l,j)$ but with both indices increasing in a "reversed" way instead (from the end of the circuit to the start).

To summarize, we have presently three "coordinate systems" to specify one of the RBS/FBS gates in this circuit, $\lambda \leftrightarrow (l,j) \leftrightarrow (\tilde{l},\tilde{j})$ (with $\lambda \in \llbracket 1, \lambda_{\text{max}}\rrbracket$, $l \in \llbracket 1, L\rrbracket, j \in \llbracket 1, J\rrbracket$, and $\tilde{l} \in \llbracket 1, L\rrbracket, \tilde{j} \in \llbracket 1, J\rrbracket$), the first two coordinate systems have indices that increase in a "forward" way while the third coordinate system has indices that increase in a "reversed" way; and the three coordinate systems are uniquely related through:
\begin{align}
    \lambda &= (l-1)J + j\,,\label{eq:depth-coordinates-correspondance-1}\\
    \lambda &= (L-(\tilde{l}-1))J - (\tilde{j}-1)\,.\label{eq:depth-coordinates-correspondance-2}
\end{align}
For a given intermediate depth $\lambda$ of the circuit, we will therefore denote by $l(\lambda),j(\lambda)$ and $\tilde{l}(\lambda),\tilde{j}(\lambda)$ the unique values of $l,j$ and $\tilde{l},\tilde{j}$ that correspond to $\lambda$, through Eqs.~\eqref{eq:depth-coordinates-correspondance-1} and \eqref{eq:depth-coordinates-correspondance-2} respectively.

Let us denote by $T$ (without any subscript) the stochastic matrix corresponding to the main pattern $U_0$ of the CPSA architecture (the one that is repeated $L$ times). Namely:
\begin{equation}\label{eq:stochastic-matrix-T-def}
    T := T_J \cdots T_2 \cdot T_1\,. 
\end{equation}
Similarly, let us denote by $\widetilde{T}$ the "reversed" main pattern: 
\begin{equation}\label{eq:stochastic-matrix-Ttilde-def}
    \widetilde{T} := T_1 \cdots T_{J-1} \cdot T_J = T^\intercal\,. 
\end{equation}

By repeated use of the recurrence relation of Eq.~\eqref{eq:stochastic-matrix-recurrence-relation}, the (forward-propagating) probability vector $\overrightarrow{Z}^{\lambda}$ may be related to the initial (left-most) one by:
\begin{equation}\label{eq:repeated-recurrence-relation-forward}
    \overrightarrow{Z}^{\lambda} = T_{\text{rest},\lambda} \cdot T^{l(\lambda)-1} \ \cdot\ \overrightarrow{Z^0}\,,
\end{equation}
where $T_{\text{rest},\lambda} := T_{j(\lambda)} \cdots T_2 \cdot T_1$.
Likewise, the repeated use of the recurrence relation of Eq.~\eqref{eq:stochastic-matrix-recurrence-relation-backward} yields:
\begin{equation}\label{eq:repeated-recurrence-relation-reverse}
    \overrightarrow{W}^{\lambda} = \widetilde{T}_{\text{rest},\lambda} \cdot \widetilde{T}^{\tilde{l}(\lambda)-1} \ \cdot\ \overrightarrow{W^{\lambda_{\text{max}}}}\,,
\end{equation}
where $\widetilde{T}_{\text{rest},\lambda} := T_{\tilde{j}(\lambda)} \cdots T_{J-1} \cdot T_J$.

Notice that Lemma \ref{lemma:VarianceHWPreserving}'s variance formula Eq.~\eqref{eq:VarianceHWPreserving-Var} may now be written more concisely as
\begin{align}\label{eq:variance-concise-1}
    \mathrm{Var}_{\bm{\theta}}[\partial_{\theta_\lambda} \mathcal{C}] &=  2 \sum_{(u,v)} \left[ \big(\overrightarrow{Z^\lambda}\big)_{\!u} + \big(\overrightarrow{Z^\lambda}\big)_{\!v} \right] \cdot \left[ \big(\overrightarrow{W^\lambda}\big)_{\!u} + \big(\overrightarrow{W^\lambda}\big)_{\!v} \right]\,.
\end{align}
As we will detail in the next section, the four quantities $\big(\overrightarrow{Z^\lambda}\big)_{\!u},\big(\overrightarrow{Z^\lambda}\big)_{\!v},\big(\overrightarrow{W^\lambda}\big)_{\!u}$ and $\big(\overrightarrow{W^\lambda}\big)_{\!v}$ in fact all converge towards the value $1/d_k$ as the depth $\lambda$ goes to infinity. In anticipation of this fact, we suggestively re-write these four terms in Eq.~\eqref{eq:variance-concise-1} as: 
\begin{align}\label{eq:variance-concise-2}
    \mathrm{Var}_{\bm{\theta}}[\partial_{\theta_\lambda} \mathcal{C}] &=  2 \sum_{(u,v)} \left[ \left(\frac{1}{d_k} + \epsilon^{(l-1,j,u)}\right) + \left(\frac{1}{d_k} + \epsilon^{(l-1,j,v)}\right) \right] \cdot \left[ \left(\frac{1}{d_k} + \tilde{\epsilon}^{(\tilde{l}-1,\tilde{j},u)}\right) + \left(\frac{1}{d_k} + \tilde{\epsilon}^{(\tilde{l}-1,\tilde{j},v)}\right) \right]\,,
\end{align}
where we dropped the dependency on $\lambda$ of $l$ and $\tilde{l}$ to simplify the notation.

We will also obtain, in the next section, upper-bounds on the absolute values of all the $|\epsilon^{(\cdot,\cdot,\cdot)}|$ terms (which we refer to as the error terms) that only depend on the number of (forward or reversed) repetitions ($l$ or $\tilde{l}$). So for now, suppose that we have such bounds, i.e. suppose that for all $j$ and all $u$:
\begin{align}
    |\epsilon^{(l-1,j,u)}| &\leq \mathcal{E}_{l-1}\,,\label{eq:variance-epsilon-commonbound-1}\\
    |\tilde{\epsilon}^{(\tilde{l}-1,\tilde{j},u)}| &\leq \mathcal{E}_{\tilde{l}-1}\,,\label{eq:variance-epsilon-commonbound-2}
\end{align}
 for some $\mathcal{E}_{l-1},\mathcal{E}_{\tilde{l}-1} >0$.
 Injecting the bounds of Eqs.~\eqref{eq:variance-epsilon-commonbound-1} and \eqref{eq:variance-epsilon-commonbound-2} into the variance expression of Eq.~\eqref{eq:variance-concise-2}, simplifying (note that the outer sum is over $\binom{n-2}{k-1}$ terms) and applying some triangle inequalities, leads to:
 \begin{equation}\label{eq:variance-difference-with-idealvariance-general-upper-bound}
     \left| \mathrm{Var}_{\bm{\theta}}[\partial_{\theta_\lambda} \mathcal{C}] - \frac{1}{d_k} \frac{8 k (n-k)}{n(n-1)} \right| \leq \frac{8 k (n-k)}{n(n-1)} \left( \mathcal{E}_{l-1} + \mathcal{E}_{\tilde{l}-1} + 2 \, d_k \, \mathcal{E}_{l-1} \, \mathcal{E}_{\tilde{l}-1} \right)\,.
 \end{equation}

In the following, we will denote for all $p \in [1,\infty]$, the \textit{entry-wise} $p$-norms and the \textit{Schatten} $p$-norms on square matrices, by $\lVert \cdot \rVert_{\mathrm{ew},p}$ and $\lVert \cdot \rVert_{\mathrm{Sc},p}$ respectively. Recall that these norms are defined for $p \in [1,\infty[$ as $\lVert M \rVert_{\mathrm{ew},p} :=  \big(\sum_{ij}\lvert M_{ij} \rvert^p\big)^{1/p}$ and $\lVert M \rVert_{\mathrm{Sc},p} :=  \big(\sum_{i} (\sigma_{i}(M)^p)\big)^{1/p}$, and for $p=\infty$ as $\lVert M \rVert_{\mathrm{ew},\infty} :=  \max_{ij}\lvert M_{ij} \rvert$ and $\lVert M \rVert_{\mathrm{Sc},\infty} :=  \max_{i} \left(\sigma_{i}(M)\right)$. In the previous expressions, $\sigma_{i}(M)$ denotes the $i^{\mathrm{th}}$ singular value of the matrix $M$.

\begin{lem}[General bound on variance error terms]
\label{lemma:variance-error-terms-first-general-bound}
For all $j$, all $u$, it holds that for all $l \geq1$:
\begin{align}
    |\epsilon^{(l-1,j,u)}| &\leq d_k \, \lVert T^{l-1} - T_\infty \rVert_{\mathrm{Sc},2} \,,\label{eq:variance-error-terms-first-general-bound-1}\\
    |\tilde{\epsilon}^{(\tilde{l}-1,\tilde{j},u)}| &\leq d_k \, \lVert T^{l-1} - T_\infty \rVert_{\mathrm{Sc},2} \,.\label{eq:variance-error-terms-first-general-bound-2}
\end{align}
Here, $T_\infty$ denotes here the $d_k \times d_k$ matrix with all coefficients equal to $1/d_k$, and $T$ is the stochastic matrix of \cref{eq:stochastic-matrix-T-def}.
\end{lem}
\begin{proof}
    We begin by showing the following claim. For $S$ any $d_k \times d_k$ complex matrix, $\overrightarrow{Y}$ any $d_k \times d_k$ probability vector, and for any $u \in [d_k]$, the following holds:
    \begin{equation}\label{eq:variance-error-terms-first-general-bound-subclaim-1}
        \left| (S \cdot \overrightarrow{Y})_u - \frac{1}{d_k}\right| \leq  \lVert S - T_\infty \rVert_{\mathrm{ew},\infty}\,.
    \end{equation}
    Indeed:
    \begin{align*}
        \left| (S \cdot \overrightarrow{Y})_u - \frac{1}{d_k}\right|
        &= \left| \overrightarrow{[S]_{u \bullet}} \cdot \overrightarrow{Y} - \frac{1}{d_k} \ \right|\\
        &= \left| \overrightarrow{[S]_{u \bullet}} \cdot \overrightarrow{Y} - \overrightarrow{[T_\infty]_{u \bullet}} \cdot \overrightarrow{Y} \ \right|\\
        &= \left| \left(\overrightarrow{[S]_{u \bullet}} - \overrightarrow{[T_\infty]_{u \bullet}}\right) \cdot \overrightarrow{Y} \ \right|\\
        &\leq \lVert \overrightarrow{[S]_{u \bullet}} - \overrightarrow{[T_\infty]_{u \bullet}} \rVert_\infty \  \,\lVert \overrightarrow{Y} \rVert_1\\
        &\leq \lVert S - T_\infty \rVert_{\mathrm{ew},\infty}\,.
    \end{align*}
    In the above, we used the bullet to denote a dummy index, so that for instance $\overrightarrow{[S]_{u \bullet}}$ is a vector of size $d_k$, whose entries are the $u^{\mathrm{th}}$ row of the matrix $S$. The second equality holds because $\overrightarrow{Y}$ is a probability vector, the third equality is the $(\infty,1)$-Hölder inequality for vectors, and the last inequality results from the definition of the matrix norm $\lVert \cdot \rVert_{\mathrm{ew},\infty}$ and from $\overrightarrow{Y}$ being a probability vector.

    Choosing for the matrix $S$ the stochastic matrix $T_{\text{rest},\lambda} \cdot T^{l(\lambda)-1}$ (from Eq.~\eqref{eq:repeated-recurrence-relation-forward}), and for the probability vector $\overrightarrow{Y}$ the vector $\overrightarrow{Z^0}$  associated to the initial state (Eq.~\eqref{eq:proba-vector-forward-lambda}), the claim of Eq.~\eqref{eq:variance-error-terms-first-general-bound-subclaim-1} yields:
    \begin{align}
        \left| (T_{\text{rest},\lambda} \cdot T^{l(\lambda)-1} \cdot \overrightarrow{Z^0})_u - \frac{1}{d_k}\right|
        &\leq  \lVert T_{\text{rest},\lambda} \cdot T^{l(\lambda)-1} - T_\infty \rVert_{\mathrm{ew},\infty}\label{eq:variance-error-terms-first-general-bound-nextineqs1}\\
        &\leq  \lVert T_{\text{rest},\lambda} \cdot T^{l(\lambda)-1} - T_\infty \rVert_{\mathrm{Sc},2}\label{eq:variance-error-terms-first-general-bound-nextineqs2}\\
        &\leq  \lVert T_{\text{rest},\lambda} \rVert_{\mathrm{Sc},2} \ \, \lVert T^{l(\lambda)-1} - T_\infty \rVert_{\mathrm{Sc},2}\label{eq:variance-error-terms-first-general-bound-nextineqs3}\\
        &\leq d_k \, \lVert T^{l(\lambda)-1} - T_\infty \rVert_{\mathrm{Sc},2}\,.\label{eq:variance-error-terms-first-general-bound-nextineqs4}
    \end{align}
In the above, the first inequality is the mentioned application of the claim of Eq.~\eqref{eq:variance-error-terms-first-general-bound-subclaim-1}, the second inequality holds because of the relation $\lVert \cdot \rVert_{\mathrm{ew},\infty} \leq \lVert \cdot \rVert_{\mathrm{ew},2} = \lVert \cdot \rVert_{\mathrm{Sc},2}$ between matrix-norms, the third inequality is due to the sub-multiplicativity of Schatten $p$-norms, and the fourth inequality is due to the norm inequalities $\lVert \cdot \rVert_{\mathrm{Sc},2} \leq d_k\lVert \cdot \rVert_{\mathrm{Sc},\infty}$ along with the fact that the moduli of eigenvalues of stochastic matrices are less or equal to one (as mentioned in \cref{subchap:StochasticMatrices}).

Thus Eq.~\eqref{eq:variance-error-terms-first-general-bound-1} is established. By applying again the claim of Eq.~\eqref{eq:variance-error-terms-first-general-bound-subclaim-1}, but this time choosing for the matrix $S$ the stochastic matrix $\widetilde{T}_{\text{rest},\lambda} \cdot \widetilde{T}^{\tilde{l}(\lambda)-1}$ (from Eq.~\eqref{eq:repeated-recurrence-relation-reverse}), and for the probability vector $\overrightarrow{Y}$ the vector $\overrightarrow{W^{\lambda_{\text{max}}}}$  associated to the target state (Eq.~\eqref{eq:proba-vector-reverse-lambda})
one gets (by following the same reasoning as \cref{eq:variance-error-terms-first-general-bound-nextineqs1,eq:variance-error-terms-first-general-bound-nextineqs2,eq:variance-error-terms-first-general-bound-nextineqs3,eq:variance-error-terms-first-general-bound-nextineqs4}):
\begin{align}
        \left| (\widetilde{T}_{\text{rest},\lambda} \cdot \widetilde{T}^{\tilde{l}(\lambda)-1} \cdot \overrightarrow{W^{\lambda_{\text{max}}}})_u - \frac{1}{d_k}\right|
        &\leq d_k \, \lVert \widetilde{T}^{l(\lambda)-1} - T_\infty \rVert_{\mathrm{Sc},2}\,.\label{eq:variance-error-terms-first-general-bound-analogous-ineq}
    \end{align}
But in the right-hand-side, $\lVert \widetilde{T}^{l(\lambda)-1} - T_\infty \rVert_{\mathrm{Sc},2} = \lVert T^{l(\lambda)-1} - T_\infty \rVert_{\mathrm{Sc},2}$ (due to the fact that $\widetilde{T}=T^{\intercal}$, that $T_\infty$ is symmetric, and that Schatten-$p$ norms are invariant under transposition), so Eq.~\eqref{eq:variance-error-terms-first-general-bound-2} is established.
\end{proof}

\subsection{\emph{Irreducible} and \emph{primitive} stochastic matrices}
\label{subchap:irreducible-and-primitive-stochmats}

Let us first introduce elementary notions about graphs. In what follows, $N$ denotes any integer such that $N\geq 2$.

By a \emph{directed graph} $\Gamma$, we mean a pair $\Gamma=(V,E)$ where $V$ is any finite set (its elements are the \textit{vertices} of $\Gamma$) and $E$ is any subset of $V^2=V \times V$ (its elements are the \textit{directed edges} of $\Gamma$).
We may denote a directed edge $(i,j)\in E$ by $i \to j$.
Importantly, note that in the above definition of directed graphs, we have allowed them to have \emph{self-loops}, i.e. directed edges $i\to i$ from a vertex to itself.

The adjacency matrix $A(\Gamma)$ of a directed graph $\Gamma=(\llbracket1, N\rrbracket,E)$ over $N$ vertices is the $N \times N$ matrix defined, for all $(i,j) \in \llbracket1, N\rrbracket^2$, by $[A(\Gamma)]_{ij} = 1$ if the graph possesses the directed edge $i \to j$, and $0$ otherwise.

A directed graph $\Gamma=(V,E)$ is said to be \emph{strongly-connected} if for every ordered pair of vertices $(i,j)\in V^2$ there exists a path in the graph from $i$ to $j$ (i.e. on a drawing of the graph one can go from $i$ to $j$ by following arrows). Importantly, one can notice that $\Gamma=(\llbracket1, N\rrbracket,E)$ is strongly-connected if and only if its adjacency matrix $A(\Gamma)$ has the property:
\begin{equation}\label{eq:directed-graph-SC-caract}
\forall (i,j) \in \llbracket1, N\rrbracket^2 \,\ \,\  \exists p_{(i,j)} \geq 1 \,\ \,\  [A(\Gamma)^{p_{(i,j)}}]_{ij}>0\,. 
\end{equation}

Now, Let $T$ be an $N \times N$ stochastic matrix.

The directed graph $\Gamma(T)$ of $T$ is defined as the directed graph $\Gamma=(\llbracket1, N\rrbracket,E)$  over $N$ vertices such that for all $(i,j) \in \llbracket1, N\rrbracket^2$, the graph possesses a directed edge $i \to j$ if and only if $[T]_{ij}>0$.

$T$ is said to be \emph{irreducible} if its directed graph $\Gamma(T)$ is strongly-connected.

Furthermore, $T$ is said to be \emph{primitive} if for a certain power of $p$, all the matrix coefficients of $T^p$ are positive. (It may be checked that if $p$ is such a power, than all subsequent matrix powers $q\geq p$ keep remain with positive coefficients as well.)
Because this property may be written as 
\begin{equation}\label{eq:stoch-mat-primitive}
\exists p \geq 1  \,\ \,\   \forall (i,j) \in \llbracket1, N\rrbracket^2 \,\ \,\  [A(\Gamma(T))^{p}]_{ij}>0\,, 
\end{equation}
notice (by comparing with \cref{eq:directed-graph-SC-caract}) that $T$ being primitive is in general a stronger property than $T$ being irreducible.

\subsection{\emph{Connected} RBS/FBS patterns, and properties of their associated stochastic matrices}\label{subchap:connected-stochmats-props}

In this section, we first phrase in a precise manner what the assumption of $U_0(\bm{\theta})$ being connected (as it was stated in the main text's \cref{def:CPSA}) means, we then prove that such an assumption indeed implies that the graphs of the associated stochastic matrices $T$ (at all Hamming-weights $k$) are strongly-connected, and we furthermore show that those $T$ are in fact primitive. Lastly, we give a sufficient condition on $U_0$ so that $T$ is symmetric.

Let $U_0=U_0(\bm{\theta})$ be a pattern of RBS/FBS gates on $n$ qubits, taken formally as an ordered list of triples $((i_1,j_1,\theta_1),(i_2,j_2,\theta_2),\dots,(i_J,j_J,\theta_J))$, where each entry $(i_k,j_k,\theta_k)$ indicates the presence of an RBS/FBS gate placed from qubit $i_k$ to qubit $j_k$, and set at angle $\theta_k$, and where the ordering of the list corresponds to time.

The \emph{graph $\Gamma(U_0)$} associated to the pattern $U_0$ is defined as the directed graph $\Gamma=(\llbracket1, N\rrbracket,E)$ over $N$ vertices whose directed edges indicate the presence of an RBS/FBS gate in the above ordered list of $U_0$, i.e. $E=\{ (i_1,j_1),(i_2,j_2),\dots,(i_J,j_J) \}$. In words, $\Gamma(U_0)$ may be thought of being the result of taking the $n$-qubit circuit depiction of $U_0$, "flattening out" the time/depth axis, and adding arrow tips on each side of every vertical line that represented a gate.

In this section, $U_0^k$ will denote the unitary matrix of the pattern $U_0(\bm{\theta})$ of RBS/FBS gates in the subspace of Hamming weight $k$. We will denote by $T^{(U_0^k)}$ the $d_k \times d_k$ stochastic matrix associated to that pattern of gates $U_0^k$  (\cref{eq:stochastic-matrix-T-def}, where it was denoted $T$).
Recall from the previous section that $\Gamma(T^{(U_0^k)})$ then denotes the graph over $d_k$ vertices with directed edges $u\to v$ exactly when $[T^{(U_0^k)}]_{uv}>0$.

\begin{definition}[Connected pattern of RBS/FBS gates]
\label{def:connected-pattern}
The pattern $U_0$ of RBS/FBS gates is said to be \emph{connected} if its associated directed graph $\Gamma(U_0)$ is strongly-connected. 
\end{definition}

The following lemma will be useful later in this section:
\begin{lem}
\label{lem:connected-section-lem1}
Let $T$ be an $N \times N$ stochastic matrix. Let $A,B\geq0$ and let $S_1,\dots,S_A,S'_{1},\dots,S'_B$ be arbitrary $N \times N$ matrices that all have nonnegative entries, and only positive entries on their diagonals.
Then, for all $(u,v) \in \llbracket1, N\rrbracket^2$:
\begin{equation}
    [T]_{uv}>0 \implies [S_1\cdots S_A \,
T S'_{1}\,\cdots S'_B]_{uv}>0\,.
\end{equation}
\end{lem}
\begin{proof}
The case $A=1,B=0$ is readily shown, since, if $[T]_{uv}>0$, then
\begin{equation}
    [S\,T]_{uv}
    = \sum_{k=1}^N S_{uk} T_{kv}
    = S_{uu} T_{uv} + \sum_{\substack{k=1\\k\neq u}}^N S_{uk} T_{kv}\,,
\end{equation}
and thus this expression is positive, as the first term $S_{uu} T_{uv}$ is positive by assumption and the rest of the summed terms are all nonnegative.

The case $A=0,B=1$ is shown similarly. Then, the cases of general $A$ and $B$ may be shown to follow by induction.
\end{proof}

Let us denote by $\operatorname{Involved}_k(i,j)$ the pairs of indices of basis vectors $B_k^n$ of Hamming-weight $k$ that would be "involved" together in a rotation if an RBS/FBS gate was applied between qubits $i$ and $j$. Explicitly: 
\begin{equation}
\operatorname{Involved}_k(i,j) := \big\{ (u,v)\in \llbracket1, d_k\rrbracket^2 
\ \big| \ (e_u)^i = 1 \mathrm{\;and\;} (e_v)^j = 0\mathrm{\ \;\;or\ \;\;} (e_u)^i = 0 \mathrm{\;and\;} (e_v)^i = 1 \big\}\,,
\end{equation}
where we used notation $(e_u)^i$ for the $i^{\mathrm{th}}$ bit of the $n$-bitstring $e_u$. Note that this set has $2{n-2 \choose k-1}$ elements.
\begin{lem}
\label{lem:connected-section-lem2}
If $(i,j)$ is a directed edge in the graph $\Gamma(U_0)$, then for any $k \in \llbracket 1, n-1\rrbracket$,  all the elements of $\operatorname{Involved}_k(i,j)$ are directed edges in the graph $\Gamma(T^{(U_0^k)})$.
\end{lem}
\begin{proof}
Suppose that the graph $\Gamma(U_0)$ contains the edge $i\to j$. This means that an RBS/FBS gate between qubits $i$ and $j$ is present in the pattern $U_0$, and therefore for every $k$, its associated stochastic matrix for the Hamming-weight $k$, $T^{(U_0^k)}$, is a product (see \cref{eq:stochastic-matrix-T-def}) of $J$ "elementary" stochastic matrices associated to single RBS/FBS gates (\cref{eq:stochastic-matrix-Tlambda-def}), and one of them corresponds to a gate between qubits $i$ and $j$. Let us write this as 
\begin{equation}
T^{(U_0^k)} = T_1\cdots T_A \,
\,T^{(i,j)}\, T'_{1}\,\cdots T'_B\,.
\end{equation}
But by the definition \cref{eq:stochastic-matrix-Tlambda-def} of these elementary stochastic matrices in this product, one has firstly that they all have positive diagonal coefficients everywhere (the $u^{\mathrm{th}}$ diagonal coefficient is either $1/2$ if $u$ is involved in the gate's rotation and $1$ otherwise), and secondly that for all $(u,v) \in \operatorname{Involved}_k(i,j)$, $[T^{(i,j)}]_{ij}=[T^{(i,j)}]_{ji}=1/2$. 
Therefore, the previous \cref{lem:connected-section-lem1} readily applies, to give, for all $(u,v) \in \operatorname{Involved}_k(i,j)$:
\begin{equation}
[T^{(U_0^k)}]_{uv}>0\,.
\end{equation}
\end{proof}

\begin{lem}
\label{lem:connected-section-lem3}
If the graph $\Gamma(U_0)$ is strongly-connected, then for all $k \in \llbracket 1, n-1\rrbracket$ the graph $\Gamma(T^{(U_0^k)})$ is strongly-connected as well.
\end{lem}
\begin{proof}
Fix a $k \in \llbracket 1, n-1\rrbracket$, and let $(u,v)\in \llbracket1, d_k\rrbracket^2$. Let us show that there exists a path $\mathcal{P}=(u \to \cdots \to v)$ of directed edges in the graph $\Gamma(T^{(U_0^k)})$ joining vertex $u$ to vertex $v$.

Let $(h_1^u,\dots,h_k^u)$ and $(h_1^v,\dots,h_k^v)$ be the indices that hold the values $1$ in the $n$-bitstrings $e_u$ and $e_v$, respectively, and  let $(i_1^u,\dots,i_{k'}^u)$ and $(i_1^v,\dots,i_{k'}^v)$ be the respective subsets of those for which none of the remaining indices are shared between the first and second tuples (i.e. all first indices $i_1^u,\dots,i_{k'}^u$ are different from all the second indices $i_1^v,\dots,i_{k'}^v$).

For all $s \in \llbracket 1, k'\rrbracket$, use the assumption that $\Gamma(U_0)$ is strongly-connected to get the existence of a path $\mathcal{P}_s=(i_s^u \to \cdots \to i_s^v)$ of directed edges in $\Gamma(U_0)$.

Now, one constructs the desired path $\mathcal{P}=(u \to \cdots \to v)$, by successively invoking \cref{lem:connected-section-lem2} along each whole path $\mathcal{P}_s$ in $\Gamma(U_0)$ to get a corresponding path $\mathcal{P}^k_s$ in $\Gamma(T^{(U_0^k)})$, and by concatenating the obtained paths $\mathcal{P}^k_1 , \mathcal{P}^k_2 ,\dots, \mathcal{P}^k_{k'}.$
\end{proof}

\begin{cor}
\label{cor:connected-section-cor1}
If the pattern $U_0$ of RBS/FBS gates is connected (\cref{def:connected-pattern}), then for every Hamming weight $k \in \llbracket 1, n-1\rrbracket$, the associated stochastic matrix $T^{(U_0^k)}$ is primitive.
\end{cor}
\begin{proof}
Fix a Hamming weight $k \in \llbracket 1, n-1\rrbracket$, and denote $T:=T^{(U_0^k)}$. If $U_0$ is connected, then by \cref{lem:connected-section-lem3} the stochastic matrix $T$ is irreducible, meaning that the property of \cref{eq:directed-graph-SC-caract} holds for $\Gamma=\Gamma(T)$. In general, a stochastic matrix $T$ being irreducible does not imply that it is primitive, but in our case this actually follows. Indeed, for each $(u,v)\in \llbracket1, d_k\rrbracket^2$, the characterization of irreducibility of $T$ of \cref{eq:directed-graph-SC-caract} gives a $p_{(u,v)}\geq1$ such that $[T^{p_{(u,v)}}]_{uv}>0$. Now, for this stochastic matrix $T^{p_{(u,v)}}$, applying \cref{lem:connected-section-lem1} to it yields (recursively, with $S_1:=T$) that for all subsequent powers $q\geq p_{(u,v)}$, one still has $[T^{q}]_{uv}>0$.
Therefore, taking $p:=\max_{(u,v)\in \llbracket1, d_k\rrbracket^2}(p_{(u,v)})$, it follows that $T^q$ has positive entries whenever $q\geq p$, i.e. we have established that $T$ is primitive.
\end{proof}

\begin{lem}\label{lem:T-is-doubly-stoch}
For every Hamming weight $k \in \llbracket 1, n-1\rrbracket$, the stochastic matrix $T^{(U_0^k)}$ associated to any pattern of RBS/FBS gates $U_0$ is always doubly-stochastic.
\end{lem}
\begin{proof}
    Denote $T:=T^{(U_0^k)}$.
    $T$ is constructed as a product (c.f. \cref{eq:stochastic-matrix-T-def}) of "elementary" stochastic matrices (\cref{eq:stochastic-matrix-Tlambda-def}) that are doubly-stochastic (since those are in fact symmetric, by definition). Hence, recalling that the set of doubly-stochastic matrices is closed under product, $T$ is doubly-stochastic as well.
\end{proof}

\begin{lem}
\label{lem:connected-section-palindrome}
If the pattern $U_0$ if RBS/FBS gates is a palindrome, then for every Hamming weight $k \in \llbracket 1, n-1\rrbracket$, the associated stochastic matrix $T^{(U_0^k)}$ is symmetric.

By the pattern being a palindrome, we mean here that the ordered list defining the pattern is of the form
\begin{equation*}
   \Big((i_1,j_1,\;\theta_1),\dots,\;(i_{M-1},j_{M-1},\theta_{M-1}),\;(i_M,j_M,\theta_M),\;(i_{M-1},j_{M-1},\theta_{M+1}),\;\dots,(i_1,j_1,\theta_{2M-1})\Big)\,.
\end{equation*}
\end{lem}
\begin{proof}
Indeed, in this case, the stochastic matrix $T:=T^{(U_0^k)}$ is of form
\begin{align} 
    T 
    &= T_{1} \, T_{2} \cdots T_{M-1} \, T_{M} \, T_{M-1} \cdots T_{2} \, T_{1}\\
    &= T_{1} \, T_{2} \cdots T_{M-1} \, T_{M} \, T_{M}\, T_{M-1} \cdots T_{2} \, T_{1}\\
    &= T_{(B)} T_{(F)}
\end{align}
where we denoted $T_{(F)}:=T_{M}\, T_{M-1} \cdots T_{2} \, T_{1}$ and $T_{(B)}:=T_{1} \, T_{2} \cdots T_{M-1} \, T_{M}$. In the second inequality, $T_M = T_M^2$ was used, as it can indeed by checked that all stochastic matrices corresponding to single RBS/FBS gates square to themselves (from their definition of \cref{eq:stochastic-matrix-Tlambda-def}).
But we have
\begin{align} 
    T_{(F)}^\intercal
    &= (T_{M} \, T_{M-1} \cdots T_{2} \, T_{1})^\intercal\\
    &= T_{1}^\intercal \, T_{2}^\intercal \cdots T_{M-1}^\intercal \, T_{M}^\intercal\\
    &= T_{1} \, T_{2} \cdots T_{M-1} \, T_{M}\\
    &= T_{(B)}
\end{align}
where the third equality is because all stochastic matrices corresponding to single RBS/FBS gates are symmetric (this also stems from their definition of \cref{eq:stochastic-matrix-Tlambda-def}).
Hence,
    \begin{align}
    T = T_{(F)}^\intercal T_{(F)}\,,
    \end{align}
which establishes that $T$ is symmetric.
\end{proof}

\subsection{Convergence of powers of a \emph{fixed} stochastic matrix $T$}\label{subchap:convergence-fixed}

\begin{thm}[Exponential convergence of $(T^l)_{l \in \mathbb{N}}$]
\label{thm:stoch-mat-convergence}
Let $N\geq1$ be fixed, and let $T \in \mathbb{R}^{N \times N}$ be a column-stochastic matrix. If $T$ is \emph{primitive}, then the following points hold:
\begin{enumerate}[align=left]
    \item[1. (Convergence to rank-one matrix)] The sequence of matrix powers $(T^l)_{l \in \mathbb{N}}$ converges to a certain matrix $T_{\infty}$, whose columns are all identical and equal to some probability vector $\vec{\pi}=(\pi_1,\dots,\pi_N)^\intercal$.

    \item[2. (Upper-bound on rate of convergence)]
    There exists constants $l_0 \in \mathbb{N}$ and $A,B>0$ (depending on $T$) such that for all $l \geq l_0$,
    \begin{equation}\label{eq:stoch-mat-convergence}
    \lVert T^l - T_{\infty} \rVert_{\mathrm{ew},\infty} \leq \frac{A}{\exp(B\, l)}\,,
    \end{equation}

    \item[3. (Double-stochastic case)] If $T$ is furthermore doubly stochastic, then $\vec{\pi}=(1/N,\dots,1/N)^\intercal$, i.e.
    \begin{equation}
    T_\infty = \begin{pmatrix} 
    1/N & \dots  & 1/N\\
    \vdots & \ddots & \vdots\\
    1/N & \dots  & 1/N 
    \end{pmatrix}\,.
    \end{equation}
\end{enumerate}
\end{thm}
\begin{proof}
$ $\newline
\begin{enumerate}[align=left]
    \item[1.] This is part of the content of the Perron–Frobenius theorem (for the special case of primitive stochastic matrices), see for instance \cite[Chapter 8]{Horn-MatrixAnalysis-2013}.

    \item[2.] Recall that all eigenvalues $\lambda$ of $T$ are contained in the complex closed unit disk. One of the other points of the Perron–Frobenius theorem for primitive stochastic matrices $T$ is that, if $\lambda$ is any eigenvalue of $T$ different from $1$, then $|\lambda|<1$. Let $\lambda_2$ denote any eigenvalue of $T$ that achieves the highest modulus value $|\lambda|$, among all eigenvalues $\lambda$ besides $1$. We will now explicitly show how this implies that $(T^l)_{l \in \mathbb{N}}$ converges exponentially with a rate $B$ governed by this largest eigenvalue modulus. We do so in pedagogical detail, notably because later on we will remark how these methods succeed or fail to be conclusive in the more generalized setting of starting with not one but a \emph{sequence} of stochastic matrices.
    
    Firstly, suppose it is the case that $T$ is \emph{normal}, i.e. $T^\dagger T = T T^\dagger$, as it is the most intuitive case. By the spectral theorem, this is equivalent to the existence of a \emph{unitary} matrix $P$ such that $T = P D P^{-1}$, where $D:=\text{diag}(1,\lambda_2,\dots,\lambda_k)$ and $1,\lambda_2,\dots,\lambda_k$ are the eigenvalues of $T$ (repeated with multiplicity). It then follows that $T^l = P D^l P^{-1}$, from which taking the limit $l \to \infty$ on both sides gives (by the previous point of the current theorem) $T_\infty = P E_1 P^{-1}$, with $E_1:=\text{diag}(1,0,\dots,0)$. Therefore, we have
    \begin{equation}\label{eq:stoch-mat-convergence-normal-eq1}
    \lVert T^l - T_{\infty} \rVert_{\mathrm{Sc},2} = \lVert P D^l P^{-1} - P E_1 P^{-1} \rVert_{\mathrm{Sc},2} = \lVert P (D^l - E_1) P^{-1} \rVert_{\mathrm{Sc},2}\,,
    \end{equation}
    But since $P$ is unitary, we have by unitary invariance of the Schatten $p$-norms that
    \begin{equation}\label{eq:stoch-mat-convergence-normal-eq2}
    \lVert P (D^l - E_1) P^{-1} \rVert_{\mathrm{Sc},2} = \lVert D^l - E_1 \rVert_{\mathrm{Sc},2}\,.
    \end{equation}
    Furthermore, we have
    \begin{align}
    \lVert D^l - E_1 \rVert_{\mathrm{Sc},2}
    &= \lVert \text{diag}(0,\lambda_2^l,\lambda_3^l,\dots,\lambda_N^l) \rVert_{\mathrm{Sc},2}\nonumber\\
    &= \sqrt{|\lambda_2|^{2l} + |\lambda_3|^{2l} + \cdots + |\lambda_N|^{2l}}
    \leq \sqrt{N-1}\,|\lambda_2|^{l}\,.\label{eq:stoch-mat-convergence-normal-eq3}
    \end{align}
    Therefore, combining \cref{eq:stoch-mat-convergence-normal-eq1,eq:stoch-mat-convergence-normal-eq2,eq:stoch-mat-convergence-normal-eq3}, along with the norm inequality $\lVert \cdot \rVert_{\mathrm{ew},\infty} \leq \lVert \cdot \rVert_{\mathrm{ew},2} = \lVert \cdot \rVert_{\mathrm{Sc},2}$, yields the claim of \cref{eq:stoch-mat-convergence} (assuming $T$ to be normal) with
    \begin{equation}
    l_0=1,\quad A=\sqrt{N-1},\quad \text{and }  B=\ln(1/|\lambda_2|).
    \end{equation}

    Without an assumption of normality of the matrix $T$, it is still possible to show the exponential convergence with $l$, by making use of the Jordan canonical form (which applies to any square matrix) of $T$. Indeed, it provides the existence of an \emph{invertible} matrix $P$ such that $T = P J P^{-1}$, where $J$ is a matrix of the form
    \begin{equation}
    J = 
        \begin{pmatrix}
        1&0&\;&\;&\;\\
        \;&\lambda_{2}&\bullet&\;&\;\\
        \;&\;&\ddots&\ddots&\;\\
        \;&\;&\;&\ddots&\bullet\\
        \;&\;&\;&\;&\lambda_{N}
        \end{pmatrix}\,,
    \end{equation}
    where the elements on the diagonal are the eigenvalues of $T$ (repeated with algebraic multiplicity), where the elements on the superdiagonal take values $0$ or $1$ (in some manner that depends on the geometric multiplicity of the eigenvalue to the left of it), and where all other entries are zero. A precise statement may be found in \cite[Section 3.1]{Horn-MatrixAnalysis-2013}, but the important takeaway is that even if $J$ is not exactly diagonal, its powers $J^l$ will still converge towards $E_1$ in the same asymptotic fashion (i.e. exponentially fast with the rate being governed by $|\lambda_2|$). Indeed, it can be readily shown (e.g. from \cite[Section 3.2.5]{Horn-MatrixAnalysis-2013}) that its powers verify the property that for all $l \geq N$,
    \begin{equation}
    \lVert J^l - E_1 \rVert_{\mathrm{ew},\infty} \leq \frac{N}{s^{N-1}} \ \, l^{N-1} |\lambda_2|^l\,,
    \end{equation}
    where $s(T)>0$ denotes the smallest positive modulus of an eigenvalue of $T$.
    But since
    \begin{equation}
    l^{N-1} |\lambda_2|^l = l^{N-1} \exp\left(-\ln(1/|\lambda_2|) \, l\right) \in \underset{l\to\infty}{\mathcal{O}}\left( \exp\left(-\frac{1}{2}\ln(1/|\lambda_2|) \, l\right) \right)\,,
    \end{equation}
    there exists some $l_0(N)\geq1$ such that for all $l\geq l_0(N)$,
    \begin{equation}\label{eq:stoch-mat-convergence-nonnormal-eq0}
    \lVert J^l - E_1 \rVert_{\mathrm{ew},\infty} \leq \frac{N}{s^{N-1}} \ \exp\left(-\frac{1}{2}\ln(1/|\lambda_2|) \, l\right)\,\,.
    \end{equation}

    Besides, just as in \cref{eq:stoch-mat-convergence-normal-eq1}, we also have
    \begin{equation}\label{eq:stoch-mat-convergence-nonnormal-eq1}
    \lVert T^l - T_{\infty} \rVert_{\mathrm{Sc},2} = \lVert P J^l P^{-1} - P E_1 P^{-1} \rVert_{\mathrm{Sc},2} = \lVert P (J^l - E_1) P^{-1} \rVert_{\mathrm{Sc},2}\,,
    \end{equation}
    however here $P$ is not a priori unitary, so we may merely invoke sub-multiplicativity of Schatten or entry-wise norms (and not unitary invariance) which gives
    \begin{equation}\label{eq:stoch-mat-convergence-nonnormal-eq2}
    \lVert P (J^l - E_1) P^{-1} \rVert_{\mathrm{Sc},2} \leq \lVert P \rVert_{\mathrm{Sc},2} \, \lVert P ^{-1}\rVert_{\mathrm{Sc},2} \,\,  \lVert J^l - E_1 \rVert_{\mathrm{Sc},2}:=\mathrm{cond}(P)\,\lVert J^l - E_1 \rVert_{\mathrm{Sc},2}\,.
    \end{equation}
    where $\mathrm{cond}(P):=\lVert P \rVert_{\mathrm{Sc},2} \, \lVert P ^{-1}\rVert_{\mathrm{Sc},2}$ is the so-called \emph{condition number} of the matrix $P$ (with respect to $\lVert \cdot \rVert_{\mathrm{Sc},2}$).
    
    Hence, combining \cref{eq:stoch-mat-convergence-nonnormal-eq0,eq:stoch-mat-convergence-nonnormal-eq1,eq:stoch-mat-convergence-nonnormal-eq2} yields this time the claim of \cref{eq:stoch-mat-convergence} (without assuming $T$ to be normal) with
    \begin{equation}
    l_0=l_0(N),\quad A=\frac{N \, \mathrm{cond}(P)}{s(T)^{N-1}},\quad \text{and }  B=\frac{1}{2}\ln(1/|\lambda_2|).
    \end{equation}

    \item[3.] If $T$ is doubly stochastic, then so is $T^l$ for all $l\geq1$, and therefore as mentioned above the limit $T_\infty$ is also doubly stochastic. Since $T_\infty$ is a doubly stochastic matrix of rank 1 (as its columns are all equal), it 
    is necessarily equal to
    \begin{equation}
    T_\infty = \begin{pmatrix} 
    1/N & \dots  & 1/N\\
    \vdots & \ddots & \vdots\\
    1/N & \dots  & 1/N 
    \end{pmatrix}\,.
\end{equation}
    Indeed, if a matrix is row-stochastic and rank-1, all of its rows are equal --- due to the fact that because its first column is nonzero (since it is a probability vector) rank-1 implies that all its other columns are scalar multiples of the first, and hence must be equal to the first (since they must all be probability vectors). Likewise, if a matrix is column-stochastic and rank-1, all of its columns are equal. It follows that if a matrix is doubly stochastic and rank-1, all of its entries are equal, and hence equal to $1/N$.
\end{enumerate}
\end{proof}

The next theorem, taken from the literature of mixing times of Markov chains, gives qualitatively the same result (with different constants involved), but it turns out that in the next section, where the constants will become sequences, only this result will be able to be converted into our more general setting of interest (while the previous \cref{thm:stoch-mat-convergence} won't be usable in general).
\begin{thm}[{Adapted from \cite[Theorem 1.2]{Jerison-GeneralMixing-2013}}]
\label{thm:fixed-stoch-mat-Jeriso-convergence}
Let $N\geq1$ be fixed, let $T \in \mathbb{R}^{N \times N}$ be a column-stochastic matrix, and suppose that $T$ is \emph{primitive}. Denote by $\Delta:=1-|\lambda_2|$ the \emph{spectral gap} of $T$, and introduce the quantities 
\begin{align}
    A&:=\frac{2}{e}  \exp\Big(\big[ \ln(1/\Delta) + 2(1+\ln(2))\big] N \Big)\,,\\[9pt]
    B&:=\frac{\Delta}{2}\,.
\end{align}
Then, for all $l\in \mathbb{N}$ such that $A/\exp(B \, l) < 1$:
\begin{equation}\label{eq:fixed-stoch-mat-Jeriso-convergence-stoch-mat-convergence}
    \lVert T^l - T_{\infty} \rVert_{\mathrm{ew},\infty} \leq \frac{A}{\exp(B \, l)}\,.
\end{equation}
Here, $T_{\infty}$ denotes a certain  $N \times N$ stochastic matrix of rank $1$.
\end{thm}
\begin{proof}
Since $T$ is primitive, we know it converges to some rank-1 column-stochastic matrix $T_\infty$ (see the proof of \cref{thm:stoch-mat-convergence}). Therefore, the result of \cite[Theorem 1.2]{Jerison-GeneralMixing-2013} applies -- and it is straight-forward manipulation of inequalities to recast it into our above statement.
\end{proof}

\subsection{Convergence of powers of a \textit{sequence} of stochastic matrices $T_n$, and a spectral gap conjecture}\label{subchap:convergence-sequence-and-conjecture}

\begin{thm}
\label{thm:stoch-mat-sequence-convergence}
Let $(T_n)_{n\geq2}$ be a sequence of $N_n \times N_n$ stochastic matrices that are doubly-stochastic and primitive. Furthermore, suppose that

\begin{equation}\label{eq:stoch-mat-sequence-convergence-poly-assumptions}
    \Delta_n \in \Omega\big(1/\mathrm{poly}(n)\big)\,,
\end{equation}
where $\Delta_n:=1-|\lambda_2(T_n)|$ denotes the spectral gap of the stochastic matrix $T_n$.

Then, for any sequence $l_n \in \Omega( \Delta_n^{-1} \, N_n \, n)$, there exists a constant $c>0$ such that:
\begin{equation}\label{eq:stoch-mat-Jeriso-seuquence-convergence}
    \lVert T_n^{l(n)} - T_{n,\infty} \rVert_{\mathrm{ew},\infty} \in \mathcal{O}\left(\frac{1}{\exp(c\, n)}\right)\,.
\end{equation}
where $T_{n,\infty}$ denotes the $N_n \times N_n$ matrix with all entries equal to $1/N_n$.

\end{thm}
\begin{proof}

Denote
\begin{align}
    A_n&:=\frac{2}{e}  \exp\Big(\big[ \ln(1/\Delta_n) + 2(1+\ln(2))\big] N_n \Big)\,,\\[9pt]
    B_n&:=\frac{\Delta_n}{2}\,.
\end{align}

First, we claim that there exists $c_A >0$ such that
\begin{equation}\label{eq:stoch-mat-sequence-convergence-epsilon-claim1}
    A_n \in \mathcal{O}\left(\exp(c_A\,N_n\, \ln(n))\right)\,.
\end{equation}
Indeed, the assumption $(1/\Delta_n) \in \mathcal{O}(\mathrm{poly}(n)$ implies that $\big[ \ln(1/\Delta_n) + 2(1+\ln(2))\big] \in \mathcal{O}\left(\ln(n)\right)$, and hence $\big[ \ln(1/\Delta_n) + 2(1+\ln(2))\big]N_n \in \mathcal{O}\left(N_n\,\ln(n)\right)$, which indeed implies that there exists a constant $c_A>0$ such that \cref{eq:stoch-mat-sequence-convergence-epsilon-claim1} holds.

Second, we claim that for any choice of sequence $(l_n) \in \Omega( \Delta_n^{-1} \, N_n \, n)$, there exists a constant $c_B>0$ such that
\begin{equation}\label{eq:stoch-mat-sequence-convergence-epsilon-claim2}
    \frac{1}{\exp(B_n \, l_n)} \in \mathcal{O}\left(\frac{1}{\exp(c_B \, N_n \, n)}\right)\,.
\end{equation}

Indeed, picking any $l_n \in \Omega( \Delta_n^{-1} \, N_n \, n)$ 
implies that $(B_n \, l_n) \in \Omega(N_n \, n)$, which indeed implies that there exists a constant $c_B>0$ such that \cref{eq:stoch-mat-sequence-convergence-epsilon-claim2} holds.

Therefore, \cref{eq:stoch-mat-sequence-convergence-epsilon-claim1,eq:stoch-mat-sequence-convergence-epsilon-claim2} together give that
\begin{equation}\label{eq:stoch-mat-sequence-convergence-epsilon-intermeq1}
    \frac{A_n}{\exp(B_n \, l_n)} \in \mathcal{O}\left(\frac{1}{\exp\left( 
N_n\left[ c_B \, n - c_A \, \ln(n) \right] \right)}\right)\,.
\end{equation}
But since
\begin{equation}
    N_n\left[ c_B \, n - c_A \, \ln(n) \right]
    \geq \left[ c_B \, n - c_A \, \ln(n) \right]
    = n\left( c_B - c_A \frac{\ln(n)}{n} \right)
    \in \Omega \left( n \right)\,,
\end{equation}
it holds that
\begin{equation}\label{eq:stoch-mat-sequence-convergence-epsilon-intermeq2}
     N_n\left[ c_B \, n - c_A \, \ln(n) \right] \in \Omega \left( n \right)\,,
\end{equation}
and thus \cref{eq:stoch-mat-sequence-convergence-epsilon-intermeq1,eq:stoch-mat-sequence-convergence-epsilon-intermeq2} imply that there exists some $c>0$ such that 
\begin{equation}\label{eq:stoch-mat-sequence-convergence-epsilon-request-bound}
    \frac{A_n}{\exp(B_n \, l_n)} \in \mathcal{O}\left(\frac{1}{\exp\left(c \, n\right)}\right)\,.
\end{equation}

We have thus shown that there for any sequence $(l_n) \in \Omega( \Delta_n^{-1} \, N_n \, n)$ there exists a $c>0$ such that \cref{eq:stoch-mat-sequence-convergence-epsilon-request-bound} holds.
Now, take any such $(l_n)$ and apply, for each $n\geq2$, \cref{thm:fixed-stoch-mat-Jeriso-convergence} to the stochastic matrix $T:=T_n$, and to the power $l:=l_n$. Because \cref{eq:stoch-mat-sequence-convergence-epsilon-request-bound} implies that there exists an $n_0\geq1$ such that ${A_n}/{\exp(B_n \, l_n)} \leq 1$, these applications of \cref{thm:fixed-stoch-mat-Jeriso-convergence} give us, for all $n\geq n_0$, the result that
\begin{equation}\label{eq:stoch-mat-Jeriso-seuquence-convergence-applied-to-seq}
    \lVert T_n^{l(n)} - T_{n,\infty} \rVert_{\mathrm{ew},\infty} \leq \frac{A_n}{\exp(B_n\, l_n)}\,.
\end{equation}
Combining \cref{eq:stoch-mat-sequence-convergence-epsilon-request-bound,eq:stoch-mat-Jeriso-seuquence-convergence-applied-to-seq} yields
\begin{equation}\label{eq:stoch-mat-Jeriso-seuquence-convergence-applied-proved}
    \lVert T_n^{l(n)} - T_{n,\infty} \rVert_{\mathrm{ew},\infty} \in \mathcal{O}\left(\frac{1}{\exp\left(c \, n\right)}\right)\,.
\end{equation}
\end{proof}

Note that in the case where the matrices $T_n$ are normal (this is for instance the case when the pattern $U_0$ is a palindrome, due to \cref{lem:connected-section-palindrome}), it is possible to prove the result of \cref{thm:stoch-mat-sequence-convergence} more simply, by relying on the previous spectral theorem argument detailed in the proof of \cref{thm:stoch-mat-convergence} instead of on the result of \cref{thm:fixed-stoch-mat-Jeriso-convergence}, and by additionally assuming that $N_n \in \mathcal{O}\left(\mathrm{poly}(n)\right)$. 
Indeed, it yields for any sequence $(l_n)$ that for all $n\geq2$,
\begin{equation}
\lVert T_n^{l_n} - T_{n,\infty} \rVert_{\mathrm{ew},\infty} \leq \frac{\sqrt{N_n -1}}{\exp(\,\ln(1/\lambda_2(T_n))\, l_n\,)}\,,
\end{equation}
and thus, since $N_n \in \mathcal{O}\left(\mathrm{poly}(n)\right)$, taking $l_n \in \Omega\left(n/\ln(1/\lambda_2(T_n))\right)$ suffices to obtain $\lVert T_n^{l_n} - T_{n,\infty} \rVert_{\mathrm{ew},\infty} \in \mathcal{O}(1/c\,\exp(n))$ for some $c>0$; and because $\ln(1/\lambda_2(T_n)) \geq 1/\Delta_n$ (for all $n\geq2$) and $1/\Delta_n \in \Omega\left(1/\mathrm{poly}(n)\right)$, such an $l_n$ can be chosen to be in $\mathcal{O}\left(\mathrm{poly}(n)\right)$ as well.  

However, in the case where the matrices $T_n$ are not normal, the Jordan canonical form argument that was given as well in the proof of \cref{thm:stoch-mat-convergence} cannot be successfully employed to prove the result of \cref{thm:stoch-mat-sequence-convergence}, as the obtained bound would involve a condition number $\mathrm{cond}(P_n)$ (with $P_n$ the change of basis matrix that converts $T_n$ into its Jordan canonical form), over which we do not have any control of its scaling behavior with $n$.

\paragraph{A spectral gap conjecture}

As per the assumption \cref{eq:stoch-mat-sequence-convergence-poly-assumptions} in the previous theorem, we will need, in order to arrive at our conclusion of absence of Barren Plateaus, to make a conjecture on the size of the spectral gaps of the relevant stochastic matrices. We state this in \cref{conj:spectral-gap}, and we we provide numerical evidence that this conjecture holds, which we defer to \cref{subsec:numerical-evidence-spectral-gap}. The actual proof of \cref{conj:spectral-gap} is left for future work.

\begin{conj}[Spectral gaps of connected RBS/FBS patterns are inverse-polynomially large]
\label{conj:spectral-gap}

Let $(U_{0,n})_{n\geq2}$ be a sequence of connected patterns of RBS/FBS gates (\cref{def:connected-pattern}), where each $U_{0,n}$ is such a pattern over $n$ qubits.
If the number $J_n$ of gates in the pattern $U_{0,n}$ satisfies $J_n \in \mathcal{O}\left(\mathrm{poly}(n)\right)$, then for any fixed Hamming weight $k \in \mathcal{O}(1)$, the associated $d_{k,n} \times d_{k,n}$ stochastic matrix $T_n:=T^{(U_{0,n}^k)}$ (\cref{eq:stochastic-matrix-T-def}) satisfies
\begin{equation}
    \Delta_n \in \Omega\left(1/\mathrm{poly}(n)\right)\,,
\end{equation}
where $\Delta_n:=1-|\lambda_2(T_n)|$ denotes the spectral gap of the stochastic matrix $T_n$.
\end{conj}

\subsection{Precised version of \cref{thm:NoBPPSA}, and proof}\label{subchap:precised-theorem-2}

Putting it all together, we finally obtain:

\begin{thm}[Absence of Barren Plateaus]
\label{thm:concluding-back-to-variance}
Let $(U_{0,n})_{n\geq2}$ be a sequence of connected patterns of RBS/FBS gates (\cref{def:connected-pattern}), where each $U_{0,n}$ is such a pattern over $n$ qubits.
Assume that the number of gates $J_n \geq 1$ in the pattern $U_{0,n}$ satisfies $J_n \in \mathcal{O}\left(\mathrm{poly}(n)\right)$, and assume any fixed Hamming weight $k \in \mathcal{O}(1)$.

Then, for any integer sequences $(L_n)_{n\geq2}$ and $(l_n)_{n\geq2}$ satisfying $1 \leq l_n \leq L_n$ (for all $n\geq2$) as well as
\begin{equation}\label{eq:concluding-back-to-variance-hyp-Ln}
L_n\,,\;\;l_n\,,\;\;(L_n - l_n) \;\;\;\in\;\;\; \Omega\left( \Delta_n^{-1} \, d_{k,n} \, n\right)\,,
\end{equation}
the quantum circuit comprised of $L_n$ repetitions of the RBS/FBS patten $U_{0,n}$ has -- for any initial and target states of Hamming weight $k$ -- a cost function whose gradient for the parameter of the $j^{\mathrm{th}}$ gate in the $l_{n}^{\,\mathrm{th}}$ repetition (c.f. \cref{eq:periodic-structure-appendix}) has, for any $j\in\llbracket 1, J_n \rrbracket$, a variance of inverse-polynomial order, i.e.:
\begin{equation}
    \mathrm{Var}_{\bm{\theta}}[\partial_{\theta_{\lambda(l_n,j)}} \mathcal{C}]
    \in
    \Theta\left( 1/\mathrm{poly}(n) \right)\,.
\end{equation}
Here, $\Delta_n:=1-|\lambda_2(T_n)|$ denotes the spectral gap of the stochastic matrix $T_n:=T^{(U_{0,n}^k)}$ (\cref{eq:stochastic-matrix-T-def}), and $d_{k,n}:={n \choose k}$.

Assuming that \cref{conj:spectral-gap} holds, there exists sequences $(L_n)$ that simultaneously satisfy $L_n \in \Omega\left( \Delta_n^{-1} \, d_{k,n} \, n\right)$ and $L_n \in \mathcal{O}\left(\mathrm{poly}(n)\right)$.

In particular, letting $q_\Delta\in\mathbb{N}$ be the lowest integer such that $(1/\Delta_n) \in \mathcal{O}(n^{q_\Delta})$, and letting $q_k := \min(k,n-k)$ (the lowest integer such that $d_{k,n} \in \mathcal{O}(n^{q_k})$), the choices
\begin{equation}
    L_n := n^{q_\Delta + q_k + 1}
    \qquad\mathrm{and}\qquad\;\;
    l_n := \lfloor \alpha \, L_n \rfloor  \,,
\end{equation}
for any fixed constant $\alpha$ such that $0<\alpha<1$,
satisfy the assumptions of \cref{eq:concluding-back-to-variance-hyp-Ln}, and thus one can say that there is an absence of Barren Plateaus for CPSA ansätze (\cref{def:CPSA}) with $L_n = n^{q_\Delta + q_k + 1}$ repetitions, for angles located at any constant fraction of the depth.
\end{thm}

\begin{proof}
Given choices of sequences $(L_n)$ and $(l_n)$ that satisfy the assumptions of the theorem, define the third sequence $(\tilde{l}_n)$ in accordance to the different "coordinates systems" $(l,j)\leftrightarrow(\tilde{l},\tilde{j})$ discussed around \cref{eq:depth-coordinates-correspondance-1,eq:depth-coordinates-correspondance-2}, i.e. by:
\begin{equation}
    \tilde{l}_n := L_n - l_n +1\,.
\end{equation}
Due to \cref{eq:concluding-back-to-variance-hyp-Ln}, the sequences $(l_n)$ and $(\tilde{l}_n)$ are both in $\Omega( \Delta_n^{-1} \, d_{k,n} \, n)$, and hence so are the sequences $(l_n -1)_{n\geq2}$ and $(\tilde{l}_n -1)_{n\geq2}$.
Furthermore, since for each $n$ the pattern $U_{0,n}$ is assumed to be connected, the associated stochastic matrices $T_n:=T^{(U_{0,n}^k)}$ are all primitive by \cref{cor:connected-section-cor1}, and they are all doubly-stochastic as well by \cref{lem:T-is-doubly-stoch}.

Therefore, one can apply \cref{thm:stoch-mat-sequence-convergence} to $(T_n)_{n\geq2}$, and with either sequences of powers $(l_n -1)_{n\geq2}$ or $(\tilde{l}_n -1)_{n\geq2}$. Doing so separately, using both of them, yields respectively constants $c,\tilde{c}>0$ such that
\begin{align}
    \lVert T_n^{l(n)-1} - T_{n,\infty} \rVert_{\mathrm{ew},\infty} 
    &\in \mathcal{O}\left(1/\exp(c\, n)\right)\,,\\
    \lVert T_n^{\tilde{l}(n)-1} - T_{n,\infty} \rVert_{\mathrm{ew},\infty} 
    &\in \mathcal{O}\left(1/\exp(\tilde{c}\, n)\right)\,.
\end{align}  
Thus, letting $c':=\min(c,\tilde{c})$:
\begin{equation}\label{eq:concluding-back-to-variance-proof-eq1}
    \lVert T_n^{l(n)-1} - T_{n,\infty} \rVert_{\mathrm{ew},\infty},\;\;\lVert T_n^{\tilde{l}(n)-1} - T_{n,\infty} \rVert_{\mathrm{ew},\infty}
    \;\in\; \mathcal{O}\left(1/\exp(c'\, n)\right)\,.
\end{equation}

Combining \cref{eq:concluding-back-to-variance-proof-eq1} with \cref{eq:variance-difference-with-idealvariance-general-upper-bound,lemma:variance-error-terms-first-general-bound}, and with $\lVert \cdot \rVert_{\mathrm{Sc},2} \leq d_{k,n}\lVert \cdot \rVert_{\mathrm{ew},\infty}$ , one obtains:
\begin{equation}\label{eq:variance-difference-with-idealvariance-general-upper-bound--re}
     \left| \mathrm{Var}_{\bm{\theta}}[\partial_{\theta_{\lambda(l_n,j)}} \mathcal{C}] - \frac{1}{d_{k,n}} \frac{8 k (n-k)}{n(n-1)} \right| 
     \in
     \mathcal{O}\left( \frac{8 k (n-k)}{n(n-1)} \left[ \frac{1}{\exp(c'\,n)} + \frac{1}{\exp(c'\,n)} + \frac{2 \, d_{k,n}}{\exp(2\,c'\,n)} \right] \right)\,,
 \end{equation}
and hence, since 
\begin{equation}
    \frac{8 k (n-k)}{n(n-1)}  \in \mathcal{O}(1)
\end{equation}
and
\begin{equation}
    \left[ \frac{1}{\exp(c'\,n)} + \frac{1}{\exp(c'\,n)} + \frac{2 \, d_{k,n}}{\exp(2\,c'\,n)} \right]  \in \mathcal{O}\left(\frac{1}{\exp(c'\,n)}\right)\,,
\end{equation}
one gets 
\begin{equation}\label{eq:variance-difference-with-idealvariance-general-upper-bound--last-proof-eq}
     \left| \mathrm{Var}_{\bm{\theta}}[\partial_{\theta_{\lambda(l_n,j)}} \mathcal{C}] - \frac{1}{d_{k,n}} \frac{8 k (n-k)}{n(n-1)} \right| 
     \in
     \mathcal{O}\left(\frac{1}{\exp(c'\,n)}\right)\,.
 \end{equation}
And thus, since
 \begin{equation}
     \frac{1}{d_{k,n}} \frac{8 k (n-k)}{n(n-1)} \;\in\;
     \Theta\left(\frac{1}{\mathrm{poly}(n)}\right)\,,
 \end{equation}
 \cref{eq:variance-difference-with-idealvariance-general-upper-bound--last-proof-eq} implies that
 \begin{equation}
     \mathrm{Var}_{\bm{\theta}}[\partial_{\theta_{\lambda(l_n,j)}} \mathcal{C}]
     \;\in\;
     \Theta\left(\frac{1}{\mathrm{poly}(n)}\right)\,.
 \end{equation}

Lastly, if \cref{conj:spectral-gap} holds, it implies (since $d_{k,n}\in \mathcal{O}\left(\mathrm{poly}(n)\right)$ that $\left( \Delta_n^{-1} \, d_{k,n} \, n\right) \in \mathcal{O}\left(\mathrm{poly}(n)\right)$, implying that indeed $\Omega\left( \Delta_n^{-1} \, d_{k,n} \, n\right) \,\cap\, \mathcal{O}\left(\mathrm{poly}(n)\right)$ is non-empty, thereby justifying the existence of sequences $(L_n)$ being in both $\Omega\left( \Delta_n^{-1} \, d_{k,n} \, n\right)$ and $\mathcal{O}\left(\mathrm{poly}(n)\right)$.
\end{proof}

\subsection{Numerical evidence supporting \cref{conj:spectral-gap}}\label{subsec:numerical-evidence-spectral-gap}

We consider three sequences of connected patterns $(U_{0,n})_{n\geq2}$ here, labeled \texttt{line-down}, \texttt{line-up}, and \texttt{pyramid}.
The circuit of \texttt{line-down} consists of a cascading line of $J=n-1$ RBS gates, going downwards and rightwards (the first gate RBS gate connects qubit n$^{\circ}$1 to qubit n$^{\circ}$2, and so on).
The circuit of \texttt{line-downup} consists of the previous cascading line of $n-1$ RBS gates, followed by a second cascading line of $n-2$ RBS gates, this time going back upwards and rightwards, for a total of $J=2n -3$ gates.
The circuit of \texttt{pyramid} is an arrangement of $J=n(n-1)/2$ RBS gates into a "triangle" (see e.g. \cite[Section 2.3.2]{Landman2022}).

These three RBS patterns are all connected (\cref{def:connected-pattern}). Note that there is no need to consider FBS gates at all because they have identical stochastic matrices than those of RBS gates.

For all three of these RBS patterns $(U_{0,n})_{n\geq2}$, for the Hamming weight values $k=2,3$, and for qubit counts $n\in\llbracket4,50\rrbracket$, we numerically construct their associated stochastic matrices $T_n:=T^{(U_{0,n}^k)}$ (\cref{eq:stochastic-matrix-T-def}), and we numerically evaluate the eigenvalues of $T_n$, from which we deduce their spectral gap values $\Delta_n=1-|\lambda_2|$. (All of this is performed using \textit{Numpy}, in double precision.)

As a side remark, note that the patterns \texttt{line-downup} and \texttt{pyramid} are both palindromes (\cref{lem:connected-section-palindrome}), however the pattern  \texttt{line-down} is not.
In fact we checked with the symbolic computation software \textit{Mathematica} that in general the stochastic matrices $T_n$ associated to the \texttt{line-down} pattern are neither symmetric, nor normal, and not even diagonalizable. (Eigenvalues are still well-defined, of course, even for non-diagonalizable matrices.)

We then plot the obtained $\Delta_n$ values as a function of the number of qubits $n$. Our \cref{conj:spectral-gap} is claiming that $\Delta_n$ should be decaying at most polynomially fast (and not faster), which corresponds to the claim that graphically, on a "loglog" plot (where both the $x$ and $y$ axes have logarithmic scaling), $\Delta_n$ should vanish "at most in a straight descending line" (and not faster).
If however, on a "semilog" plot (where the $x$ axis has a regular scaling and the $y$ axis has a logarithmic scaling), we were to observe that $\Delta_n$ vanishes "in a straight descending line" (or faster), then it would indicate an exponentially-vanishing trend.
Hence, we perform two types of linear regressions, corresponding to both the semilog and loglog plot types just described, to respectively assess how good does a polynomial decay fit to the data, and how bad does an exponential decay fit to the data.
Since we are trying to evaluate the \emph{asymptotic} nature of the decays anyways, we offset the start of the fitted region to $n=20$ (in hopes of better matching the asymptotic regime of the data, but while still keeping a good amount of data points).
The $r^2$ value quantifies how well the respective model fits the data (the closer to $1$, the better of a fit).

\begin{figure}[h!]
    \centering
    \includegraphics[width=1\textwidth]{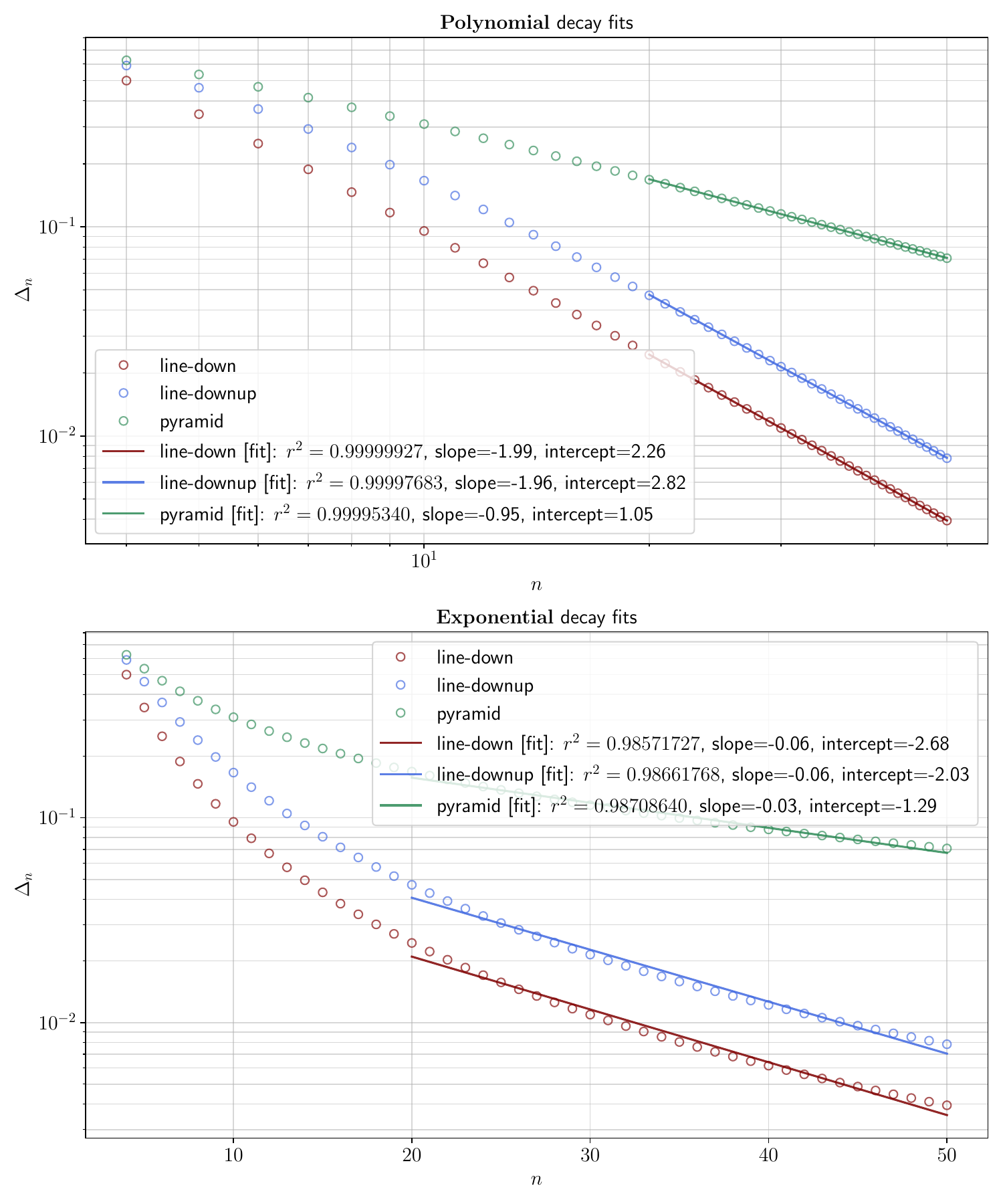}
    \caption{Numerical evidence for the inverse-polynomial largeness of the spectral gap $\Delta_n$; Hamming-weights of $k=1$ and $k=2$ both produce this exact figure.}
    \label{fig:numerical-evidence-spectral-gap-k=2}
\end{figure}

The results are presented in \cref{fig:numerical-evidence-spectral-gap-k=2}. In fact, the plots we obtain for $k=1$ or $k=2$ have no perceivable difference at all -- numerically, we observe differences in data point values of order $10^{-13}$. Therefore, we only plot one of them to avoid an unnecessary "duplicate" figure, but the whole \cref{fig:numerical-evidence-spectral-gap-k=2} (including the shown parameters of the fitting results) is to be taken for both cases of $k=1$ and $k=2$.\footnote{Perhaps we could hence add to \cref{conj:spectral-gap} that either the spectral gaps values $\Delta^{(k)}_n$ are strictly independent of $k$, or that their differences for different values of $k$ vanishes exponentially with $n$.}
As a side note, we observe numerically for these two cases that, even though the second largest eigenvalue moduli $|\lambda_2|$ have practically identical values (up to some $10^{-13}$) throughout all the values $n$, the third largest eigenvalue moduli $|\lambda_3|$ still differ significantly for the first few values of $n$.

For each of the three patterns (and for both $k=1,2$), the $r^2$ values obtained indicate that a polynomial decay trend fits the spectral gap data much better than an exponential decay trend (as the polynomial $r^2$ values are closer to $1$ than the exponential $r^2$ values are, by multiple orders of magnitude), thereby supporting our \cref{conj:spectral-gap}. The better fitting of the polynomial decay regressions may also be appreciated visually on these plots.

\subsection{A priori differences between $2$-design arguments and \cref{thm:concluding-back-to-variance}}\label{subchap:design-discussion}
For the purpose of discussion, let us recall some definitions (c.f. e.g. \cite{Larocca2021}).
A VQC $U(\bm{\theta})$ (with some given probability distribution $\mu_\Theta$ on the angles $\bm{\theta}$), defined over a $d$-dimensional quantum system, is said to be an \emph{$\epsilon$-approximate $2$-design} if
\begin{equation}
\lVert \mathcal{A}_{U} \rVert_{\mathrm{Sc},\infty}\leq \epsilon\,,
\end{equation}
where we introduced the linear operator
\begin{equation}
\mathcal{A}_{U} := 
\mathbb{E}_{U \sim \mu_{\mathrm{Haar}}}[ U^{\otimes 2} \otimes  \overline{U}\,^{\otimes 2} ] - \mathbb{E}_{\bm\theta \sim \mu_\Theta}[ U(\bm{\theta})^{\otimes 2} \otimes  \overline{U(\bm{\theta})}\,^{\otimes 2} ]\,,
\end{equation}
with $\overline{U}$ denoting the entry-wise complex-conjugation of the matrix $U$, and $\mu_{\mathrm{Haar}}$ the Haar measure on $\mathrm{U}(d)$.

Since our result \cref{thm:concluding-back-to-variance} of absence of Barren Plateaus requires a number of repetitions $L_n$ that is polynomially large in the qubit number $n$ (at least of order $\Delta_n^{-1} \, d_{k,n} \, n$) to hold, one may wonder if this polynomial repetition number $L_n$ is already enough to guarantee the overall circuit to be an \emph{$\epsilon_n$-approximate $2$-design} with $\epsilon_n\in\mathcal{O}\left( 1/\exp(\alpha\,n) \right)$ (for some $\alpha>0$). As, if that is the case, then (by definition of being an $\epsilon_n$-approximate $2$-design), the variance quantity we study, being the variance of an observable expectation cost, can be approximated to order $\mathcal{O}(\epsilon_n)$ by the corresponding variance taken over the Haar ensemble of $d_k \times d_k$ unitaries (see e.g. \cite[Appendix D]{holmes2022connecting}). Since the latter exact Haar variance, which can be calculated using formulas derived from Weingarten calculus of the unitary Haar measure (see e.g. \cite[Appendix E.1]{holmes2022connecting}), ought to coincide with our asymptotic (polynomially-vanishing) variance value of
\begin{equation}
\frac{1}{d_{k,n}} \frac{8 k (n-k)}{n(n-1)}\,,
\end{equation}
one would be able to derive that our variance quantity of study lies in $\Theta( \frac{1}{d_{k,n}} \frac{8 k (n-k)}{n(n-1)} )$ just from the fact that the polynomial repetition number of the pattern leads to an $\mathcal{O}( 1/\exp(\alpha\,n))$-approximate $2$-design -- without a need to resort to our \cref{thm:concluding-back-to-variance}.

However, it is not at all immediate if $L_n \in \Omega(\Delta_n^{-1} \, d_{k,n} \, n)$ repetitions of a connected RBS/FBS ansatz $U_{0,n}$ suffices to guarantee that the total unitary is an $\mathcal{O}( 1/\exp(\alpha\,n))$-approximate $2$-design (for some $\alpha>0$).
In fact, the existence of a repetition number $L_n \in O(\mathrm{poly}(n))$ such that the total unitary is an $\mathcal{O}( 1/\exp(\alpha\,n))$-approximate $2$-design (for some $\alpha>0$) is equivalent (due to \cite[Section 4.1]{Larocca2021}) to:
\begin{equation}\label{eq:design-discussion-exp-2-design-underlying-hyp}
\ln\left( \frac{1}{\lVert \mathcal{A}_{U_{0,n}} \rVert_{\mathrm{Sc},\infty}} \right) \in \Omega\left( \frac{1}{\mathrm{poly}(n)} \right)\,.
\end{equation}
It is not immediate to theoretically prove that \cref{eq:design-discussion-exp-2-design-underlying-hyp} holds in our setting, and even assessing its numerical validity may not be so straightforward, as constructing the operator $\mathcal{A}_{U_{0,n}}$ numerically could be costly. In fact, we are not aware of any existing literature exploring numerically the validity of \cref{eq:design-discussion-exp-2-design-underlying-hyp} for any setting of VQC ansätze.

In contrast, in this work's \cref{thm:concluding-back-to-variance}, it is the assumption
\begin{equation}\label{eq:design-discussion-our-hyp}
\Delta_n \in \Omega\left( \frac{1}{\mathrm{poly}(n)} \right)\,
\end{equation}
(i.e. our spectral gap \cref{conj:spectral-gap}) that guarantees that the variance lies in $\Theta( \frac{1}{d_{k,n}} \frac{8 k (n-k)}{n(n-1)} )$.
This spectral gap quantity $\Delta_n$ is conceptually simpler (for instance, it does not directly involve any probability measures, like $\mathcal{A}_{U_{0,n}}$ does), it is simpler to evaluate numerically (up to the difficulty of numerically evaluating eigenvalues of $d_{k,n}$-sized matrices), and doing so we were able to offer direct numerical evidence for the validity of \cref{eq:design-discussion-our-hyp} in the previous \cref{subsec:numerical-evidence-spectral-gap}, for several RBS/FBS ansätze.

Exploring whether the assumption responsible for the fast-enough convergence of second moments to that of the Haar measure (\cref{eq:design-discussion-exp-2-design-underlying-hyp}), and the assumption responsible for the fast-enough convergence of stochastic matrix powers (\cref{eq:design-discussion-our-hyp}), are actually equivalent for given RBS/FBS patterns, would be an interesting future direction of work.

Lastly, let us note that our \cref{thm:NoBPgeneralcase} makes no assumptions on the number of repetitions $L$, and so its result may not be obtained in any way using closeness to unitary $2$-design assumptions.

\section{Proof of Theorem~\ref{thm:NoBPgeneralcase}}\label{chap:proof_BP_general}

We recall the Theorem~\ref{thm:NoBPgeneralcase}:

\NoBPgeneralcase*
\begin{proof}
    
According to Lemma~\ref{lemma:VarianceHWPreserving}, we have:
\begin{equation}
    \mathbb{E}_{\bm{\theta}}[\partial_{\theta_i} \mathcal{C}] = 0 \, \textrm{,}
\end{equation}
\begin{equation}
    \mathrm{Var}_{\bm{\theta}}[\partial_{\theta_i} \mathcal{C}] =  2 \sum_{(l,j)} \left(\int_{\bm{\theta} \in \Theta} (\frac{1}{2\pi})^D (\zeta^{\lambda}_l)^2 + (\zeta^{\lambda}_j)^2 d\bm{\theta}  \right) \cdot \left(\int_{\bm{\theta} \in \Theta} (\frac{1}{2\pi})^D (\tilde{y}^{\lambda}_l)^2 + (\tilde{y}^{\lambda}_j)^2 d\bm{\theta} \right) \, \textrm{.}
\end{equation}

By the assumption on the input state $\zeta^0$'s distribution, we have:
    \begin{equation}
        \forall r \in [d_k], \quad \mathbb{E}_{\zeta^0,y} \left[\int_{\bm{\theta} \in \Theta} (\frac{1}{2\pi})^D (\zeta_r^{0})^2 d\bm{\theta}\right] = \frac{1}{d_k} \, \textrm{.}
    \end{equation}
From the recurrence relation given by Eq.~\eqref{eq:Inductive_relation} of Lemma~\ref{lemma:InductiveRelation}, it follows that:
    \begin{equation}\label{eq:thmNoBPgeneralcase-proof-eq-1}
        \forall \lambda \in \llbracket 0, \lambda_{\max} \rrbracket, \;\forall r \in [d_k], \quad \mathbb{E}_{\zeta^0,y} \left[\int_{\bm{\theta} \in \Theta} (\frac{1}{2\pi})^D (\zeta_r^{\lambda})^2 d\bm{\theta} \right] = \frac{1}{d_k} \, \textrm{.}
    \end{equation}

Indeed, to be more explicit, we have using the notations of \cref{subchap:variance-formula-recast} (\cref{eq:repeated-recurrence-relation-forward}):
\begin{align}
    \overrightarrow{Z}^{\lambda} &= T_{\text{rest},\lambda} \cdot T^{l(\lambda)-1} \ \cdot\ \overrightarrow{Z^0}\,,\\
\intertext{and hence}
    \mathbb{E}_{\zeta^0,y}\big[\overrightarrow{Z}^{\lambda}\big]
    &= \big(T_{\text{rest},\lambda} \cdot T^{l(\lambda)-1}\big) \ \cdot\ \mathbb{E}_{\zeta^0,y}\big[\overrightarrow{Z^0}]\\
     &= \big(T_{\text{rest},\lambda} \cdot T^{l(\lambda)-1}\big) \ \cdot\ (1/d_k,\dots,1/d_k)^\intercal\\
     &= (1/d_k,\dots,1/d_k)^\intercal\,,
\end{align}
where the last equality follows from the fact that $\big(T_{\text{rest},\lambda} \cdot T^{l(\lambda)-1}\big)$ is row-stochastic (c.f. \cref{lem:T-is-doubly-stoch}).

As explained in Section \ref{subchap:InductiveRelation}, the recurrence relation given by Eq.~\eqref{eq:Inductive_relation} can also be applied for the backpropagation of the target state $\tilde{y}$, and so analogously as above, we find, due to the assumption on the target state $y$'s distribution, that:
    \begin{equation}\label{eq:thmNoBPgeneralcase-proof-eq-2}
        \forall \lambda \in \llbracket 0, \lambda_{\max} \rrbracket, \;\forall r \in [d_k], \quad \mathbb{E}_{\zeta^0,y} \left[\int_{\bm{\theta} \in \Theta} (\frac{1}{2\pi})^D (\tilde{y}_r^{\lambda})^2 d\bm{\theta} \right] = \frac{1}{d_k} \, \textrm{.}
    \end{equation}

Using Eqs.~\eqref{eq:thmNoBPgeneralcase-proof-eq-1} and \eqref{eq:thmNoBPgeneralcase-proof-eq-2}, Lemma \ref{lemma:VarianceHWPreserving} yields, for all $\lambda$:
\begin{equation}
    \mathrm{Var}_{\bm{\theta}}[\partial_{\theta_{\lambda}} \mathcal{C}(\bm{\theta})] = \mathbb{E}_{\zeta^0,y}  \left[ 2 \sum_{l,j} \left( \frac{1}{\left(2 \pi\right)^D} \int_{\bm{\theta}} (\zeta_l^{\lambda})^2 + (\zeta_j^{\lambda})^2 d\bm{\theta} \right)
    \cdot \left( \frac{1}{\left(2 \pi\right)^D} \int_{\bm{\theta}} (\tilde{y}_l^{\lambda})^2 + (\tilde{y}_j^{\lambda})^2 d\bm{\theta} \right) \right] = 2 \sum_{l,j} \frac{4}{d_k^2}\,.
\end{equation}

Each $(l,j)$ represents the indices of two basis states that are involved in the rotations created by the RBS/FBS gate corresponding to the inner layer $\lambda$. Considering $n$ qubits and a Hamming weight of $k$, there are $\binom{n-2}{k-1}$ such different pairs which are involved in rotations. And since
    \begin{equation}
        \binom{n-2}{k-1} = \frac{(n-2)!}{(k-1)! (n-1-k)!} = \frac{k(n-k)}{n(n-1)} \binom{n}{k} = \frac{k(n-k)}{n(n-1)} d_k \,,
    \end{equation}
we can conclude that for any  $\lambda \in \llbracket0, \lambda_{\mathrm{max}}\rrbracket$:
    \begin{equation}
        \begin{aligned}
            \mathbb{E}_{\zeta^0,y} \mathrm{Var}_{\bm{\theta}}[\partial_{\theta_{\lambda}} \mathcal{C}(\bm{\theta})] &= \frac{k(n-k)}{n(n-1)}\frac{8}{d_k}\,.
        \end{aligned}
    \end{equation}
\end{proof}

\end{document}